\documentclass[11pt]{article}
\pdfoutput=1
\usepackage{jheppub}
\usepackage{graphicx}
\usepackage{amssymb}
\usepackage{amsmath}
\usepackage{amsfonts}
\usepackage{latexsym}
\usepackage{hyperref}
\usepackage{physics}
\usepackage{bbm}
\usepackage{empheq}
\usepackage{tikz}
\usepackage{cancel}
\usepackage{hhline}
\usepackage{caption}
\usepackage{soul}
\usepackage{bbold} %new

\usepackage[T1]{fontenc}

\usetikzlibrary{arrows.meta}
\usetikzlibrary{decorations.markings}
\tikzset{->-/.style={decoration={
			markings,
			mark=at position #1 with {\arrow{>}}},postaction={decorate}}}

\newcommand\nn{\nonumber}
\newcommand\fft[2]{\frac{#1}{#2}}
\newcommand\mn{\mathfrak n}
\newcommand\mm{\mathfrak m}
\newcommand\mbDelta{\mathbb{\Delta}}

\newcommand\mA{\mathcal{A}}
\newcommand\mO{\mathcal{O}}
\newcommand\mN{\mathcal{N}}

\newcommand\mB{\mathcal{B}}
\newcommand\mC{\mathcal{C}}
\newcommand\mW{\mathcal{W}}
\newcommand\mI{\mathcal{I}}
\newcommand\mS{\mathcal{S}}
\newcommand\mZ{\mathcal{Z}}
\newcommand\mR{\mathcal{R}}
\newcommand\ri{\mathfrak{i}}

\newcommand\mfg{\mathfrak{g}}
\newcommand\mfs{\mathfrak{s}}
\newcommand\U{\text{U}}
\newcommand\hu{\hat{u}}
\newcommand\hv{\hat{v}}
\newcommand\hbu{\hat{\bu}}
\newcommand\hbv{\hat{\bv}}
\newcommand\bu{\bar{u}}
\newcommand\bv{\bar{v}}

\newcommand\bS{\overline{S}}

\newcommand\bDelta{\overline{\Delta}}
\newcommand\bmZ{\overline{\mZ}}
\newcommand\bmA{\overline{\mA}}
\newcommand\bmB{\overline{\mB}}
\newcommand\bmC{\overline{\mC}}
\newcommand\bvarphi{\bar{\varphi}}

\newcommand\bmS{\overline{\mS}} %%%%%new
\newcommand\bphi{\bar{\phi}} %%%%%new
\newcommand\bmbDelta{\overline{\mbDelta}} %%%%%new
\newcommand\tih{\tilde{h}}

\newcommand\tmm{\tilde{\mathfrak{m}}}
\newcommand\tDelta{\widetilde{\Delta}}
\newcommand\trho{\widetilde{\rho}}
\newcommand\tu{\tilde{u}}
\newcommand\tf{\tilde{f}}

\newcommand\tQ{\widetilde{Q}} %%%%%new
\newcommand\tb{\tilde{b}} %%%%%new
\newcommand\tv{\tilde{v}} %%%%%new
\newcommand\htv{\hat{\tv}} %%%%%new
\newcommand\htu{\hat{\tu}} %%%%%new
\newcommand\tphi{\tilde{\phi}} %%%%%new
\newcommand\tomega{\tilde{\omega}} %%%%%new
\newcommand\tvarphi{\tilde{\varphi}} %%%%%new
\newcommand\tmA{\widetilde{\mathcal{A}}} %%%%%new
\newcommand\tmB{\widetilde{\mathcal{B}}} %%%%%new
\newcommand\tmC{\widetilde{\mathcal{C}}} %%%%%new
\newcommand\boldmm{\boldsymbol{\mm}}
\newcommand\boldmn{\boldsymbol{\mn}}
\newcommand\boldz{\boldsymbol{z}}
\newcommand\boldm{\boldsymbol{m}}
\newcommand\boldh{\boldsymbol{h}}
\newcommand\boldtih{\boldsymbol{\tih}}
\newcommand\boldu{\boldsymbol{u}}
\newcommand\boldbu{\boldsymbol{\bu}}
\newcommand\boldhu{\boldsymbol{\hu}}
\newcommand\boldhbu{\boldsymbol{\hbu}}
\newcommand\boldv{\boldsymbol{v}}
\newcommand\boldbv{\boldsymbol{\bv}}
\newcommand\boldhv{\boldsymbol{\hv}}
\newcommand\boldhbv{\boldsymbol{\hbv}}
\newcommand\boldlambda{\boldsymbol{\lambda}}
\newcommand\boldtmm{\boldsymbol{\tmm}}
\newcommand\boldDelta{\boldsymbol{\Delta}}

\newcommand\boldxi{\boldsymbol{\xi}}

\newcommand\boldk{\boldsymbol{k}}
\newcommand\boldvarphi{\boldsymbol{\varphi}}
\newcommand\boldbvarphi{\boldsymbol{\bvarphi}}
\newcommand\boldhtu{\boldsymbol{\htu}} %%%%%new
\newcommand\boldhtv{\boldsymbol{\htv}} %%%%%new
\newcommand\boldtvarphi{\boldsymbol{\tvarphi}} %%%%%new
\newcommand\boldD{\boldsymbol{D}} %%%%%new
\newcommand{\boldmbDeltaaux}[2]{%
	\ooalign{$#1\mbDelta$\cr\kern0.4pt$#1\mbDelta$\cr\kern0.7pt$#1\mbDelta$\cr}}
\newcommand{\boldmbDelta}{\mathpalette\boldmbDeltaaux\relax} %%%%%new
\makeatletter
\newcommand*{\rom}[1]{\expandafter\@slowromancap\romannumeral #1@}
\makeatother

\begin{document}
	
%\preprint{KIAS-P24009}
\title{Towards OSV in AdS}
	
\author[a]{Nikolay Bobev,}
\author[b]{Sunjin Choi,}
\author[c]{Junho Hong,}
\author[d]{and Valentin Reys}

\affiliation[a]{Institute for Theoretical Physics and Leuven Gravity Institute, KU Leuven\,,\\ Celestijnenlaan 200D, B-3001 Leuven, Belgium}
\affiliation[b]{Kavli Institute for the Physics and Mathematics of the Universe (WPI), \\ 
The University of Tokyo Institutes for Advanced Study, The University of Tokyo, \\
Kashiwa, Chiba 277-8583, Japan}
\affiliation[c]{Department of Physics \& Center for Quantum Spacetime, Sogang University\,,\\ 35 Baekbeom-ro, Mapo-gu, Seoul 04107, Republic of Korea}
\affiliation[d]{Laboratoire de Physique de l’Ecole normale sup\'erieure\,\\ 
CNRS, PSL Research University and Sorbonne Universit\'es, 24 rue Lhomond, 75005 Paris, France}

\emailAdd{nikolay.bobev@kuleuven.be}
\emailAdd{sunjin.choi@ipmu.jp}
\emailAdd{junhohong@sogang.ac.kr}
\emailAdd{valentin.reys@phys.ens.fr}

%%%%%
	
\abstract{We use supersymmetric localization for 3d $\mathcal{N}=2$ SCFTs to establish a relation of the form $Z_{S^1\times S^2} \sim |Z_{S^3_b}|^2$ between the superconformal index and the squashed three-sphere partition function. This applies to general 3d $\mathcal{N}=2$ SCFTs and  is derived using a saddle point approximation in the Cardy-like limit of small $S^1$ radius and large squashing parameter. We also show a similar relation between the topologically twisted index and the squashed sphere partition function. In the context of holography our results lead to a relation between the partition function of supersymmetric asymptotically AdS$_4$ black holes and the large $N$ limit of $Z_{S^3_b}$ of the dual SCFT. Since the latter encodes the gauged supergravity prepotential, this result is akin to the OSV conjecture for asymptotically flat black holes. We confirm this relation in detail for SCFTs arising from M2-branes using recent large $N$ results from supersymmetric localization. We also briefly discuss a similar relation for 5d SCFTs and its implications for asymptotically AdS$_6$ black holes.}
	
\maketitle \flushbottom

\newpage

%%%%%
\section{Introduction}
\label{sec:intro}
%%%%%

Supersymmetric QFTs provide a rich arena on which to establish, test and explore dualities by leveraging the constraints of supersymmetry to explicitly calculate physical observables. Supersymmetric localization for QFTs on compact Euclidean manifolds is a powerful tool to facilitate such explicit calculations which has led to numerous new insights into the structure of dualities. Our goal here is to use supersymmetric localization to establish relations between three seemingly different partition functions of 3d $\mathcal{N}=2$ SCFTs placed on compact Euclidean manifolds. When applied to large $N$ holographic SCFTs these relations lead to new insights into holography and the physics of supersymmetric asymptotically AdS$_4$ black holes.

Two of the partition functions of interest here are the superconformal index (SCI), $\mI_{\rm SCI}$, and the squashed $S^3$ partition function, $Z_{S^3_b}$.  The SCI can be defined as a Euclidean supersymmetric path integral on $S^1\times S^2$ that depends on various continuous parameters compatible with supersymmetry that can be interpreted as chemical potentials when the index is viewed as a trace over the Hilbert space of $\frac{1}{16}$-BPS states, i.e. states preserving two real supercharges, in the SCFT in radial quantization. Importantly, the SCI is a non-trivial function of the radius $\beta$ of the $S^1$ and in much of the discussion below we will study the SCI in the Cardy-like limit of small $\beta$. The supersymmetric $S^3$ partition function of the SCFT can be deformed by squashing and real mass parameters and we will consider the $\U(1)\times \U(1)$ invariant squashing deformation in the large squashing limit. Using the supersymmetric localization matrix models for these partition functions we show that the following relation holds
\begin{equation}\label{eq:IZZbintro}
	\mI_{\rm SCI}(\beta;\boldxi) \sim Z(b;\boldm)\,\overline{Z}(b;\boldm)\,,
\end{equation}
where $\boldxi$ and $\boldm$ denote the flavor symmetry fugacities and real mass parameters, respectively. The symbol $\sim$ in~\eqref{eq:IZZbintro} needs clarification. We show that~\eqref{eq:IZZbintro} holds to all orders in the small $\beta$ and small $b$ perturbative expansion\footnote{Note that $Z_{S^3_b}$ is symmetric under the transformation $b \to 1/b$, therefore both the limits of small and large $b$ correspond to large squashing deformations of $S^3$.} provided we identify $\beta = \pm\pi \ri b^2$, appropriately map the parameters $\boldxi$ to $\boldm$ and carefully define the conjugate $\overline{Z}(b;\boldm)$. To show this relation we use a saddle point approximation in the limit of small $\beta$ and the right hand side of~\eqref{eq:IZZbintro} may need to be summed over saddle points. To be concrete and explicit we show~\eqref{eq:IZZbintro} for 3d $\mathcal{N}=2$ asymptotically free theories that can be formulated in the UV in terms of standard Lagrangians and that flow to the SCFT in the IR. 

We test and illustrate the relation~\eqref{eq:IZZbintro} on three different SCFTs. For a theory of a single $\mathcal{N}=2$ chiral superfield with general R-charge assignment we can directly evaluate both sides of the relation in the limit of small $\beta$ and $b$ and explicitly show its validity. For the $\U(N)_k\times \U(N)_{-k}$ $\mathcal{N}=6$ ABJM theory and the $\U(N)$ $\mathcal{N}=4$ SYM theory coupled to one adjoint and $N_f$ fundamental hypermultiplets (known also as the ADHM model) our calculations are more involved.  To evaluate the SCI for these theories arising from M2-branes we focus on the ``M-theory'' limit of large $N$ and fixed $k$ and $N_f$. We then use the results of~\cite{Bobev:2022wem} to find the first two orders of the SCI in the small $\beta$ expansion to all orders in the perturbative $1/N$ expansion. This result is then successfully compared with the right hand side of~\eqref{eq:IZZbintro} which is evaluated by employing the conjectured Airy form of the squashed $S^3$ partition function discussed in~\cite{Bobev:2025ltz}. 

The topologically twisted index (TTI) is a different supersymmetric partition function one can define on $S^1\times S^2$ distinguished by the presence of a partial topological twist by the superconformal R-symmetry along the $S^2$. We provide evidence that this partition function also enjoys a relation of the schematic form
\begin{equation}\label{eq:ITTIZZbintro}
	\mI_{\rm TTI}(\beta;\boldxi) \sim Z(b;\boldm)\,\widetilde{Z}(b;\boldm)\,,
\end{equation}
in the small $\beta$ limit accompanied by a large squashing expansion of $Z(b;\boldm)$. Notably, the conjugation rule that defines $\widetilde{Z}(b;\boldm)$ in terms of $Z(b;\boldm)$ in~\eqref{eq:ITTIZZbintro} is not the same as the one used for $\overline{Z}(b;\boldm)$ in~\eqref{eq:IZZbintro}. This small difference leads to important modifications in the structure of the final answer for the TTI. For vanishing angular fugacity on the $S^2$ the TTI is independent of the circle radius $\beta$ and therefore for~\eqref{eq:ITTIZZbintro} to hold there need to be non-trivial cancellations in the $b$-dependence of the product of the two squashed sphere partition functions. Indeed, the relation~\eqref{eq:ITTIZZbintro} can be explicitly confirmed for the chiral superfield, ABJM, and ADHM theories. While the relation for the chiral superfield can be shown by direct calculation, the ABJM and ADHM relations are more involved. Nevertheless,~\eqref{eq:ITTIZZbintro} can be confirmed to all orders in the $1/N$ perturbative expansion by using the precise determination of the TTI reported in~\cite{Bobev:2022jte,Bobev:2022eus,Bobev:2023lkx} together with the Airy conjecture for $Z(b;\boldm)$~\cite{Bobev:2025ltz}. Notably, this comparison includes the order $N^0$ term in the large $N$ expansion which, as we discuss further below, is a non-trivial consistency check of the results we report here as well as those in~\cite{Bobev:2022jte,Bobev:2022eus,Bobev:2023lkx} and~\cite{Bobev:2025ltz}. 

While our results are derived using supersymmetric localization methods in field theory they admit an interesting interpretation in the context of AdS/CFT. For holographic SCFTs the SCI is dual to a Euclidean asymptotically AdS$_4$ supergravity solution that can be viewed as a Euclidean continuation of the Kerr-Newman black hole. Similarly, the TTI describes holographically a Euclidean supergravity saddle point related to supersymmetric static dyonic black holes in AdS$_4$. The relations in~\eqref{eq:IZZbintro} and~\eqref{eq:ITTIZZbintro} then translate into an expression for the partition functions of these black holes of the schematic form
\begin{equation}\label{eq:ZBHZ2}
	Z_{\rm BH} \sim |Z_{S^3}|^2\,.
\end{equation}
This schematic relation clearly resembles the OSV conjecture~\cite{Ooguri:2004zv} for 4d asymptotically flat supersymmetric black holes arising from Type II string theory compactifications on CY$_3$ manifolds, see~\cite{Pioline:2006ni,Guica:2007wd} for reviews. In the OSV relation, however, the right hand side contains the square of the topological string partition function on the CY$_3$ manifold, while in the AdS$_4$ context what transpires from the relations we derive is a square of the $S^3$ partition function of the holographically dual SCFT. In hindsight, this apparent difference is perhaps not too surprising. First, the reason for the appearance of the topological string partition function on the right hand side of the OSV relation is due to the fact that it determines the prepotential of the 4d $\mathcal{N}=2$ ungauged supergravity associated to a given CY$_3$ compactification to all orders in the derivative expansion~\cite{Bershadsky:1993cx,Antoniadis:1993ze}. Similarly, the $S^3$ free energy of a holographic SCFT as a function of real mass parameters is related to the prepotential of the dual gauged supergravity. This has been shown at the two-derivative level in~\cite{Zan:2021ftf} and conjectured to be true also in the presence of higher-derivative corrections, see~\cite{Bobev:2021oku,Bobev:2022eus,Hristov:2021qsw,Hristov:2022lcw,Hristov:2022plc,BenettiGenolini:2026qdm}. In addition, for SCFTs arising from M2-branes, there are intriguing connections between $Z_{S^3}$ and topological strings on non-compact CY manifolds, see~\cite{Marino:2016new} for a review, and their equivariant generalizations~\cite{Cassia:2025aus,Cassia:2025jkr,Hristov:2026zjh}.

Our ability to check the relation~\eqref{eq:ZBHZ2} from the bulk string/M-theory is limited due to the familiar difficulties in calculating path integrals on non-trivial flux backgrounds. Nevertheless, progress can be made and we successfully test the relation~\eqref{eq:ZBHZ2} against available results in supergravity including higher-derivative, loop, and non-perturbative corrections. At the two-derivative supergravity level the relation can be demonstrated universally in minimal 4d $\mathcal{N}=2$ gauged supergravity without appealing to a particular embedding in string/M-theory, see~\cite{Bobev:2017uzs,Bobev:2019zmz} and references therein. Similarly, we check that the relation~\eqref{eq:ZBHZ2} is obeyed in 4-derivative minimal 4d $\mathcal{N}=2$ gauged supergravity by using the results in~\cite{Bobev:2020egg,Bobev:2021oku}. For AdS$_4$ vacua arising from M2-branes one can also confirm the validity of~\eqref{eq:ZBHZ2} at order $\log N$ in the large $N$ expansion by studying 1-loop corrections to 11d supergravity using the results in~\cite{Bhattacharyya:2012ye,Liu:2017vbl,PandoZayas:2020iqr,Bobev:2023dwx}. A final non-trivial consistency check can be performed at the non-perturbative level by showing that the recent results in~\cite{Gautason:2025per} for the leading $e^{-2\pi\sqrt{2N/k}}$ correction to $\log Z_{\rm BH}$, obtained by a semiclassical quantization of M2-brane instantons, are in perfect agreement with~\eqref{eq:ZBHZ2}. 

It is worth stressing that relations of the form~\eqref{eq:IZZbintro},~\eqref{eq:ITTIZZbintro}, and~\eqref{eq:ZBHZ2} were studied previously in the literature from various perspectives. The validity of~\eqref{eq:IZZbintro} and~\eqref{eq:ITTIZZbintro} in the leading and subleading orders in the Cardy-like expansion was shown using QFT methods in~\cite{Choi:2019zpz,Choi:2019dfu} and~\cite{Bobev:2022wem,Bobev:2024mqw}, respectively. It could also have been anticipated from the holomorphic block decomposition of Euclidean 3d $\mathcal{N}=2$ QFT path integrals described in~\cite{Pasquetti:2011fj,Beem:2012mb,Hwang:2012jh}. From the AdS$_4$ bulk perspective, the relation~\eqref{eq:ZBHZ2} has been discussed in 4d supergravity at the two-derivative~\cite{Azzurli:2017kxo,Bobev:2017uzs,Bobev:2019zmz,Hosseini:2019iad} and 4-derivative level~\cite{Bobev:2020egg,Bobev:2021oku,Bobev:2020zov} and conjectured to hold to higher orders in the supergravity derivative expansion~\cite{Hristov:2024cgj,Hristov:2021qsw,Hristov:2022lcw,Hristov:2022plc,BenettiGenolini:2026qdm} as well as the 1-loop perturbative level~\cite{Hristov:2021zai,Bobev:2023dwx}. Many of these previous results are limited to the first one or two leading orders in the Cardy limit or are based on educated guesses and conjectures. Our new results amount to a more rigorous derivation of~\eqref{eq:IZZbintro},~\eqref{eq:ITTIZZbintro}, and corresponding checks of~\eqref{eq:ZBHZ2}, to all orders in the Cardy-like/large squashing limits and all orders in the $1/N$ expansion together with non-perturbative corrections. 

We continue in the next section by introducing the SCI and the squashed $S^3$ partition functions of interest. In Section~\ref{sec:fact} we study the Cardy-like limit of the SCI and show that it can be expressed as a product of two $Z_{S^3_b}$ in the large $b$ limit. In Section~\ref{sec:ex} we illustrate this relation on three explicit examples of 3d $\mathcal{N}=2$ SCFTs. Section~\ref{sec:TTI} is devoted to exploring a similar factorization formula for the TTI in terms of a product of two $Z_{S^3_b}$ partition functions. In Section~\ref{sec:holo} we analyze the consequences of our field theory results for the dual holographic description and black hole physics and the analogy with the OSV conjecture. Some open questions and possible generalizations are discussed in Section~\ref{sec:discussion}. The two appendices contain some technical results on certain special functions as well as some comments on a similar relation between the 5d superconformal index and squashed sphere partition function.

%%%%%
\section{Supersymmetric partition functions}
\label{sec:susyloc}
%%%%%
In this section, we summarize the matrix models arising from supersymmetric localization for two of the partition functions of interest in this work:  
the $\U(1)\times\U(1)$ invariant squashed 3-sphere $S^3_b$ partition function and the $S^1\times S^2$ superconformal index. For the sake of concreteness, we focus on supersymmetric gauge theories admitting a UV Lagrangian description.

%%%%%
\subsection{Squashed 3-sphere partition function}
\label{sec:susyloc:S3b}
%%%%%
On the squashed 3-sphere preserving $\U(1)\times\U(1)$ isometry, the supersymmetric partition function of an $\mN=2$ Chern-Simons(CS)-matter quiver gauge theory can be expressed as a matrix model via supersymmetric localization \cite{Hama:2011ea,Imamura:2011wg}, in close analogy with the round-sphere case \cite{Kapustin:2009kz,Kapustin:2010xq,Jafferis:2010un}. The resulting expression reads
\begin{align}
	Z(b,\boldk,\boldhv)&=\fft{1}{|\mW|}\int_{-1/2}^{1/2} d\boldhu\,e^{-\pi\ri \sum_{\ell,n=1}^{r_G}k^{\ell n}\hu_\ell \hu_n-\pi\ri \sum_{x,y=1}^{r_F}k^{xy}\hv_x\hv_y-2\pi\ri \sum_{\ell=1}^{r_G}\sum_{x=1}^{r_F}k^{\ell x}\hu_\ell \hv_x}\nn\\
	&\quad\times\prod_{\alpha\in\mR[\mfg]}s_b\bigg(\fft{\ri Q}{2}-\alpha(\boldhu)\bigg)^{-1} \prod_\Psi\prod_{\rho_\Psi}\prod_{\trho_\Psi}s_b\bigg(\fft{\ri Q}{2}(1-r_\Psi)-\rho_\Psi(\boldhu)-\trho_\Psi(\boldhv)\bigg)\,,\label{S3b}
\end{align}
where the double sine function $s_b(x)$ is introduced in Appendix \ref{app:special}. The notation appearing in this expression is
summarized below. 
\begin{itemize}
	\item The background is specified by the squashing parameter $b$ and $Q$ denotes the symmetric combination $Q\equiv b+b^{-1}$.
	
	\item The CS-matter theory has gauge group $G$ and flavor group $F$ of ranks $r_G$ and $r_F$ respectively. The symbol $\alpha$ runs over the roots associated with the gauge algebra $\mfg$, denoted $\mR[\mfg]$. The matter content consists of $\mN=2$ chiral multiplets $\Psi$ with $R$ charges $r_\Psi$. For each chiral multiplet, $\rho_\Psi$ and $\trho_\Psi$ run over the weights of the corresponding gauge and flavor representations. The variables $\hu_\ell$ and $\hv_x$ denote vacuum expectation values of scalar fields in $\mN=2$ gauge and background vector multiplets respectively. Under localization, $\hu_\ell$ parametrize the supersymmetric locus, while $\hv_x$ act as real mass parameters~\cite{Freedman:2013oja}. Bold symbols collectively represent these
	quantities, \emph{e.g.} $d\boldhu\equiv\prod_{\ell=1}^{r_G}d\hu_\ell$. Lastly, $|\mW|$ denotes the order of the Weyl group acting on the gauge zero modes $\boldhu$.
	
	\item The collective symbol $\boldk$ denotes the set of Chern–Simons levels
	$\{k^{\ell n},\,k^{xy},\,k^{\ell x}\}$, corresponding to gauge–gauge,
	flavor–flavor, and gauge–flavor mixed couplings, respectively.\footnote{There exists an overall sign convention difference in the CS contribution across the literature.  Here we follow the convention of \cite{Kapustin:2009kz,Hama:2010av,Willett:2016adv,Chester:2021gdw}, as opposed to that of \cite{Martelli:2011qj,Nosaka:2015iiw,Bobev:2025ltz}. This difference does not affect the factorization properties discussed in this work.\label{foot:CS}}
\end{itemize}
%

%%%%%
\subsection{Superconformal index}
\label{sec:susyloc:S1S2}
%%%%%
The 3d superconformal index (SCI) on $S^1\times S^2$ is defined as \cite{Bhattacharya:2008zy,Bhattacharya:2008bja,Kim:2009wb}
\begin{align}
	\mI_{{\rm SCI}\,(2j_3)}(\beta,\boldlambda)=\Tr\Big[e^{2\pi\ri j_3}e^{-\beta(R+2j_3)}\prod_{x=1}^{r_F}e^{\ri\lambda_xF_x}\Big]\,,\label{SCI-tr:1st}
\end{align}
where the fermion number operator is specified as $F=2j_3$ following the prescription discussed in~\cite{Aharony:2013dha,Aharony:2013kma}. Here $j_3$ is the third component of the angular momentum on $S^2$, $\Delta$ is the energy in radial quantization, $R$ is the superconformal $R$ charge, and $F_x$ are the flavor charges of a given CS-matter theory. The trace is taken over the Hilbert space of the radially quantized CFT on $S^2$ and the index receives contributions from the BPS states satisfying
\begin{equation}
	\Delta-R-j_3=0\ .
\end{equation}
The $S^1\times S^2$ SCI depends on the chemical
potentials $\beta$ and $\lambda_x$, where the latter are collectively denoted by $\boldlambda$. For a convergent SCI, we assume
\begin{equation}
	\Re[\beta]>0\,, \label{beta:posi}
\end{equation}
throughout this paper. The geometric interpretation of $\beta$ is discussed in~\cite{Kim:2009wb,Imamura:2011su, Bobev:2022wem,BenettiGenolini:2023rkq}. One can also turn on the flavor magnetic fluxes through $S^2$ to deform the Hilbert space, which defines the ``generalized'' SCI. Throughout this paper, we shall discuss this generalized SCI.

\medskip

To analyze the factorization of the SCI, it is more convenient to work on the ``2nd sheet'' obtained by shifting $\beta\to\beta-\pi\ri$. We refer the reader to~\cite{ArabiArdehali:2025bub} for a recent analysis of the behavior of the 3d SCI on different sheets, motivated by earlier studies of the 4d superconformal index~\cite{Choi:2018hmj,Cassani:2021fyv}. We also discuss the alternative convention $\beta\to\beta+\pi\ri$ adopted in \cite{Choi:2019dfu}, which is not necessarily equivalent to the shift $\beta\to\beta-\pi\ri$ in theories involving non-integer $R$ charges. In this work, we therefore investigate the properties of the following SCI:
\begin{align}
	\mI_{{\rm SCI}\,(\pm)}(\beta,\boldlambda)=\Tr\Big[e^{\mp\pi\ri R}e^{-\beta(R+2j_3)}\prod_{x=1}^{r_F}e^{\ri\lambda_xF_x}\Big]=\mI_{{\rm SCI}\,(2j_3)}(\beta\pm\pi\ri,\boldlambda)\,.\label{SCI-tr:2nd}
\end{align}
Supersymmetric localization yields a matrix model representation of the (generalized) SCI~\eqref{SCI-tr:2nd}, given by~\cite{Imamura:2011su,Imamura:2011uj,Krattenthaler:2011da,Kapustin:2011jm,Aharony:2013dha,Aharony:2013kma,Choi:2019dfu}
\begin{align}
	\mI_{{\rm SCI}\,(\pm)}(\beta,\boldk,\boldlambda,\boldmn)&=\fft{1}{|\mW|}\sum_{\boldmm\in Q^\vee(\mathfrak g)}\int_0^{2\pi} \fft{d\boldh}{(2\pi)^{r_G}}\prod_{\ell=1}^{r_G}z_\ell^{\sum_{n=1}^{r_G}k^{\ell n}\mm_n+\sum_{x=1}^{r_F}k^{\ell x}\mn_x}\prod_{x=1}^{r_F}\xi_x^{\sum_{y=1}^{r_F}k^{xy}\mn_y+\sum_{\ell=1}^{r_G}k^{x\ell}\mm_\ell}\nn\\
	&\quad\times\prod_{\alpha\in\mR[\mfg]}(e^{\ri\alpha(\boldh)}q^{-1}e^{\pm\pi\ri})^{\fft12|\alpha(\boldmm)|}\fft{(e^{\ri\alpha(\boldh)}q^{|\alpha(\boldmm)|};q^2)}{(e^{-\ri\alpha(\boldh)}q^{2+|\alpha(\boldmm)|};q^2)}\nn\\
	&\quad\times\prod_{\Psi}\prod_{\rho_\Psi}\prod_{\trho_\Psi}\Big(e^{-\ri\rho_\Psi(\boldh)-\ri\trho_\Psi(\boldlambda)}q^{1-r_\Psi}e^{\mp\pi\ri(1-r_\Psi)}\Big)^{\fft12|\rho_\Psi(\boldmm)+\trho_\Psi(\boldmn)|}\nn\\
	&\qquad\times\fft{(e^{\pm\pi\ri r_\Psi}e^{-\ri\rho_\Psi(\boldh)-\ri\trho_\Psi(\boldlambda)}q^{2-r_\Psi+|\rho_\Psi(\boldmm)+\trho_\Psi(\boldmn)|};q^2)}{(e^{\mp\pi\ri r_\Psi}e^{\ri\rho_\Psi(\boldh)+\ri\trho_\Psi(\boldlambda)}q^{r_\Psi+|\rho_\Psi(\boldmm)+\trho_\Psi(\boldmn)|};q^2)}\,,\label{SCI:2nd}
\end{align}
where we use the same conventions as in Section~\ref{sec:susyloc:S3b} for the $S^3_b$ partition function along with the $\infty$-Pochhammer symbol introduced in Appendix~\ref{app:special}. We also introduce the gauge holonomy $z_\ell=e^{\ri h_\ell}$ together with other fugacities $q=e^{-\beta}$ and $\xi_x=e^{\ri\lambda_x}$. The symbols $\mm_\ell$ and $\mn_x$ denote the gauge and flavor magnetic fluxes through $S^2$ respectively, which take values in the co-root lattice $Q^\vee$ of the gauge and flavor algebra; throughout this manuscript we take these magnetic fluxes to be integer-quantized, identifying $Q^\vee(\mfg)\simeq \mathbb{Z}^{r_G}$. In the SCI expression~\eqref{SCI:2nd}, we explicitly include the dependence on the Chern–Simons levels and flavor magnetic fluxes in the argument of $\mI_{{\rm SCI}\,(\pm)}$, which was implicit in the trace formula \eqref{SCI-tr:2nd}.

%%%%%
\section{Superconformal index and $Z_{S^3_b}$}
\label{sec:fact}
%%%%%
In this section, we study the factorization of the $S^1\times S^2$ SCI (\ref{SCI:2nd}) in terms of the $S^3_b$ partition function (\ref{S3b}) to all orders in the Cardy-like limit
\begin{align}
	\Re[\beta]\to0^+\qquad\text{with~~fixed}~~~\arg[\beta]\label{Cardy}
\end{align}
up to non-perturbative corrections, improving the results of~\cite{Choi:2019dfu,Bobev:2022wem,Bobev:2024mqw,Bobev:2025ltz} restricted to the first two leading orders.

\subsection{Factorizing the SCI}
\label{sec:fact:SCI}
As a first step, we eliminate the absolute value signs imposed on the magnetic fluxes by utilizing the identity of the $\infty$-Pochhammer symbol presented in (\ref{poch:identity}). We then introduce the following complexified parameters:
\begin{equation}
	\begin{alignedat}{2}
		u_\ell&=\ri h_\ell+\beta\mm_\ell\,,&\qquad v_x&=\ri\lambda_x+\beta\mn_x\,,\\
		\bu_\ell&=-\ri h_\ell+\beta\mm_\ell\,,&\qquad \bv_x&=-\ri\lambda_x+\beta\mn_x\,. \label{complexification}
	\end{alignedat}
\end{equation}
With this reparametrization, the SCI (\ref{SCI:2nd}) can be rewritten as
\begin{align}
	&\mI_{{\rm SCI}\,(\pm)}(\beta,\boldk,\boldlambda,\boldmn) \label{SCI:2nd:1}\\
	&=\fft{1}{|\mW|}\sum_{\boldmm\in\mathbb{Z}^{r_G}}\int_0^{2\pi} \fft{d\boldh}{(2\pi)^{r_G}}e^{\sum_{\ell,n=1}^{r_G}\fft{k^{\ell n}(u_\ell u_n-\bu_\ell\bu_n)}{4\beta}+\sum_{\ell=1}^{r_G}\sum_{x=1}^{r_F}\fft{k^{\ell x}(u_\ell v_x-\bu_\ell\bv_x)}{2\beta}+\sum_{x,y=1}^{r_F}\fft{k^{xy}(v_xv_y-\bv_x\bv_y)}{4\beta}}\nn\\
	&\quad\times\prod_{\alpha\in\mR[\mfg]}e^{-\fft{\alpha(\boldu)^2-\alpha(\boldbu)^2}{8\beta}-\fft{(\beta\pm\pi\ri)(\alpha(\boldu)+\alpha(\boldbu))}{4\beta}}\fft{(e^{\alpha(\boldu)};q^2)}{(e^{\alpha(\boldbu)}q^2;q^2)}\nn\\
	&\quad\times\prod_{\Psi}\prod_{\rho_\Psi}\prod_{\trho_\Psi}\Big(e^{-\fft{\rho_\Psi(\boldu)-\rho_\Psi(\boldbu)}{2}-\fft{\trho_\Psi(\boldv)-\trho_\Psi(\boldbv)}{2}}e^{-(\beta\pm\pi\ri)(1-r_\Psi)}\Big)^{-\fft{\rho_\Psi(\boldu)+\rho_\Psi(\boldbu)+\trho_\Psi(\boldv)+\trho_\Psi(\boldbv)}{4\beta}}\nn\\
	&\kern7em\times\fft{(e^{\rho_\Psi(\boldbu)+\trho_\Psi(\boldbv)+(\beta\pm\pi\ri) r_\Psi}q^2;q^2)}{(e^{\rho_\Psi(\boldu)+\trho_\Psi(\boldv)-(\beta\pm\pi\ri)r_\Psi};q^2)}\,. \nn
\end{align}
The integrand in the expression (\ref{SCI:2nd:1}) can now be factorized manifestly as  
\begin{align}
	&\mI_{{\rm SCI}\,(\pm)}(\beta,\boldk,\boldlambda,\boldmn)\nn\\
	&=\fft{1}{|\mW|}\sum_{\boldmm\in\mathbb{Z}^{r_G}}\int_0^{2\pi} \fft{d\boldh}{(2\pi)^{r_G}} \label{SCI:2nd:2}\\
	&\quad\times e^{\fft{\sum_{\ell,n=1}^{r_G}k^{\ell n}u_\ell u_n+2\sum_{\ell=1}^{r_G}\sum_{x=1}^{r_F}k^{\ell x}u_\ell v_x+\sum_{x,y=1}^{r_F}k^{xy}v_xv_y}{4\beta}}\times\prod_{\alpha\in\mR[\mfg]} e^{\fft{\pi\ri}{2}(-\fft{(\beta\pm\pi\ri+\alpha(\boldu))^2}{4\pi\ri\beta}+\fft{\fft{\beta}{\pi\ri}+\fft{\pi\ri}{\beta}}{12})}(e^{\alpha(\boldu)};q^2)\nn\\
	&\quad\times\prod_{\Psi}\prod_{\rho_\Psi}\prod_{\trho_\Psi} e^{-\fft{\pi\ri}{2}(-\fft{(\beta\pm\pi\ri+\rho_\Psi(\boldu)+\trho_\Psi(\boldv)-(\beta\pm\pi\ri)r_\Psi)^2}{4\pi\ri\beta}+\fft{\fft{\beta}{\pi\ri}+\fft{\pi\ri}{\beta}}{12})}\fft{1}{(e^{\rho_\Psi(\boldu)+\trho_\Psi(\boldv)-(\beta\pm\pi\ri)r_\Psi};q^2)}\nn\\
	&\quad\times e^{-\fft{\sum_{\ell,n=1}^{r_G}k^{\ell n}\bu_\ell \bu_n+2\sum_{\ell=1}^{r_G}\sum_{x=1}^{r_F}k^{\ell x}\bu_\ell \bv_x+\sum_{x,y=1}^{r_F}k^{xy}\bv_x\bv_y}{4\beta}}\times\prod_{\alpha\in\mR[\mfg]} e^{-\fft{\pi\ri}{2}(-\fft{(-(\beta\pm\pi\ri)+\alpha(\boldbu))^2}{4\pi\ri\beta}+\fft{\fft{\beta}{\pi\ri}+\fft{\pi\ri}{\beta}}{12})}(e^{\alpha(\boldbu)};q^{-2})\nn\\
	&\quad\times\prod_{\Psi}\prod_{\rho_\Psi}\prod_{\trho_\Psi} e^{\fft{\pi\ri}{2}(-\fft{(-(\beta\pm\pi\ri)+\rho_\Psi(\boldbu)+\trho_\Psi(\boldbv)+(\beta\pm\pi\ri)r_\Psi)^2}{4\pi\ri\beta}+\fft{\fft{\beta}{\pi\ri}+\fft{\pi\ri}{\beta}}{12})}\fft{1}{(e^{\rho_\Psi(\boldbu)+\trho_\Psi(\boldbv)+(\beta\pm\pi\ri)r_\Psi};q^{-2})}\,.\nn
\end{align}

\medskip

To derive the factorization of the full SCI, not merely its integrand, we begin by applying the Euler-Maclaurin formula to approximate the infinite sums over gauge magnetic fluxes in~(\ref{SCI:2nd:2}) by integrals as \cite{Pasquetti:2019uop,Choi:2019zpz,Choi:2019dfu,Nian:2019pxj}
\begin{align}
	\fft{1}{|\mW|}\sum_{\boldmm\in\mathbb{Z}^{r_G}}\int_0^{2\pi} \fft{d\boldh}{(2\pi)^{r_G}}e^{-\mS(\boldh,\boldmm;\beta)} = \fft{1}{|\mW|}\int_{-\infty}^\infty d\boldmm \int_0^{2\pi} \fft{d\boldh}{(2\pi)^{r_G}}e^{-\mS(\boldh,\boldmm;\beta)}+\mO(\beta^0)\,,\label{EM}
\end{align}
where the boundary corrections from the Euler-Maclaurin formula are
encapsulated within $\mO(\beta^0)$, since they do not scale with
$\beta$ in the Cardy-like limit \cite{Bobev:2022wem,Bobev:2024mqw}. The first two leading terms in the Cardy-like expansion of the SCI are insensitive to the boundary corrections of order $\mO(\beta^0)$. Moreover, on the 2nd sheet where the asymptotic behavior of the SCI is exponentially diverging ($\sim e^{\#/\beta}$), the $\mO(\beta^0)$ corrections are expected to be exponentially suppressed relative to the dominant term. We therefore proceed with the continuation (\ref{EM}), neglecting the boundary terms, and focus on the factorization of the full SCI up to nonperturbative corrections. For an exact treatment of the factorization based on Poisson resummation over the gauge magnetic fluxes, which becomes essential for understanding the structure beyond the perturbative regime in the Cardy-like limit, we refer the reader to~\cite{ArabiArdehali:2025bub,Choi:2026}.

\medskip

Next, we use~\eqref{complexification} and perform the following change of integration variables
\begin{align}
	\int_{-\infty}^\infty d\mm_\ell\int_0^{2\pi}\fft{dh_\ell}{2\pi}=\int \fft{du_\ell  \wedge d\bu_\ell}{-4\pi\ri\beta}\,,\label{change}
\end{align}
where the integration domain is given by the image of the linear transformation (\ref{complexification}) acting on the original variables $\mm_\ell\in(-\infty,\infty)$ and $h_\ell\in(0,2\pi)$. Importantly, the resulting integral over $u_\ell$ and $\bu_\ell$ is \emph{not} factorized into two independent complex contour integrals, while the original real integrals over $\mm_\ell$ and $h_\ell$ can be treated separately. This distinction will play a key role in determining the order of the Weyl group $|\mW|$ for a particular saddle point in the next subsection.

\medskip

Applying the approximation~(\ref{EM}) together with the change of integration variables (\ref{change}) to the SCI (\ref{SCI:2nd:2}), we obtain
\begin{align}
	\mI_{{\rm SCI}\,(\pm)}(\beta,\boldk,\boldlambda,\boldmn)&=\fft{1}{|\mW|}\int (\prod_{\ell=1}^{r_G}du_\ell\wedge d\bu_\ell)\, e^{-S_{(\pm)}(\boldu;\beta,\boldk,\boldv)}e^{-\bS_{(\pm)}(\boldbu;-\beta,\boldk,\boldbv)}\,,\label{SCI:2nd:3}
\end{align}
where we have defined the holomorphic and anti-holomorphic effective actions as
\begin{subequations}
	\begin{align}
		e^{-S_{(\pm)}(\boldu;\beta,\boldk,\boldv)}&\equiv \fft{1}{(-4\pi\ri\beta)^{r_G/2}}\,e^{\fft{\sum_{\ell,n=1}^{r_G}k^{\ell n}u_\ell u_n+2\sum_{\ell=1}^{r_G}\sum_{x=1}^{r_F}k^{\ell x}u_\ell v_x+\sum_{x,y=1}^{r_F}k^{xy}v_xv_y}{4\beta}}\\
		&\quad\times\prod_{\alpha\in\mR[\mfg]} e^{\fft{\pi\ri}{2}(-\fft{(\beta\pm\pi\ri+\alpha(\boldu))^2}{4\pi\ri\beta}+\fft{\fft{\beta}{\pi\ri}+\fft{\pi\ri}{\beta}}{12})}(e^{\alpha(\boldu)};q^2)\nn\\
		&\quad\times\prod_{\Psi}\prod_{\rho_\Psi}\prod_{\trho_\Psi} \fft{e^{-\fft{\pi\ri}{2}(-\fft{(\beta\pm\pi\ri+\rho_\Psi(\boldu)+\trho_\Psi(\boldv)-(\beta\pm\pi\ri)r_\Psi)^2}{4\pi\ri\beta}+\fft{\fft{\beta}{\pi\ri}+\fft{\pi\ri}{\beta}}{12})}}{(e^{\rho_\Psi(\boldu)+\trho_\Psi(\boldv)-(\beta\pm\pi\ri)r_\Psi};q^2)}\,,\nn\\
		\bS_{(\pm)}&\equiv 	\text{obtained by replacing all $\ri$'s in}~S_{(\pm)}~\text{with $-\ri$}\,.
	\end{align}\label{S:bS}%
\end{subequations}
The expression in~(\ref{SCI:2nd:3}) at first sight provides a manifestly factorized form of the SCI. As discussed above, however, the integration over holomorphic and anti-holomorphic coordinates cannot be treated independently. In this sense, the factorization of the full SCI remains incomplete at this stage. In the next subsection, we employ the saddle point approximation in the Cardy-like limit to resolve this issue and thereby obtain a complete factorization of the $S^1\times S^2$ SCI in terms of $S^3_b$ partition functions. 

%%%%%
\subsection{Saddle point approximation}\label{sec:fact:saddle}
%%%%%
The saddle point equations for the matrix integral representation of the SCI (\ref{SCI:2nd:3}) in the Cardy-like limit~\eqref{Cardy} are decomposed into holomorphic and anti-holomorphic parts as
\begin{align}
	\fft{\partial S_{(\pm)}}{\partial u_\ell}\quad=\quad\fft{\partial \bS_{(\pm)}}{\partial \bu_\ell}\quad=\quad0\,.\label{saddle:eqn}
\end{align}
To evaluate the SCI (\ref{SCI:2nd:3}) via the saddle point approximation, one must, in principle, identify the complete set of relevant saddle points that satisfy \eqref{saddle:eqn} and contribute to the integral, and then sum over their corresponding contributions. In this work, we do not attempt to explicitly determine all such saddle points. Instead, we analyze the resulting structure of the SCI expressed in terms of these saddle-point contributions, which naturally leads to a factorized form in terms of $S^3_b$ partition functions.

\medskip

We begin by noting that any saddle point $\boldu^\star$ that is not acted upon freely by a Weyl reflection $w_\alpha\in\mW$ associated with a root $\alpha\in\mR[\mfg]$, namely 
\begin{align}
	w_\alpha\cdot\boldu^\star=\boldu^\star\qquad\Leftrightarrow \qquad\alpha(\boldu^\star)=0\,,
\end{align}
does not contribute to the matrix integral (\ref{SCI:2nd:3}) in the saddle-point approximation, even if it solves the saddle-point equation (\ref{saddle:eqn}). The reason is that the vector multiplet contribution in the integrand of (\ref{SCI:2nd:3}) contains the factor $(e^{\alpha(\boldu)};q^2)$, which vanishes at such a saddle point. In the effective action language, this zero acts as a repulsive interaction that forbids gauge-enhancing saddles with $\alpha(\boldu)=0$. This is nothing but the effect of the Vandermonde determinant in Hermitian matrix models that prevents eigenvalue coincidence. For a recent discussion of the screening of gauge-enhancing saddles in the 3d SCI, see \cite{ArabiArdehali:2025bub}. 

\medskip

We now analyze the contributions to the SCI (\ref{SCI:2nd:3}) from saddle points that are freely acted upon by Weyl reflections within the saddle point approximation. To this end, we first consider the holomorphic and anti-holomorphic matrix integrals separately, defined as 
\begin{subequations}
	\begin{align}
		\mZ_{(\pm)}(\beta,\boldk,\boldv)&\equiv\fft{1}{|\mW|}\int \prod_{\ell=1}^{r_G}du_\ell \, e^{-S_{(\pm)}(\boldu;\beta,\boldk,\boldv)}\,,\\
		\bmZ_{(\pm)}(-\beta,\boldk,\boldbv)&\equiv\fft{1}{|\mW|}\int \prod_{\ell=1}^{r_G}d\bu_\ell\, e^{-\bS_{(\pm)}(\boldbu;-\beta,\boldk,\boldbv)}\,.
	\end{align}\label{Z:bZ}%
\end{subequations}
Applying the saddle point approximation to each integral gives
\begin{subequations}
	\begin{align}
		\mZ_{(\pm)}(\beta,\boldk,\boldv)&=\sum_{\boldu^\star\in \mS_{\mZ}}\underbrace{e^{-S_{(\pm)}(\boldu^\star;\beta,\boldk,\boldv)}\times(\text{loop-corrections})}_{\equiv \mZ^{(\boldu^\star)}_{(\pm)}(\beta,\boldk,\boldv)}\,,\\
		\bmZ_{(\pm)}(-\beta,\boldk,\boldbv)&=\sum_{\boldbu^\star\in \mS_{\bmZ}}\underbrace{e^{-\bS_{(\pm)}(\boldbu^\star;-\beta,\boldk,\boldbv)}\times(\text{loop-corrections})}_{\equiv \bmZ^{(\boldbu^\star)}_{(\pm)}(-\beta,\boldk,\boldbv)}\,,
	\end{align}\label{Z:bZ:saddle-approx}%
\end{subequations}
where the summations run over the following sets of saddle points
\begin{subequations}
	\begin{align}
		\mS_{\mZ}&=\bigg\{\boldu\,\bigg|\,\fft{\partial S_{(\pm)}}{\partial u_\ell}=0\,,~~w\cdot\boldu\neq \boldu\,,~\forall w\in\mW\bigg\}\big/\mW\,,\\
		\mS_{\bmZ}&=\bigg\{\boldbu\,\bigg|\,\fft{\partial \bS_{(\pm)}}{\partial \bu_\ell}=0\,,~~w\cdot\boldbu\neq \boldbu\,,~\forall w\in\mW\bigg\}\big/\mW\,.
	\end{align}\label{Z:bZ:saddle}%
\end{subequations}
As explained above, saddle points that are fixed under Weyl reflections are excluded. Moreover, we restrict our attention to cases in which both the holomorphic and anti-holomorphic effective actions are invariant under Weyl reflections. Consequently, the contributions from Weyl-related saddles --- say $\boldu_a$ and $\boldu_b$ satisfying $\boldu_a=w\cdot\boldu_b$ for some $w\in\mW$ --- are supposed to be identical. Instead of counting each Weyl-equivalent contribution separately (which would cancel the prefactor $|\mW|^{-1}$ in (\ref{Z:bZ})), we quotient the set of saddle points by the Weyl group as in (\ref{Z:bZ:saddle}) and drop the $|\mW|^{-1}$ factor accordingly. 

\medskip

Building on the notation introduced in (\ref{Z:bZ:saddle-approx}), the saddle point approximation of the SCI~(\ref{SCI:2nd:3}) takes the factorized form
\begin{align}
	\mI_{{\rm SCI}\,(\pm)}(\beta,\boldk,\boldlambda,\boldmn)&=\sum_{(\boldu^\star,\boldbu^\star)\in \mS_{\mI}}\mI^{(\boldu^\star,\boldbu^\star)}_{{\rm SCI}\,(\pm)}(\beta,\boldk,\boldlambda,\boldmn)\nn\\
	&=\sum_{(\boldu^\star,\boldbu^\star)\in \mS_{\mI}}\mZ^{(\boldu^\star)}_{(\pm)}(\beta,\boldk,\boldv)\bmZ^{(\boldbu^\star)}_{(\pm)}(-\beta,\boldk,\boldbv)\,,\label{SCI:2nd:4}
\end{align}
where $(\boldv,\boldbv)=(\ri\boldlambda+\beta\boldmn,-\ri\boldlambda+\beta\boldmn)$ from the complexification (\ref{complexification}) and the set of contributing saddles is given by the product space $\mS_{\mI}=\mS_{\mZ}\times \mS_{\bmZ}$ in terms of (\ref{Z:bZ:saddle}). It is worth highlighting that the number of Weyl-equivalent saddles contributing to the SCI (\ref{SCI:2nd:3}) is $|\mW|$, rather than $|\mW|^2$, since $u_\ell$ and $\bu_\ell$ should not be regarded as independent contour integration variables as discussed in Section~\ref{sec:fact:SCI}. Owing to this subtlety, we did not identify the SCI~(\ref{SCI:2nd:3}) directly with the product of the separated matrix integrals (\ref{Z:bZ}); instead, we employed their saddle point contributions in (\ref{Z:bZ:saddle-approx}) to derive the factorized form of the SCI (\ref{SCI:2nd:4}). 

%%%%%
\subsection{SCI in terms of $S_b^3$ partition function}\label{sec:fact:SCI-S3b}
%%%%%
To investigate the relation between the SCI obtained from the saddle-point approximation~(\ref{SCI:2nd:4}) and the $S^3_b$ partition function (\ref{S3b}), we compare the latter with the $\mZ_{(\pm)}$ factor (\ref{Z:bZ}) that governs the former. For this purpose, it is convenient to express the $\infty$-Pochhammer symbols appearing in $\mZ_{(\pm)}$ in terms of the double sine function using the identity (\ref{sb:poch}), which holds under the assumption (\ref{beta:posi}). The resulting expression takes the form 
\begin{subequations}
	\begin{align}
		\mZ_{(\pm)}(\beta,\boldk,\boldv)&=\fft{e^{\ri\phi_{(\pm)}}}{|\mW|}\int d\boldhu\,e^{\mp\pi\ri\big(\sum_{\ell,n=1}^{r_G}k^{\ell n}\hu_\ell \hu_n+2\sum_{\ell=1}^{r_G}\sum_{x=1}^{r_F}k^{\ell x}\hu_\ell \hv_x+\sum_{x,y=1}^{r_F}k^{xy}\hv_x\hv_y\big)}\nn\\
		&\quad\times\prod_{\alpha\in\mR[\mfg]} \fft{(e^{\pm\fft{\pi\ri\alpha(\boldu)}{\beta}-\fft{2\pi^2}{\beta}};e^{-\fft{2\pi^2}{\beta}})}{s_b(\fft{\ri Q}{2}-\alpha(\boldhu))}\nn\\
		&\quad\times\prod_{\Psi}\prod_{\rho_\Psi}\prod_{\trho_\Psi}\fft{s_b(\fft{\ri Q}{2}(1-r_\Psi)-\rho_\Psi(\boldhu)-\trho_\Psi(\boldhv))}{(e^{\pm\fft{\pi\ri(\rho_\Psi(\boldu)+\trho_\Psi(\boldv)-(\beta\pm\pi\ri)r_\Psi)}{\beta}-\fft{2\pi^2}{\beta}};e^{-\fft{2\pi^2}{\beta}})}\,, \\[0.5em]
		\bmZ_{(\pm)}(-\beta,\boldk,\boldbv)&=\fft{e^{-\ri\bphi_{(\pm)}}}{|\mW|}\int d\boldhbu\,e^{\pm\pi\ri\big(\sum_{\ell,n=1}^{r_G}k^{\ell n}\hbu_\ell \hbu_n+2\sum_{\ell=1}^{r_G}\sum_{x=1}^{r_F}k^{\ell x}\hbu_\ell \hbv_x+\sum_{x,y=1}^{r_F}k^{xy}\hbv_x\hbv_y\big)}\nn\\
		&\quad\times\prod_{\alpha\in\mR[\mfg]} \fft{(e^{\pm\fft{\pi\ri\alpha(\boldbu)}{\beta}};e^{-\fft{2\pi^2}{\beta}})^{-1}}{s_b(\fft{\ri Q}{2}-\alpha(\boldhbu))}\nn\\
		&\quad\times\prod_{\Psi}\prod_{\rho_\Psi}\prod_{\trho_\Psi}\fft{s_b(\fft{\ri Q}{2}(1-r_\Psi)-\rho_\Psi(\boldhbu)-\trho_\Psi(\boldhbv))}{(e^{\pm\fft{\pi\ri(\rho_\Psi(\boldbu)+\trho_\Psi(\boldbv)+(\beta\pm\pi\ri)r_\Psi)}{\beta}};e^{-\fft{2\pi^2}{\beta}})^{-1}}\,,
	\end{align}\label{Z:bZ:squash}%
\end{subequations}
where the chemical potential $\beta$ is related to the squashing parameter $b$ via
\begin{align}
	\beta=\pm\pi\ri b^2 \,, \label{beta:b}
\end{align}
and the remaining parameters of the SCI are rescaled into those of the $S^3_b$ partition function as
\begin{subequations}
	\begin{align}
		u_\ell=\mp 2\pi b\hu_\ell\qquad\&\qquad v_x=\mp 2\pi b\hv_x\,, \\
		\bu_\ell=\pm 2\pi b\hbu_\ell\qquad\&\qquad \bv_x=\pm 2\pi b\hbv_x\,. 
	\end{align}\label{identification}%
\end{subequations}
The overall phases $e^{\ri\phi_{(\pm)}}$ and $e^{\ri\bphi_{(\pm)}}$ do not have individual physical significance, as the $\mZ_{(\pm)}$ factors (\ref{Z:bZ}) are defined with respect to a specific decomposition of the SCI (\ref{SCI:2nd:3}) in the first place. When combined, they yield 
\begin{align}
	e^{\ri(\phi_{(\pm)}-\bphi_{(\pm)})}=\begin{cases}
		(-1)^{r_G} & (\text{upper sign}) \\
		1 & (\text{lower sign})
	\end{cases}\,,\label{phase}
\end{align}
which still exhibits no nontrivial dependence on the chemical potentials and magnetic fluxes $(\beta,\boldlambda,\boldmn)$ governing the SCI.

\medskip

It is now manifest that the $\mZ_{(\pm)}$ factors (\ref{Z:bZ:squash}) and the $S^3_b$ partition function (\ref{S3b}) coincide up to overall phase factors and the $\infty$-Pochhammer symbols that contain the factor of $e^{-2\pi^2/\beta}$, which become exponentially suppressed in the Cardy-like limit (\ref{Cardy}). This observation naturally leads to the conclusion that the saddle point contributions to those two matrix integrals --- the $\mZ_{(\pm)}$ factors (\ref{Z:bZ:squash}) and the $S^3_b$ partition function (\ref{S3b}) --- agree up to exponentially suppressed corrections as 
\begin{subequations}
	\begin{align}
		\mZ^{(\boldu^\star)}_{(\pm)}(\beta,\boldk,\boldv) &\approx e^{\ri\phi_{(\pm)}} Z^{(\boldhu^\star)}(b,\pm\boldk,\boldhv) \,, \\
		\bmZ^{(\boldbu^\star)}_{(\pm)}(-\beta,\boldk,\boldbv) &\approx e^{-\ri\bphi_{(\pm)}} Z^{(\boldhbu^\star)}(b,\mp\boldk,\boldhbv) \,.
	\end{align}\label{Z:to:S3b}%
\end{subequations}
Throughout this paper, the symbol ``$\approx$'' denotes equality up to exponentially suppressed terms, namely
\begin{align}
	A\approx B\qquad\Leftrightarrow\qquad A-B=\mO(e^{-\#/|\beta|})
\end{align}
with $\#$ a positive constant that does not scale with $\beta$ in the Cardy-like limit. In the relation~(\ref{Z:to:S3b}), the parameters of the SCI and those of the $S^3_b$ partition function are related as specified in (\ref{beta:b}) and (\ref{identification}). In particular, the saddle points for the $\mZ$ factors, $\boldu^\star$ and $\boldbu^\star$, are related to those for the $S^3_b$ partition functions, $\boldhu^\star$ and $\boldhbu^\star$, through the identification (\ref{identification}).

\medskip

Finally, substituting the relation between saddle point contributions (\ref{Z:to:S3b}) into the factorized SCI expression (\ref{SCI:2nd:4}) yields our main result 
\begin{align}
	\mI_{{\rm SCI}\,(\pm)}(\beta,\boldk,\boldlambda,\boldmn)\approx e^{\ri(\phi_{(\pm)}-\bphi_{(\pm)})}\sum_{(\boldu^\star,\boldbu^\star)\in \mS_{\mI}} Z^{(\boldhu^\star)}(b,\pm\boldk,\boldhv)Z^{(\boldhbu^\star)}(b,\mp\boldk,\boldhbv)\,.\label{SCI:2nd:5}
\end{align}
Here, the overall phase factor has been extracted due to its independence from the saddle points, and the parameters can be mapped to each other through (\ref{complexification}), (\ref{beta:b}), and (\ref{identification}). The formula (\ref{SCI:2nd:5}) establishes an explicit relation between the SCI and the $S^3_b$ partition function through the saddle point approximation, valid to all orders in the Cardy-like expansion (\ref{Cardy}). This result extends previous analyses, which were limited to the first two leading orders~\cite{Choi:2019dfu,Bobev:2022wem,Bobev:2024mqw}.

\medskip

In the above derivation, the non-perturbative suppression of the $\infty$-Pochhammer symbols involving the $e^{-2\pi^2/\beta}$ factor plays a crucial role in identifying the $\mZ_{(\pm)}$ factors, which arise from the decomposition of the SCI, with the $S^3_b$ partition function in the Cardy-like limit (\ref{Cardy}), as expressed in (\ref{Z:to:S3b}). In particular, they should be suppressed along the steepest descent contour determined by Picard-Lefschetz theory so that saddle point approximations of the SCI and the $S^3_b$ partition function coincide. Determining the steepest descent contour is highly nontrivial and depends on the details of the theory. We do not address this subtle issue here --- apart from a brief comment on the simplest example, the chiral multiplet, given below --- and leave its general analysis to future work. We now proceed to discuss some explicit examples of SCFTs where we can illustrate the validity of~(\ref{Z:to:S3b}).

%%%%%
\section{Examples}\label{sec:ex}
%%%%%
In this section, we demonstrate the factorization formula of the SCI (\ref{SCI:2nd:5}) by applying it to concrete examples.

%%%%%
\subsection{Chiral multiplet}\label{sec:ex:chiral}
%%%%%
According to the localization formula (\ref{S3b}), a single chiral multiplet $\Psi$ with $R$ charge $r_\Psi$ has the following $S^3_b$ partition function
\begin{align}
	Z^{\Psi}(b,\boldhv) = \prod_{\trho_\Psi} s_b\bigg( \fft{\ri Q}{2}(1-r_\Psi) - \trho_\Psi(\boldhv) \bigg)\,. \label{chiral:S3b}
\end{align}
Its SCI can be extracted from (\ref{SCI:2nd}) and reads
\begin{equation}
	\mI_{{\rm SCI}\,(\pm)}^{\Psi} (\beta,\boldlambda) = \prod_{\trho_\Psi}\frac{(e^{-\ri\trho_\Psi(\boldlambda)+(\beta\pm\pi \ri)r_\Psi} q^2;q^2)}{(e^{\ri\trho_\Psi(\boldlambda)-(\beta\pm\pi \ri)r_\Psi};q^2)} \,.\label{chiral:SCI}
\end{equation}

To relate the two partition functions, we map the parameters in the SCI to those in the $S_b^3$ partition function using (\ref{complexification}), (\ref{beta:b}), and (\ref{identification}) and the fact that there are no gauge fields to find
\begin{equation}
	\begin{split}
		\ri\lambda_x&=v_x=\mp 2\pi b\hv_x\\
		&=-\bv_x=\mp 2\pi b\hbv_x
	\end{split}\qquad\&\qquad \beta=\pm\pi\ri b^2\,,\quad \Re[\beta]>0\,. \label{chiral:identification}
\end{equation}
We now consider the Cardy-like limit (\ref{Cardy}) while keeping $v_x$ finite. Using the identification~(\ref{chiral:identification}) and the formula (\ref{sb:poch}) with an arbitrary sign choice $\mfs\in\{\pm1\}$, the $S^3_b$ partition function (\ref{chiral:S3b}) can be rewritten in terms of the SCI parameters as
\begin{align}
	Z^{\Psi}(b,\boldhv) &= \prod_{\trho_\Psi} e^{\mp\fft{\mfs\pi\ri}{2}(-\fft{(\ri\trho_\Psi(\boldlambda)+(\beta\pm\pi\ri)(1-r_\Psi))^2}{4\pi\ri\beta}+\fft{\fft{\beta}{\pi\ri}+\fft{\pi\ri}{\beta}}{12})} \fft{(e^{\fft{\mfs\pi\ri}{\beta} (\ri\trho_\Psi(\boldlambda)-(\beta\pm\pi\ri)r_\Psi) \mp\frac{2\mfs\pi^2}{\beta}};e^{\mp\frac{2\mfs\pi^2}{\beta}})}{(e^{\pm\mfs(\ri\trho_\Psi(\boldlambda)-(\beta\pm\pi\ri)r_\Psi)};e^{\mp 2\mfs \beta})}\nn\\
	&=Z^{\Psi}(b,\boldhbv) \,.
\end{align}
Multiplying the two $S^3_b$ partition functions with opposite choices of $\mfs$ then yields
\begin{align}
	Z^{\Psi}(b,\boldhv)Z^{\Psi}(b,\boldhbv)&=\prod_{\trho_\Psi} \fft{(e^{\pm\fft{\pi\ri}{\beta} (\ri\trho_\Psi(\boldlambda)-(\beta\pm\pi\ri)r_\Psi) -\frac{2\pi^2}{\beta}};e^{-\frac{2\pi^2}{\beta}})}{(e^{\ri\trho_\Psi(\boldlambda)-(\beta\pm\pi\ri)r_\Psi};q^2)} \fft{(e^{-\ri\trho_\Psi(\boldlambda)+(\beta\pm\pi\ri)r_\Psi}q^2;q^2)}{(e^{\mp\fft{\pi\ri}{\beta} (\ri\trho_\Psi(\boldlambda)-(\beta\pm\pi\ri)r_\Psi) };e^{-\frac{2\pi^2}{\beta}})}\nn\\[0.5em]
	&\approx\prod_{\trho_\Psi} \fft{(e^{-\ri\trho_\Psi(\boldlambda)+(\beta\pm\pi\ri)r_\Psi}q^2;q^2)}{(e^{\ri\trho_\Psi(\boldlambda)-(\beta\pm\pi\ri)r_\Psi};q^2)} = \mI_{{\rm SCI}\,(\pm)}^{\Psi} (\beta,\boldlambda)\,.\label{CM:fact}
\end{align}
The result is consistent with the general factorization formula (\ref{SCI:2nd:5}), which simplifies considerably for a single chiral multiplet, due to the absence of CS levels, flavor magnetic fluxes, and complexified gauge holonomies. 

\medskip

In this simple example one can also state explicitly the condition under which the terms involving $e^{-2\pi^2/\beta}$ in \eqref{CM:fact} are non-perturbatively suppressed. The constraint reads
\begin{align}\label{chiral-con}
	0<\Re\left[\fft{r_\Psi\mp\fft{\trho_\Psi(\boldlambda)}{\pi}}{2\beta}\right]<\Re\left[\fft{1}{\beta}\right]\,.
\end{align}
While this seemingly restricts the flavor holonomies $\boldlambda$, they are defined only modulo $2\pi\mathbb{Z}$, so the factorization \eqref{CM:fact} can always be made to hold by choosing the representatives appropriately. We defer the analogous non-perturbative analysis for more complicated gauge theories to future work.

%%%%%
\subsection{ABJM theory}\label{sec:ex:ABJM}
%%%%%
We now explore the application of the factorization formula~(\ref{SCI:2nd:5}) to the ABJM theory~\cite{Aharony:2008ug}, presenting the ABJM SCI in terms of its $S_b^3$ partition function to all orders in the Cardy-like expansion.

%%%%%
\subsubsection{Supersymmetric partition functions}\label{sec:ex:ABJM:mm}
%%%%%
The matrix model for the ABJM $S_b^3$ partition function follows from the general localization formula (\ref{S3b}) and takes the form\footnote{Our Chern–Simons level convention differs by a sign from that of \cite{Bobev:2025ltz}, as explained in Footnote~\ref{foot:CS}. This difference is immaterial because the ABJM $S_b^3$ partition function is invariant under $k\to -k$ 
	as
	\begin{equation}
		Z^\text{ABJM}(b,k,\Delta_1,\Delta_2,\Delta_3,\Delta_4)=Z^\text{ABJM}(b,k,\Delta_3,\Delta_4,\Delta_1,\Delta_2)=Z^\text{ABJM}(b,-k,\Delta_1,\Delta_2,\Delta_3,\Delta_4) \,, \label{ABJM:CS}
	\end{equation}
	where we use $\Delta_a$ instead of $\hv_a$ for convenience.
}  
\begin{align}
	Z^\text{ABJM}(\omega,k,\boldvarphi)&=\fft{1}{(N!)^2}\int\left(\prod_{i=1}^Nd\mu_i d\nu_i\right)\,e^{-\pi\ri k\sum_{i=1}^N(\mu_i^2-\nu_i^2)} \prod_{i\neq j}s_b\bigg(\fft{\ri Q}{2}-\mu_{ij}\bigg)^{-1}s_b\bigg(\fft{\ri Q}{2}-\nu_{ij}\bigg)^{-1}\nn\\
	&\quad\times\prod_{i,j=1}^N\prod_{a=1}^4s_b\bigg(\fft{\ri Q}{2}(1-r_a)-\sigma_a(\mu_i-\nu_j)-\hv_a\bigg)\,,\label{ABJM:S3b}
\end{align}
where the $R$ charges at the superconformal point take the values $r_a=\fft12$ and we have introduced $\sigma_a=(1,1,-1,-1)$. In \eqref{ABJM:S3b} we have traded the original arguments $(b,\boldhv)$ for $(\omega,\boldvarphi)$ defined by\footnote{The new parameter $\omega$ will unambiguously fix the squashing parameter $b$ under $\Re[b]>0$, which corresponds to the regime of analytic continuation discussed in \eqref{ABJM:analytic-conti}. We exclude the case with $\Re[b]=0$ throughout this manuscript, which becomes pathological while analyzing the factorization of the index in terms of the squashed 3-sphere partition function.}

\begin{align}
	\omega\equiv b^2\qquad\&\qquad \varphi_a\equiv(1+b^2)\Delta_a\quad\bigg(\Delta_a\equiv r_a+\fft{2\hv_a}{\ri Q}~~\to~~\sum_{a=1}^4\Delta_a=2\bigg)\,,\label{ABJM:omegavarphi}
\end{align}
where the constraint on the summation of $\Delta_a$ reflects the invariance of the ABJM superpotential under the flavor symmetries; this repackaging both renders the Airy data below more compact and provides a simple dictionary between the $S_b^3$ parameters and the corresponding SCI parameters that will be used in the factorization later. In the large $N$ limit, the ABJM $S_b^3$ partition function~(\ref{ABJM:S3b}) is expected to exhibit an Airy behavior~\cite{Bobev:2022jte,Bobev:2022eus,Bobev:2025ltz}
\begin{align}
	Z^\text{ABJM}(\omega,k,\boldvarphi)=\mC^{-1/3}e^{\mA}\text{Ai}\Big[\mC^{-1/3}(N-\mB)\Big]\Big(1+\mO(e^{-\#\sqrt{N}})\Big)\,,\label{ABJM:S3b:Airy}
\end{align}
where $\mB$, $\mC$ are given by ($\varphi_{ab}\equiv\varphi_a+\varphi_b$)
\begin{subequations}
	\begin{align}
		\mC(\omega,k,\boldvarphi)&=\fft{2\omega^2}{\pi^2 k\varphi_1\varphi_2\varphi_3\varphi_4}\,,\\
		\mB(\omega,k,\boldvarphi)&=\fft{k}{24}+\fft{(1+\omega)^2F_1(\varphi)+(1-\omega)^2F_2(\varphi)}{48k\varphi_1\varphi_2\varphi_3\varphi_4}\,,\\
		F_1(\boldvarphi)&=\sum_{a=1}^4\varphi_a^2-\fft{(\varphi_{12}-\varphi_{34})(\varphi_{13}-\varphi_{24})(\varphi_{14}-\varphi_{23})}{\sum_{a=1}^4\varphi_a}\,,\\
		F_2(\boldvarphi)&=-2\sum_{a<b}^4\varphi_a\varphi_b\,,
	\end{align}\label{ABJM:Airy:coeffi}%
\end{subequations}
while $\mA(\omega,k,\boldvarphi)$ is not known in closed form for generic configurations of $(\omega,\boldvarphi)$. For several special cases with known analytic expressions for $\mA$, which is independent of $N$, see \cite{Bobev:2025ltz} and references therein. For later use, we record the large-$k$ expansion of $\mA(\omega,k,\boldvarphi)$, 
\begin{align}
	\mA(\omega,k,\boldvarphi)&=-\fft{\zeta(3)}{8\pi^2}\bigg[\bigg(\fft{(\sum_a\varphi_a)^2}{\varphi_{13}\varphi_{14}\varphi_{23}\varphi_{24}}-\fft{(1+\omega)^2}{\omega}\Big(\fft{1}{\varphi_{13}\varphi_{24}}+\fft{1}{\varphi_{14}\varphi_{23}}\Big)\bigg)\fft{2\prod_a\varphi_a}{\sum_a\varphi_a}\sum_a\fft{1}{\varphi_a}\nn\\
	&\hspace{4.5em}+\fft{(1+\omega)^2}{\omega}\left(1-\fft{\sum_a\varphi_a^2}{(\sum_a\varphi_a)^2}\right)\bigg]k^2-\fft16\log k+\mO(k^0)\,,\label{ABJM:mA}
\end{align}
where the leading $k^2$ term follows from the planar analysis of \cite{Bobev:2025ltz}, while the universal $\log k$ term is conjectured here on the basis of the special cases \cite{Marino:2011eh,Nosaka:2015iiw,Bobev:2025ltz}. The Airy formula \eqref{ABJM:S3b:Airy} has been analytically derived and numerically confirmed in various special cases, typically restricted to the ranges
\begin{align}
	b>0\qquad\&\qquad 0<\Delta_a<1\,,
\end{align}
see \cite{Bobev:2025ltz} and references therein. 

\medskip

The matrix model for the ABJM SCI can be obtained from the general localization formula~(\ref{SCI:2nd})  
\begin{align}
	\mI_{{\rm SCI}\,(\pm)}^\text{ABJM}(\beta,k,\boldlambda,\boldmn)&=\fft{1}{(N!)^2}\sum_{\boldmm,\boldtmm\in\mathbb{Z}^N}\int \fft{d\boldh}{(2\pi)^N}\fft{d\boldtih}{(2\pi)^N}\prod_{i=1}^{N}e^{\ri k\mm_ih_i}e^{-\ri k\tmm_i\tih_i} \nn\\
	&\quad\times\prod_{i\neq j}e^{\fft12(\beta\pm\pi\ri)|\mm_{ij}|}\Big(1-e^{\ri h_{ij}} q^{|\mm_{ij}|}\Big)e^{\fft12(\beta\pm\pi\ri)|\tmm_{ij}|}\Big(1-e^{\ri\tih_{ij}} q^{|\tmm_{ij}|}\Big)\nn\nn\\
	&\quad\times\prod_{a=1}^4\prod_{i,j=1}^N\Big(e^{-\ri\sigma_a(h_i-\tih_j)-\ri\lambda_a-(\beta\pm\pi\ri)(1-r_a)}\Big)^{\fft12|\sigma_a(\mm_i-\tmm_j)+\mn_a|}\nn\\
	&\kern6em\times\fft{(e^{-\ri\sigma_a(h_i-\tih_j)-\ri\lambda_a+(\beta\pm\pi\ri)r_a}q^{2+|\sigma_a(\mm_i-\tmm_j)+\mn_a|};q^2)}{(e^{\ri\sigma_a(h_i-\tih_j)+\ri\lambda_a-(\beta\pm\pi\ri)r_a}q^{|\sigma_a(\mm_i-\tmm_j)+\mn_a|};q^2)}\,,\label{ABJM:SCI}
\end{align}
where quantities with two subscripts are defined as $X_{ij}\equiv X_i-X_j$. The logarithm of the matrix integral (\ref{ABJM:SCI}) has been evaluated using the saddle point approximation in the Cardy-like limit, initially at leading $\beta^{-1}$ order~\cite{Choi:2019zpz} and subsequently generalized to include the first subleading $\beta^0$ order with vanishing flavor magnetic fluxes~\cite{Choi:2019dfu,Bobev:2022wem,Bobev:2024mqw}. See also~\cite{GonzalezLezcano:2022hcf,BenettiGenolini:2023rkq} for related large $N$ analyses of the ABJM SCI. Here we extend these results by generalizing the analysis to all orders in the Cardy-like expansion and by allowing for generic flavor magnetic fluxes, employing the factorization formula~\eqref{SCI:2nd:5}.

%%%%%
\subsubsection{SCI in terms of $S_b^3$ partition function}\label{sec:ex:ABJM:fact}
%%%%%
To apply the factorization formula (\ref{SCI:2nd:5}) to the ABJM theory and thereby relate the $S^3_b$ Airy formula (\ref{ABJM:S3b:Airy}) and the SCI (\ref{ABJM:SCI}), we first note that the Airy formula (\ref{ABJM:S3b:Airy}) admits a smooth analytic continuation to complex $(b,\boldDelta)$ provided
\begin{align}
	\Re[b]>0\qquad\&\qquad 0<\Re[\Delta_a]<1\,,\label{ABJM:analytic-conti}
\end{align}
thanks to the analytic properties of the double sine function \cite{Narukawa:2003ltf}. This justifies the validity of the Airy formula (\ref{ABJM:S3b:Airy}) under the parameter identification $\beta=\pm\pi\ri b^2~(\Re[\beta]>0)$ presented in (\ref{beta:b}), which underpins the factorization formula (\ref{SCI:2nd:5}). Moreover, we observe that the Airy form (\ref{ABJM:S3b:Airy}) arises from a particular saddle point $\boldhu^\star$ of the matrix model (\ref{ABJM:S3b}) modulo its orbit under Weyl reflections.\footnote{This has been demonstrated explicitly in the 't~Hooft limit with large $N$ fixed $\lambda=N/k$ \cite{Geukens:2024zmt} as well as in the M-theory limit with large $N$ fixed $k$ \cite{Drukker:2010nc}.} This allows us to specify the Airy form as a particular saddle contribution as 
\begin{align}
	Z^{\text{ABJM}\,(\boldhu^\star)}(\omega,k,\boldvarphi)=\mC^{-1/3}e^{\mA}\text{Ai}\Big[\mC^{-1/3}(N-\mB)\Big]\Big(1+\mO(e^{-\#\sqrt{N}})\Big) \,. \label{ABJM:S3b:saddle}
\end{align}%
Contributions from all remaining saddles are expected to be exponentially suppressed in the large-$N$ limit, see \cite{Bobev:2022eus,Bobev:2023lkx,Bobev:2025ltz} for related discussions.

\medskip

Now, by substituting a particular saddle contribution to the ABJM $S^3_b$ partition function~(\ref{ABJM:S3b:saddle}) into the factorization formula (\ref{SCI:2nd:5}), one obtains the corresponding saddle contribution to the ABJM SCI (\ref{ABJM:SCI}). The result reads 
\begin{align}
	\mI^{\text{ABJM}\,(\boldu^\star,\boldbu^\star)}_{(\pm)}(\beta,k,\boldlambda,\boldmn)&\approx\mC_{(\pm)}^{-1/3}e^{\mA_{(\pm)}}\text{Ai}\Big[\mC_{(\pm)}^{-1/3}(N-\mB_{(\pm)})\Big] \, \bmC_{(\pm)}^{-1/3}e^{\bmA_{(\pm)}}\text{Ai}\Big[\bmC_{(\pm)}^{-1/3}(N-\bmB_{(\pm)})\Big]\nn\\
	&\quad\times\Big(1+\mO(e^{-\#\sqrt{N}})\Big)\,,\label{ABJM:SCI:factorization}
\end{align}
where we have implicitly used the parity invariance \eqref{ABJM:CS}. Here, the Airy parameters $(\mA_{(\pm)},\mB_{(\pm)},\mC_{(\pm)})$ and $(\bmA_{(\pm)},\bmB_{(\pm)},\bmC_{(\pm)})$ are obtained from \eqref{ABJM:Airy:coeffi} by replacing $\boldvarphi$ with $\boldvarphi^{(\pm)}$ and $\boldbvarphi^{(\pm)}$ respectively, where the latter are related to the SCI parameters via (\ref{complexification}), (\ref{beta:b}), (\ref{identification}) and (\ref{ABJM:omegavarphi}) as 
\begin{equation}
\begin{alignedat}{3}
	\omega&=b^2&&&&=\pm\fft{\beta}{\pi\ri}\,,\\
	\varphi_a^{(\pm)}&=(1+b^2)\Delta_a^{(\pm)}&&=(1+b^2)\bigg(r_a+\fft{2\hv_a}{\ri Q}\bigg)&&=\bigg(1\pm\fft{\beta}{\pi\ri}\bigg)r_a\mp\fft{\ri\lambda_a+\beta\mn_a}{\pi\ri}\,,\\
	\bvarphi_a^{(\pm)}&=(1+b^2)\bDelta_a^{(\pm)}&&=(1+b^2)\bigg(r_a+\fft{2\hbv_a}{\ri Q}\bigg)&&=\bigg(1\pm\fft{\beta}{\pi\ri}\bigg)r_a\pm\fft{-\ri\lambda_a+\beta\mn_a}{\pi\ri}\,.
\end{alignedat}
\end{equation}
When this particular saddle point contribution (\ref{ABJM:SCI:factorization}) is dominant, the full ABJM SCI (\ref{ABJM:SCI}) is captured by a product of two Airy functions up to non-perturbative corrections in the Cardy-like expansion as
\begin{align}
	\mI^{\text{ABJM}}_{{\rm SCI}\,(\pm)}(\beta,k,\boldlambda,\boldmn)&\approx\mC_{(\pm)}^{-1/3}e^{\mA_{(\pm)}}\text{Ai}\Big[\mC_{(\pm)}^{-1/3}(N-\mB_{(\pm)})\Big] \, \bmC_{(\pm)}^{-1/3}e^{\bmA_{(\pm)}}\text{Ai}\Big[\bmC_{(\pm)}^{-1/3}(N-\bmB_{(\pm)})\Big]\nn\\
	&\quad\times\Big(1+\mO(e^{-\#\sqrt{N}})\Big)\,.\label{ABJM:SCI:factorization:dom}
\end{align}
Several comments on this factorized form of the ABJM SCI are in order.

\begin{itemize}
	\item The factorized expression (\ref{ABJM:SCI:factorization:dom}) significantly extends previous saddle-point analyses, which captured only the $\beta^{-1}$ and $\beta^0$ terms in $\log\mI$ and only for vanishing flavor magnetic fluxes~\cite{Bobev:2022wem,Bobev:2024mqw}. Here, the full Cardy-like expansion is determined to all orders in $\beta$ and for general flavor magnetic fluxes.
	
	\item The factorized ABJM SCI (\ref{ABJM:SCI:factorization:dom}) perfectly agrees with the expression conjectured from the supergravity side \cite{Hristov:2022lcw}, including the $N$-independent behavior governed by $\mA_{(\pm)},\bmA_{(\pm)}$ as incorporated later in \cite{Bobev:2022wem}. Note that this constitutes the first direct field-theoretic confirmation of the ABJM SCI factorization to all orders in the Cardy-like expansion.
	
	\item In the large $N$ limit, the factorization formula (\ref{ABJM:SCI:factorization:dom}) yields
	\begin{align}
		-\log\mI_{{\rm SCI}\,(\pm)}^\text{ABJM}&\approx\fft23(\mC_{(\pm)}^{-\fft12}+\bmC_{(\pm)}^{-\fft12})N^\fft32-(\mC_{(\pm)}^{-\fft12}\mB_{(\pm)}+\bmC_{(\pm)}^{-\fft12}\bmB_{(\pm)})N^\fft12 \nn \\
		&\quad + \fft12\log N + \mO(N^0)\,.\label{ABJM:SCI:factorization:largeN}
	\end{align}
	The large $N$ structure (\ref{ABJM:SCI:factorization:largeN}) has been explored extensively in the context of holographic duality. The $N^{\fft32}$ leading behavior was identified with the supergravity on-shell action of a dual supersymmetric Kerr-Newman (KN) AdS$_4$ black hole and, via Legendre transform, its entropy function~\cite{Choi:2018fdc,Cassani:2019mms,Hristov:2019mqp,Bobev:2019zmz}. The $N^{\fft12}$ term was investigated holographically only in the universal limit with vanishing flavor parameters ($\boldlambda=\boldmn=0$), when~(\ref{ABJM:SCI:factorization:largeN}) reduces to
	\begin{align}
		-\log\mI_{{\rm SCI}\,(\pm)}^\text{ABJM}\Big|_\text{universal}&\approx \fft{\pi\sqrt{2k}}{3}\bigg[\fft{(\omega+1)^2}{2\omega}\bigg(N^\fft32+\bigg(\fft{1}{k}-\fft{k}{16}\bigg)N^\fft12\bigg)-\fft{3}{k}N^\fft12\bigg]\nn\\
		&\quad+\fft12\log N+\mO(N^0)\,,
	\end{align}
	and the $N^\fft12$ term matches the 4-derivative corrections to the KN AdS$_4$ on-shell action~\cite{Bobev:2020egg,Bobev:2021oku,Bobev:2022wem}. Finally, the logarithmic behavior characterized by the coefficient $\fft12$ also agrees with the 1-loop supergravity analysis in~\cite{Bobev:2023dwx}.
	
	Previously, however, the aforementioned holographic comparisons were restricted to leading $\beta^{-1}$ order or at best subleading $\beta^{0}$ order in the Cardy-like limit~\cite{Bobev:2022wem,Bobev:2024mqw}. In other words, the coefficients of the first few leading terms in the large $N$ expansion (\ref{ABJM:SCI:factorization:largeN}) were not accessible beyond the $\beta^{0}$ order from the field theory side. The large $N$ ABJM SCI (\ref{ABJM:SCI:factorization:largeN}) derived using~(\ref{ABJM:SCI:factorization:dom}) finally resolves this limitation, allowing for precision holographic tests at all perturbative orders in the Cardy-like expansion.
	
	\item The factorized expression of the ABJM SCI in (\ref{ABJM:SCI:factorization:dom}) was derived using the grading $(-1)^F = e^{\mp \pi\ri R}$, as explained in Section~\ref{sec:susyloc:S1S2}. Nevertheless, its functional form continues to hold in the alternative grading $(-1)^F = e^{2\pi \ri j_3}$, provided one identifies $\lambda_a \mp \pi r_a$ in the former grading with $\lambda_a$ in the latter. This equivalence follows from the fact that, after this simple redefinition, the matrix-model representations of the ABJM SCI in the two gradings coincide.\footnote{More precisely, under this identification the integrands in the two gradings differ by a phase factor of the form $\prod_{i\neq j}^N e^{\fft{\pi\ri}{2}(|\mm_{ij}|+|\tmm_{ij}|)}$. This discrepancy produces effective-action terms linear in the complexified gauge holonomies $u_i,\tilde u_i$, which may shift the locations of the saddle points and modify the overall phase of the index but leave the modulus of the corresponding saddle-point contributions unchanged. The overall phase is in any case ambiguous, owing to the branch choices for the fractional exponents in the localization formula. For this reason, throughout this paper we compare our factorization formulae for the SCI and TTI with previous results~\cite{Bobev:2022eus,Bobev:2022wem,Bobev:2023lkx,Bobev:2024mqw} only at the level of the real parts of their logarithms. \label{foot:phase}} This observation clarifies why the factorization formula (\ref{ABJM:SCI:factorization:dom}), obtained in the $(-1)^F = e^{\mp \pi\ri R}$ grading, reproduces results consistent with earlier saddle-point analyses performed in the $(-1)^F = e^{2\pi \ri j_3}$ grading~\cite{Choi:2019zpz,Bobev:2022wem,Bobev:2024mqw}. It also explains how the 2nd sheet is accessed differently in the two gradings: in the $(-1)^F = e^{\mp \pi\ri R}$ grading, the characteristic $N^{\fft32}$ scaling of the second sheet is visible at vanishing flavor chemical potentials $\boldsymbol{\lambda}$, whereas in the $(-1)^F = e^{2\pi \ri j_3}$ grading one must impose the non-trivial constraint $\sum_a \lambda_a = \mp 2\pi$ in order to probe the same sector.
\end{itemize}
%

%%%%%
\subsection{ADHM theory}\label{sec:ex:ADHM}
%%%%%
In this subsection, we apply the factorization formula (\ref{SCI:2nd:5}) to the ADHM theory, also known as the $N_f$ matrix model, in close parallel with our analysis of the ABJM case. In the UV, the ADHM theory is described by a $\U(N)$ gauge theory containing three adjoint $\mN=2$ chiral multiplets together with $N_f$ pairs of fundamental and anti-fundamental chiral multiplets. For detailed discussions of the theory and its underlying ADHM construction, we refer the reader to \cite{Bobev:2022eus,Geukens:2024zmt,Mezei:2013gqa,Grassi:2014vwa,Atiyah:1978ri}, while in the following we focus on the implementation of the factorization formula.

%%%%%
\subsubsection{Supersymmetric partition functions}\label{sec:ex:ADHM:mm}
%%%%%
The $S^3_b$ partition function of the ADHM theory can be written as a matrix integral using the general localization formula (\ref{S3b}) in the form 
\begin{align}
	Z^\text{ADHM}(\omega,N_f,\boldvarphi)&=\fft{1}{N!}\int\bigg(\prod_{i=1}^Nd\mu_i e^{-2\pi\ri \mu_i \hv_T}\bigg)\times\prod_{i\neq j}s_b\bigg(\fft{\ri Q}{2}-\mu_{ij}\bigg)^{-1}\label{ADHM:S3b}\\
	&\quad\times\prod_{i,j=1}^N\prod_{I=1}^3s_b\bigg(\fft{\ri Q}{2}(1-r_I)-\mu_{ij}-\hv_I\bigg)\nn\\
	&\quad\times\prod_{i=1}^N\bigg[s_b\bigg(\fft{\ri Q}{2}(1-r_f)-\mu_i-\hv_f\bigg)s_b\bigg(\fft{\ri Q}{2}(1-r_{\tf})+\mu_i-\hv_{\tf}\bigg)\bigg]^{N_f}\,,\nn
\end{align}
where at the superconformal point the $R$ charges take values $(r_I,r_f,r_{\tf},r_T)=(\fft12,\fft12,1,\fft12,\fft12,0)$ with $I\in\{1,2,3\}$, and the mixed CS level between gauge and topological symmetries is set to unity.  As in the ABJM theory, we have replaced $(b,\boldhv)$ with $(\omega,\boldvarphi)$ defined by 
\begin{align}
	\omega\equiv b^2\qquad\&\qquad \varphi_a=(1+b^2)\mbDelta_a\,.\label{ADHM:omegavarphi}
\end{align}
Here the $\boldmbDelta$ are a useful recombination of the original $\boldDelta$ parameters
\begin{align}
	\Delta_X\equiv r_X+\fft{2\hv_X}{\ri Q}~~(X\in\{1,2,3,f,\tf,T\}) \quad\&\quad  \sum_{I=1}^3\Delta_I=\Delta_3+\Delta_f+\Delta_{\tf}=2\,,\label{ADHM:Delta}
\end{align}
which are introduced as
\begin{align}
	\mbDelta_a=(\Delta_1,\Delta_2,\fft{2-\Delta_f-\Delta_{\tf}}{2}+\fft{\Delta_T}{N_f},\fft{2-\Delta_f-\Delta_{\tf}}{2}-\fft{\Delta_T}{N_f})\,.\label{ADHM:mbDelta}
\end{align}
The ADHM $S^3_b$ partition function \eqref{ADHM:S3b} also takes the Airy form in the large $N$ limit as \cite{Bobev:2025ltz} 
\begin{align}
	Z^\text{ADHM}(\omega,N_f,\boldvarphi)=\mC^{-1/3}e^{\mA}\text{Ai}\Big[\mC^{-1/3}(N-\mB)\Big]\Big(1+\mO(e^{-\#\sqrt{N}})\Big)\,, \label{ADHM:S3b:Airy}
\end{align}
analogous to the ABJM case. The Airy parameters for the ADHM theory read ($\varphi_{ab}\equiv\varphi_a+\varphi_b$)
\begin{subequations}
	\begin{align}
		\mC(\omega,N_f,\boldvarphi)&=\fft{2\omega^2}{\pi^2 N_f\varphi_1\varphi_2\varphi_3\varphi_4}\,,\\
		\mB(\omega,N_f,\boldvarphi)&=\fft{N_f}{24}+\fft{1}{12N_f\varphi_1\varphi_2\varphi_3\varphi_4}\Big(2\omega(\varphi_1\varphi_2+\varphi_{12}\varphi_{34}+N_f^2\varphi_3\varphi_4)\\
		&\kern11em~-(1+\omega)(\varphi_1\varphi_2\varphi_{34}+N_f^2\varphi_3\varphi_4\varphi_{12})\Big)\,,\nn
	\end{align}\label{ADHM:Airy:coeffi}%
\end{subequations}
while the form of the $N$-independent parameter $\mA(\omega,N_f,\boldvarphi)$ for special ranges of the parameters can be found in~\cite{Bobev:2025ltz}. The Airy formula for the ADHM theory has been derived analytically and supported by numerical evidence in various special cases, assuming $b>0$ and $0<\mbDelta_a<1$.

\medskip

The SCI of the ADHM theory can also be written in terms of a matrix model based on the general formula (\ref{SCI:2nd}) as 
\begin{align}
	&\mI_{{\rm SCI}\,(\pm)}^\text{ADHM}(\beta,N_f,\boldlambda,\boldmn)\nn\\
	&=\fft{1}{N!}\sum_{\boldmm\in\mathbb{Z}^N}\int \fft{d\boldh}{(2\pi)^N}\prod_{i=1}^{N}e^{\ri\mn_T h_i+\ri \mm_i\lambda_T} \times\prod_{i\neq j}e^{\fft12(\beta\pm\pi\ri)|\mm_{ij}|}\Big(1-e^{\ri h_{ij}} q^{|\mm_{ij}|}\Big)\nn\\
	&\quad\times\prod_{I=1}^3\prod_{i,j=1}^N\Big(e^{-\ri h_{ij}-\ri\lambda_I-(\beta\pm\pi\ri)(1-r_I)}\Big)^{\fft12|\mm_{ij}+\mn_I|}\fft{(e^{-\ri h_{ij}-\ri\lambda_I+(\beta\pm\pi\ri)r_I}q^{2+|\mm_{ij}+\mn_I|};q^2)}{(e^{\ri h_{ij}+\ri\lambda_I-(\beta\pm\pi\ri)r_I}q^{|\mm_{ij}+\mn_I|};q^2)} \nn \\
	&\quad\times\prod_{i=1}^N\bigg[\Big(e^{-\ri h_i-\ri\lambda_f-(\beta\pm\pi\ri)(1-r_f)}\Big)^{\fft12|\mm_i+\mn_f|}\fft{(e^{-\ri h_i-\ri\lambda_f+(\beta\pm\pi\ri)r_f}q^{2+|\mm_i+\mn_f|};q^2)}{(e^{\ri h_i+\ri\lambda_f-(\beta\pm\pi\ri)r_f}q^{|\mm_i+\mn_f|};q^2)}\nn\\
	&\kern4em\times\Big(e^{\ri h_i-\ri\lambda_{\tf}-(\beta\pm\pi\ri)(1-r_{\tf})}\Big)^{\fft12|-\mm_i+\mn_{\tf}|}\fft{(e^{\ri h_i-\ri\lambda_{\tf}+(\beta\pm\pi\ri)r_{\tf}}q^{2+|-\mm_i+\mn_{\tf}|};q^2)}{(e^{-\ri h_i+\ri\lambda_{\tf}-(\beta\pm\pi\ri)r_{\tf}}q^{|-\mm_i+\mn_{\tf}|};q^2)}\bigg]^{N_f}\,.\label{ADHM:SCI}
\end{align}
The logarithm of the matrix integral (\ref{ADHM:SCI}) has been investigated in the Cardy-like limit using a saddle-point approximation, including the leading $\beta^{-1}$ and subleading $\beta^{0}$ contributions, where the latter was restricted to the case of vanishing flavor magnetic fluxes \cite{Choi:2019zpz,Bobev:2022wem,Bobev:2024mqw}. In the next subsection, we extend these results by deriving a factorized representation of the ADHM SCI based on (\ref{SCI:2nd:5}), which captures perturbatively exact information to all orders in the Cardy-like expansion.

%%%%%
\subsubsection{SCI in terms of $Z_{S_b^3}$}
\label{sec:ex:ADHM:fact}
%%%%%
As in the ABJM case, in order to apply the factorization formula (\ref{SCI:2nd:5}) to the ADHM theory, we first analytically continue the Airy formula of the ADHM $S^3_b$ partition function (\ref{ADHM:S3b:Airy}) to complex values of $(b,\boldmbDelta)$ subject to $\Re[b]>0$ and $0<\Re[\mbDelta_a]<1$, and regard the Airy form~(\ref{ADHM:S3b:Airy}) as the contribution of a particular saddle point to the matrix model. On top of that, we restrict our attention to the sector in which the flavor symmetry parameters associated with the fundamental and anti-fundamental multiplets coincide as $(\hv_f,\lambda_f,\mn_f)=(\hv_{\tf},\lambda_{\tf},\mn_{\tf})$, a choice that does not entail any essential loss of generality.\footnote{For instance, the identification $\hv_f=\hv_{\tf}$ can always be achieved by an appropriate shift of integration variables, modifying the partition function only by an overall phase \cite{Geukens:2024zmt}.} Under this assumption, the ADHM matrix models presented in the previous subsection become invariant under a sign flip of the mixed CS level between the gauge and topological symmetries.

\medskip

Substituting the Airy representation of the ADHM $S^3_b$ partition function (\ref{ADHM:S3b:Airy}) into the factorization formula (\ref{SCI:2nd:5}), and interpreting it as the contribution of a particular saddle point, the ADHM SCI (\ref{ADHM:SCI}) takes a factorized form. Concretely, one finds
\begin{align}
	\mI^{\text{ADHM}}_{{\rm SCI}\,(\pm)}(\beta,N_f,\boldlambda,\boldmn)&\approx\mC_{(\pm)}^{-1/3}e^{\mA_{(\pm)}}\text{Ai}\Big[\mC_{(\pm)}^{-1/3}(N-\mB_{(\pm)})\Big] \, \bmC_{(\pm)}^{-1/3}e^{\bmA_{(\pm)}}\text{Ai}\Big[\bmC_{(\pm)}^{-1/3}(N-\bmB_{(\pm)})\Big]\nn\\
	&\quad\times\Big(1+\mO(e^{-\#\sqrt{N}})\Big)\,,\label{ADHM:SCI:factorization:dom}
\end{align}
where we have used the invariance of the ADHM matrix model under a sign flip of the mixed CS level discussed above, and restricted to a regime in which the Airy saddle provides the dominant contribution. As in the ABJM case, the Airy parameters $(\mA_{(\pm)},\mB_{(\pm)},\mC_{(\pm)})$ and $(\bmA_{(\pm)},\bmB_{(\pm)},\bmC_{(\pm)})$ are obtained from \eqref{ADHM:Airy:coeffi} by replacing $\boldvarphi$ with $\boldvarphi^{(\pm)}$ and $\boldbvarphi^{(\pm)}$ respectively, where the latter are related to the SCI parameters via \eqref{complexification}, \eqref{beta:b}, \eqref{identification} together with the ADHM specialized relations \eqref{ADHM:omegavarphi}, \eqref{ADHM:Delta}, and \eqref{ADHM:mbDelta}. The key identifications consist of two parts
\begin{equation}
	\omega=b^2=\pm\fft{\beta}{\pi\ri}\,,\qquad \varphi_a^{(\pm)}=(1+b^2)\mbDelta_a^{(\pm)}\,,\qquad \bvarphi_a^{(\pm)}=(1+b^2)\bmbDelta_a^{(\pm)}\,, 
\end{equation}
and
\begin{equation}
	\Delta_X^{(\pm)}=r_X+\fft{2\hv_X}{\ri Q}=r_X-\fft{\ri\lambda_X+\beta\mn_X}{\beta\pm\pi\ri}\,,\quad \bDelta_X^{(\pm)}=r_X+\fft{2\hbv_X}{\ri Q}=r_X+\fft{-\ri\lambda_X+\beta\mn_X}{\beta\pm\pi\ri}\,,
\end{equation}
where $\mbDelta_a^{(\pm)}$'s and $\Delta_X^{(\pm)}$'s (the overlined parameters as well) are related to each other through the same linear combination as in \eqref{ADHM:mbDelta}. Most of the comments on the factorized ADHM superconformal index \eqref{ADHM:SCI:factorization:dom} closely parallel those for the ABJM theory, and we therefore keep the discussion brief. 

\begin{itemize}
	\item The factorized form \eqref{ADHM:SCI:factorization:dom} goes well beyond earlier saddle point treatments, which were limited to the $\beta^{-1}$ and $\beta^{0}$ contributions to $\log\mI$ \cite{Bobev:2022wem,Bobev:2024mqw}, by providing control over the SCI to all orders in the Cardy-like expansion. Especially in the large $N$ regime, the formula (\ref{ADHM:SCI:factorization:dom}) exhibits precisely the same asymptotic structure given in (\ref{ABJM:SCI:factorization:largeN}), now with coefficients that are determined perturbatively exactly in the Cardy-like limit. This refinement substantially strengthens earlier holographic tests confined to the Cardy-like limit.
	
	\item The factorized ADHM SCI \eqref{ADHM:SCI:factorization:dom} is manifestly consistent with the expression conjectured in \cite{Cassia:2025aus,Cassia:2025jkr}, which was motivated by the introduction of an equivariant topological string partition function on the Calabi–Yau fourfold supporting the stack of $N$ M2-branes. Crucially, our result \eqref{ADHM:SCI:factorization:dom} follows from a direct evaluation of the ADHM SCI matrix model (\ref{ADHM:SCI}), and in addition captures the full $N$-independent structure not studied in~\cite{Cassia:2025aus,Cassia:2025jkr}.
	
	\item Analogous to the ABJM case, the factorized expression of the ADHM SCI \eqref{ADHM:SCI:factorization:dom} can also be employed in the $(-1)^F=e^{2\pi\ri j_3}$ grading, provided that the flavor chemical potentials are shifted appropriately by their corresponding $R$-charges.
\end{itemize}
%

%%%%%
\section{Topologically twisted index and $Z_{S^3_b}$}
\label{sec:TTI}
%%%%%
In this section, we extend the factorization analysis of Section~\ref{sec:fact} to the refined topologically twisted index (TTI) on $S^1\times S^2$, establishing a relation of the form \eqref{eq:ITTIZZbintro} to all orders in the Cardy-like limit~\eqref{Cardy}. The logic closely parallels that of the SCI, so we keep the discussion concise and emphasize only the points specific to the TTI.

%%%%%
\subsection{Localization formula and factorization}
\label{sec:fact:TTI}
%%%%%
The refined TTI of a 3d $\mN=2$ gauge theory on $S^1\times S^2$ is defined as a trace over the radially quantized Hilbert space twisted along $S^2$ by the $R$-symmetry magnetic flux \cite{Benini:2015noa,Closset:2016arn,Benini:2016hjo,Closset:2017zgf,Closset:2018ghr,Closset:2019hyt,Choi:2019dfu,Hosseini:2022vho},
\begin{align}
	\mI_{{\rm TTI}\,(2\nu_R)}(\beta,\boldk,\boldlambda,\boldmn) = \Tr\bigg[(e^{(1+2\nu_R)\pi\ri})^F\,(e^{2\pi\ri\nu_R})^R\,q^{2j_3}\,\prod_{x=1}^{r_F}e^{\ri\lambda_xF_x}\bigg]\,,\label{TTI:tr}
\end{align}
where $q=e^{-\beta}$ is the fugacity for the angular momentum $j_3$ generating the rotational isometry of $S^2$, and the $R$-symmetry holonomy along $S^1$ is encoded in $\nu_R\in\mathbb{Z}/2$. Throughout this section we restrict to half-integer holonomy
\begin{equation}
	2\nu_R=\pm 1\,, \label{nuR:half}
\end{equation}
corresponding to anti-periodic fermions along the temporal circle, in close analogy with the 2nd-sheet SCI~\eqref{SCI-tr:2nd}. We focus on theories whose matter fields carry half-integer $R$-charges, for which the choice~\eqref{nuR:half} entails no loss of generality. This choice also eliminates the subtle phase factor arising from the diagonal CS contributions in the localization formula, which has been extensively discussed in~\cite{Closset:2017zgf,Closset:2018ghr,Closset:2019hyt}. 

\medskip

Supersymmetric localization yields the matrix-model representation~\cite{Benini:2015noa,Closset:2016arn,Choi:2019dfu,Closset:2018ghr,Closset:2019hyt}
\begin{align}
	&\mI_{{\rm TTI}\,(2\nu_R)}(\beta,\boldk,\boldlambda,\boldmn) \nn \\
	&=\fft{1}{|\mW|}\sum_{\boldmm\in Q^\vee(\mfg)}\oint_{\rm JK} \fft{d\boldh}{(2\pi)^{r_G}}\,\prod_{\ell=1}^{r_G}z_\ell^{\sum_{n=1}^{r_G}k^{\ell n}\mm_n+\sum_{x=1}^{r_F}k^{\ell x}\mn_x}\prod_{x=1}^{r_F}\xi_x^{\sum_{y=1}^{r_F}k^{xy}\mn_y+\sum_{\ell=1}^{r_G}k^{x\ell}\mm_\ell}\nn\\
	&\quad\times (e^{2\pi\ri\nu_R})^{|\mR_+[\mfg]|}\prod_{\alpha\in\mR[\mfg]}(e^{2\pi\ri\nu_R}q)^{-\fft{|\alpha(\boldmm)|}{2}}\Big(1-\boldz^\alpha q^{|\alpha(\boldmm)|}\Big)\label{TTI:loc}\\
	&\quad\times\prod_{\Psi}\prod_{\rho_\Psi}\prod_{\trho_\Psi}\fft{\Big(e^{\ri\rho_\Psi(\boldh)+\ri\trho_\Psi(\boldlambda)}(e^{2\pi\ri\nu_R})^{r_\Psi-1}\Big)^{\fft{\rho_\Psi(\boldmm)+\trho_\Psi(\boldmn)+(r_\Psi-1)\mn_R}{2}}}{(e^{\ri\rho_\Psi(\boldh)+\ri\trho_\Psi(\boldlambda)}(e^{2\pi\ri\nu_R})^{r_\Psi}q^{1-(\rho_\Psi(\boldmm)+\trho_\Psi(\boldmn)+(r_\Psi-1)\mn_R)};q^2)_{\rho_\Psi(\boldmm)+\trho_\Psi(\boldmn)+(r_\Psi-1)\mn_R}}\,,\nn
\end{align}
where the contour extracts the Jeffrey-Kirwan residues and the $R$-symmetry magnetic flux is fixed to $\mn_R=-1$. With the lower sign choice in \eqref{nuR:half}, the formula~\eqref{TTI:loc} reduces to the expression of~\cite{Choi:2019dfu} up to the overall phase $(e^{2\pi\ri\nu_R})^{|\mR_+[\mfg]|}$, where $|\mR_+[\mfg]|$ denotes the number of positive roots. We emphasize that such an overall phase was not unambiguously fixed in the original localization formula~\cite{Benini:2015noa} and our choice can be understood as a benign integer $RR$ contact term in the formulation of~\cite{Closset:2018ghr,Closset:2019hyt}, which identifies the $R$-charge $r_\Psi=2$ chiral-multiplet contribution with the vector-multiplet contribution as in the squashed three-sphere partition function and the SCI.

\medskip

Following the procedure detailed in Section~\ref{sec:fact:SCI} for the SCI, we eliminate the absolute value signs in~\eqref{TTI:loc} via the $\infty$-Pochhammer identity~\eqref{poch:identity}, introduce the complexified parameters~\eqref{complexification}, apply the Euler-Maclaurin approximation~\eqref{EM} to the gauge magnetic-flux sums, and change integration variables via~\eqref{change}. The result takes the manifestly factorized form
\begin{align}
	\mI_{{\rm TTI}\,(2\nu_R)}(\beta,\boldk,\boldlambda,\boldmn)=\fft{1}{|\mW|}\int\bigg(\prod_{\ell=1}^{r_G}du_\ell\wedge d(-\bu_\ell)\bigg)\,e^{-S_{(2\nu_R)}(\boldu;\beta,\boldk,\boldv)}\,e^{-S_{(2\nu_R)}(-\boldbu;-\beta,\boldk,-\boldbv)}\,,\label{TTI:fact}
\end{align}
where the holomorphic effective action reads 
\begin{align}
	e^{-S_{(2\nu_R)}(\boldu;\beta,\boldk,\boldv)}&\equiv \fft{1}{(4\pi\beta)^{r_G/2}}\,e^{\fft{\sum_{\ell,n}k^{\ell n}u_\ell u_n+2\sum_{\ell,x}k^{\ell x}u_\ell v_x+\sum_{x,y}k^{xy}v_xv_y}{4\beta}}\nn\\
	&\quad\times\prod_{\alpha\in\mR[\mfg]} \fft{e^{-\fft{\pi\ri}{2}\big(-\fft{(-(\beta-2\nu_R\pi\ri)+\alpha(\boldu))^2}{4\pi\ri\beta}+\fft{\fft{\beta}{\pi\ri}+\fft{\pi\ri}{\beta}}{12}\big)}}{(e^{\alpha(\boldu)}q^2;q^2)}\nn\\
	&\quad\times\prod_{\Psi}\prod_{\rho_\Psi}\prod_{\trho_\Psi} \fft{e^{-\fft{\pi\ri}{2}\big(-\fft{((\beta-2\nu_R\pi\ri)+\rho_\Psi(\boldu)+\trho_\Psi(\boldv)-(\beta-2\nu_R\pi\ri)r_\Psi)^2}{4\pi\ri\beta}+\fft{\fft{\beta}{\pi\ri}+\fft{\pi\ri}{\beta}}{12}\big)}}{(e^{\rho_\Psi(\boldu)+\trho_\Psi(\boldv)-(\beta-2\nu_R\pi\ri)r_\Psi};q^2)}\,.\label{TTI:S}
\end{align}
In contrast to the SCI case~\eqref{S:bS}, where the anti-holomorphic action $\bS_{(\pm)}$ is obtained from $S_{(\pm)}$ through the replacement $\ri\to-\ri$, here the anti-holomorphic contribution is governed by the same functional $S_{(2\nu_R)}$, merely evaluated on the negated conjugate variables $(-\boldbu,-\boldbv)$.

\medskip

Introducing the holomorphic and anti-holomorphic matrix integrals as in Section~\ref{sec:fact:saddle},
\begin{subequations}
\begin{align}
	\mZ_{(2\nu_R)}(\beta,\boldk,\boldv)&\equiv\fft{1}{|\mW|}\int\prod_{\ell=1}^{r_G}du_\ell\,e^{-S_{(2\nu_R)}(\boldu;\beta,\boldk,\boldv)}\,,\\
	\mZ_{(2\nu_R)}(-\beta,\boldk,-\boldbv)&\equiv\fft{1}{|\mW|}\int\prod_{\ell=1}^{r_G}d(-\bu_\ell)\,e^{-S_{(2\nu_R)}(-\boldbu;-\beta,\boldk,-\boldbv)}\,,
\end{align}
\end{subequations}
and applying the saddle point approximation, the TTI~\eqref{TTI:fact} takes the factorized form 
\begin{align}
	\mI_{{\rm TTI}\,(2\nu_R)}(\beta,\boldk,\boldlambda,\boldmn)= \sum_{(\boldu^\star,\boldbu^\star)\in\mS_I}\mZ^{(\boldu^\star)}_{(2\nu_R)}(\beta,\boldk,\boldv)\,\mZ^{(-\boldbu^\star)}_{(2\nu_R)}(-\beta,\boldk,-\boldbv)\,,\label{TTI:saddle}
\end{align}
with $(\boldv,\boldbv)=(\ri\boldlambda+\beta\boldmn,-\ri\boldlambda+\beta\boldmn)$. The contributing saddles form the product space $\mS_I=\mS_\mZ\times\bmS_\mZ$, where the holomorphic and anti-holomorphic saddle sets are defined as in~\eqref{Z:bZ:saddle},
\begin{subequations}
\begin{align}
	\mS_\mZ&=\bigg\{\boldu\,\bigg|\,\fft{\partial S_{(2\nu_R)}(\boldu;\beta,\boldk,\boldv)}{\partial u_\ell}=0\,,~~w\cdot\boldu\neq\boldu\,,~\forall w\in\mW\bigg\}\big/\mW\,,\\
	\bmS_\mZ&=\bigg\{\boldbu\,\bigg|\,\fft{\partial S_{(2\nu_R)}(-\boldbu;-\beta,\boldk,-\boldbv)}{\partial \bu_\ell}=0\,,~~w\cdot\boldbu\neq\boldbu\,,~\forall w\in\mW\bigg\}\big/\mW\,.
\end{align}
\end{subequations}

\medskip

To identify the $\mZ_{(2\nu_R)}$ factors with the $S^3_b$ partition function~\eqref{S3b}, we introduce distinct parameter mappings for the holomorphic and anti-holomorphic parts: 
\begin{subequations}
	\begin{align}
		\text{(holomorphic)}\quad&: \quad \beta=-2\pi\ri\nu_R\,b^2\,,&&\quad u_\ell=4\pi\nu_R b\,\hu_\ell\,,&&\quad v_x=4\pi\nu_R b\,\hv_x\,, \label{TTI:b:hol}\\
		\text{(anti-holomorphic)}\quad&: \quad \beta=2\pi\ri\nu_R \tb^2\,,&&\quad \bu_\ell=-4\pi\nu_R \tb\,\htu_\ell\,,&&\quad \bv_x=-4\pi\nu_R \tb\,\htv_x\,,\label{TTI:b:ahol}
	\end{align}\label{TTI:b}%
\end{subequations}
with $b$ and $\tb$ playing the role of squashing parameters for the holomorphic and anti-holomorphic parts respectively, each fixed unambiguously by $\Re[b]>0$ and $\Re[\tb]>0$. Implementing the change of variables \eqref{TTI:b} and expressing the $\infty$-Pochhammer symbols in terms of the double sine function via \eqref{sb:poch}, the matrix integrals coincide with the $S^3_b$ partition function up to overall phases and $\infty$-Pochhammer factors involving $e^{-2\pi^2/\beta}$, which are exponentially suppressed in the Cardy-like limit~\eqref{Cardy} as discussed in the SCI case. The corresponding saddle-point contributions are therefore related as
\begin{subequations}
\begin{align}
	\mZ^{(\boldu^\star)}_{(2\nu_R)}(\beta,\boldk,\boldv) &\approx e^{\ri\phi_{(2\nu_R)}}\,Z^{(\boldhu^\star)}(b,-2\nu_R\boldk,\boldhv)\,,\\
	\mZ^{(-\boldbu^\star)}_{(2\nu_R)}(-\beta,\boldk,-\boldbv) &\approx e^{\ri\tphi_{(2\nu_R)}}\,Z^{(\boldhtu^\star)}(\tb,-2\nu_R\boldk,\boldhtv)\,,
\end{align}\label{TTI:Z:to:S3b}%
\end{subequations}
where the saddle-independent phases combine to give
\begin{align}
	e^{\ri(\phi_{(2\nu_R)}+\tphi_{(2\nu_R)})}=(-1)^{r_G}\,.\label{TTI:phase}
\end{align}
Substituting~\eqref{TTI:Z:to:S3b} into~\eqref{TTI:saddle} yields the main result of this section,
\begin{align}
	\mI_{{\rm TTI}\,(2\nu_R)}(\beta,\boldk,\boldlambda,\boldmn)\approx (-1)^{r_G}\sum_{(\boldu^\star,\boldbu^\star)\in\mS_I}Z^{(\boldhu^\star)}(b,-2\nu_R\boldk,\boldhv)\,Z^{(\boldhtu^\star)}(\tb,-2\nu_R\boldk,\boldhtv)\,,\label{TTI:main}
\end{align}
which is the TTI counterpart of the SCI factorization formula~\eqref{SCI:2nd:5} and likewise holds to all orders in the Cardy-like expansion~\eqref{Cardy} up to non-perturbative corrections.

%%%%%
\subsection{Examples}\label{sec:ex:TTI}
%%%%%
We illustrate the factorization formula~\eqref{TTI:main} for the chiral multiplet and the ABJM/ADHM theories, in close parallel with the SCI analysis of Section~\ref{sec:ex}.

%%%%%
\subsubsection{Chiral multiplet}\label{sec:ex:TTI:chiral}
%%%%%
Specializing the localization formula~\eqref{TTI:loc} to a single chiral multiplet $\Psi$ of $R$ charge $r_\Psi$ with trivial flavor magnetic fluxes, the TTI reads
\begin{equation}
	\mI_{{\rm TTI}\,(2\nu_R)}^\Psi (\beta,\boldlambda) = \prod_{\trho_\Psi} \fft{\big(e^{\ri\trho_\Psi(\boldlambda)}e^{2\pi\ri\nu_R(r_\Psi-1)}\big)^{\fft{1-r_\Psi}{2}}}{\big(e^{\ri\trho_\Psi(\boldlambda)}e^{2\pi\ri\nu_R\, r_\Psi}q^{r_\Psi};q^2\big)_{1-r_\Psi}}\,. \label{chiral:TTI}
\end{equation}
To relate~\eqref{chiral:TTI} to the chiral $S^3_b$ partition function~\eqref{chiral:S3b}, we use \eqref{complexification} and \eqref{TTI:b}, which for a single chiral multiplet without flavor magnetic fluxes reduce to
\begin{subequations}
	\begin{align}
		\text{(holomorphic)}\quad&: \quad \beta=-2\pi\ri\nu_R\,b^2\,,&&\quad v_x=\ri\lambda_x=4\pi\nu_R b\,\hv_x\,,\\
		\text{(anti-holomorphic)}\quad&: \quad \beta=2\pi\ri\nu_R \tb^2\,,&&\quad \bv_x=-\ri\lambda_x=-4\pi\nu_R \tb\,\htv_x\,,
	\end{align}\label{chiral:TTI:id}%
\end{subequations}
The identity \eqref{sb:poch} along with the mapping \eqref{chiral:TTI:id} reorganizes the two chiral multiplet $S^3_b$ partition functions as
\begin{subequations}
	\begin{align}
		Z^\Psi(b,\boldhv)&=\prod_{\trho_\Psi}e^{-\fft{\pi\ri}{2}\big(-\fft{(\ri\trho_\Psi(\boldlambda)+(\beta-2\nu_R\pi\ri)(1-r_\Psi))^2}{4\pi\ri\beta}+\fft{\fft{\beta}{\pi\ri}+\fft{\pi\ri}{\beta}}{12}\big)}\fft{(e^{-\fft{2\nu_R\pi\ri}{\beta}(\ri\trho_\Psi(\boldlambda)-(\beta-2\nu_R\pi\ri)r_\Psi)-\fft{2\pi^2}{\beta}};e^{-\fft{2\pi^2}{\beta}})}{(e^{\ri\trho_\Psi(\boldlambda)-(\beta-2\nu_R\pi\ri)r_\Psi};q^2)}\,, \\
		Z^\Psi(\tb,\boldhtv)&=\prod_{\trho_\Psi}e^{\fft{\pi\ri}{2}\big(-\fft{(\ri\trho_\Psi(\boldlambda)-(\beta+2\nu_R\pi\ri)(1-r_\Psi))^2}{4\pi\ri\beta}+\fft{\fft{\beta}{\pi\ri}+\fft{\pi\ri}{\beta}}{12}\big)}\fft{(e^{\ri\trho_\Psi(\boldlambda)+(\beta+2\nu_R\pi\ri)r_\Psi}q^2;q^2)}{(e^{\fft{2\nu_R\pi\ri}{\beta}(\ri\trho_\Psi(\boldlambda)+(\beta+2\nu_R\pi\ri)r_\Psi)};e^{-\fft{2\pi^2}{\beta}})}\,.
	\end{align}\label{chiral:TTI:ZZ}%
\end{subequations}
Combining the two partition functions and discarding the terms involving $e^{-2\pi^2/\beta}$, which are exponentially suppressed in the Cardy-like limit~\eqref{Cardy}, we obtain
\begin{align}
	Z^\Psi(b,\boldhv)\,Z^\Psi(\tb,\boldhtv)&\approx \prod_{\trho_\Psi}\fft{\big(e^{\ri\trho_\Psi(\boldlambda)}e^{2\pi\ri\nu_R(r_\Psi-1)}\big)^{\fft{1-r_\Psi}{2}}}{\big(e^{\ri\trho_\Psi(\boldlambda)}e^{2\pi\ri\nu_R\, r_\Psi}q^{r_\Psi};q^2\big)_{1-r_\Psi}}=\mI_{{\rm TTI}\,(2\nu_R)}^\Psi (\beta,\boldlambda)\,.\label{chiral:TTI:fact}
\end{align}
The result~\eqref{chiral:TTI:fact} is the simplest instance of the general factorization formula~\eqref{TTI:main}, here for a single chiral multiplet.

%%%%%
\subsubsection{ABJM theory}\label{sec:ex:TTI:ABJM}
%%%%%
We now apply the factorization formula~\eqref{TTI:main} to the ABJM theory. The ABJM TTI matrix model obtained from~\eqref{TTI:loc} reads
\begin{align}
	&\mI_{{\rm TTI}\,(2\nu_R)}^{\text{ABJM}}(\beta,k,\boldlambda,\boldmn)=\fft{1}{(N!)^2}\sum_{\boldmm,\boldtmm\in\mathbb{Z}^N}\int\fft{d\boldh}{(2\pi)^N}\fft{d\boldtih}{(2\pi)^N}\prod_{i=1}^N e^{\ri k\mm_i h_i}e^{-\ri k\tmm_i\tih_i}\label{ABJM:TTI}\\
	&\quad\times\prod_{i\neq j}(e^{2\pi\ri\nu_R}q)^{-\fft{|\mm_{ij}|}{2}}\big(1-e^{\ri h_{ij}}q^{|\mm_{ij}|}\big)(e^{2\pi\ri\nu_R}q)^{-\fft{|\tmm_{ij}|}{2}}\big(1-e^{\ri\tih_{ij}}q^{|\tmm_{ij}|}\big)\nn\\
	&\quad\times\prod_{a=1}^4\prod_{i,j=1}^N\fft{\big(e^{\ri\sigma_a(h_i-\tih_j)+\ri\lambda_a}(e^{2\pi\ri\nu_R})^{r_a-1}\big)^{\fft{\sigma_a(\mm_i-\tmm_j)+\mn_a+(r_a-1)\mn_R}{2}}}{(e^{\ri\sigma_a(h_i-\tih_j)+\ri\lambda_a}(e^{2\pi\ri\nu_R})^{r_a}q^{1-(\sigma_a(\mm_i-\tmm_j)+\mn_a+(r_a-1)\mn_R)};q^2)_{\sigma_a(\mm_i-\tmm_j)+\mn_a+(r_a-1)\mn_R}}\,, \nn
\end{align}
following the same conventions described in Subsection \ref{sec:ex:ABJM}. On the other hand, the ABJM $S^3_b$ partition function \eqref{ABJM:S3b} is given by the Airy formula in the large $N$ limit, which arises from a particular saddle point as in~\eqref{ABJM:S3b:saddle}. For presentational purposes we assume that it provides the dominant contribution in the large $N$ limit; see the more detailed discussion of this point for the SCI in Subsection~\ref{sec:ex:ABJM:fact}. Then, substituting the Airy saddle contribution \eqref{ABJM:S3b:saddle} into the factorization formula \eqref{TTI:main} yields
\begin{align}
	\mI_{{\rm TTI}\,(2\nu_R)}^{\text{ABJM}}(\beta,k,\boldlambda,\boldmn) &\approx \mC_{(2\nu_R)}^{-1/3}e^{\mA_{(2\nu_R)}}\text{Ai}\Big[\mC_{(2\nu_R)}^{-1/3}(N-\mB_{(2\nu_R)})\Big] \nn \\
	&\quad\times\tmC_{(2\nu_R)}^{-1/3}e^{\tmA_{(2\nu_R)}}\text{Ai}\Big[\tmC_{(2\nu_R)}^{-1/3}(N-\tmB_{(2\nu_R)})\Big] \Big(1+\mO(e^{-\#\sqrt{N}})\Big)\,, \label{ABJM:TTI:fact}
\end{align}
where we have used the parity invariance~\eqref{ABJM:CS}. Here, the Airy parameters $(\mA_{(2\nu_R)},\mB_{(2\nu_R)},\mC_{(2\nu_R)})$ and $(\tmA_{(2\nu_R)},\tmB_{(2\nu_R)},\tmC_{(2\nu_R)})$ are obtained from \eqref{ABJM:Airy:coeffi} by replacing $(\omega,\boldvarphi)\to(\omega,\boldvarphi^{(2\nu_R)})$ and $(\omega,\boldvarphi)\to(\tomega,\boldtvarphi^{(2\nu_R)})$ respectively, where the latter are related to the TTI parameters via \eqref{complexification}, \eqref{TTI:b}, and  \eqref{ABJM:omegavarphi} as  ($Q\equiv b+b^{-1}$, $\tQ\equiv\tb+\tb^{-1}$)
\begin{equation}
\begin{alignedat}{3}
	\omega&=b^2&&&&=-\fft{\beta}{2\pi\ri\nu_R}\,,\\
	\tomega&=\tb^2&&&&=\fft{\beta}{2\pi\ri\nu_R}\,,\\
	\varphi_a^{(2\nu_R)}&=(1+\omega)\Delta_a^{(2\nu_R)}&&=(1+\omega)\bigg(r_a+\fft{2\hv_a}{\ri Q}\bigg)&&=r_a+\fft{\ri\lambda_a}{2\pi\ri\nu_R}-\fft{\beta}{2\pi\ri\nu_R}(r_a-\mn_a)\,,\\
	\tvarphi_a^{(2\nu_R)}&=(1+\tomega)\tDelta_a^{(2\nu_R)}&&=(1+\tomega)\bigg(r_a+\fft{2\htv_a}{\ri\tQ}\bigg)&&=r_a+\fft{\ri\lambda_a}{2\pi\ri\nu_R}+\fft{\beta}{2\pi\ri\nu_R}(r_a-\mn_a)\,.
\end{alignedat}\label{ABJM:TTI:varphi}
\end{equation}
Crucially, in contrast to the SCI case~\eqref{ABJM:SCI:factorization:dom} where both Airy factors share the same $\omega$ parameter, the two ABJM TTI Airy factors in~\eqref{ABJM:TTI:fact} are evaluated at opposite values of the squashing parameter, $\omega$ and $\tomega=-\omega$.

\medskip

Several comments on the factorized expression \eqref{ABJM:TTI:fact} are in order. 

\begin{itemize}
	\item Earlier field-theory results on the ABJM TTI fall into two complementary types. For the \emph{refined} index with $q\neq 1$, the factorized structure was obtained either as a product of squashed 3-sphere partition functions in the Cardy-like limit \eqref{Cardy} to the first two leading orders~\cite{Choi:2019dfu}, or as a gravitational-block form in the strict large-$N$ limit~\cite{Hosseini:2019iad,Hosseini:2022vho}. For the \emph{unrefined} index with $q=1$, by contrast, the large-$N$ free energy is known to all perturbative orders in $1/N$ \cite{Bobev:2022jte,Bobev:2022eus} --- see also \cite{Hong:2024uns} for the extension to general Seifert manifolds, including a non-trivial $S^1$ fibration over a higher-genus Riemann surface. The factorization formula \eqref{ABJM:TTI:fact} extends this all-orders-in-$1/N$ control to the refined ABJM TTI via the Airy formula, fixing it to all orders in the Cardy-like expansion, while reproducing the results of \cite{Bobev:2022jte,Bobev:2022eus} in the unrefined limit.

	\item The factorization~\eqref{ABJM:TTI:fact} provides, up to $N$-independent multiplicative factors, the field-theory proof of the holographic prediction of~\cite{Hristov:2022lcw} to all orders in the Cardy-like expansion~\eqref{Cardy}. In that reference the refined ABJM TTI appears as a mixed product of one $\text{Ai}$ and one $\text{Bi}$, whereas~\eqref{ABJM:TTI:fact} is a product of two $\text{Ai}$'s evaluated at opposite squashing parameters $\omega$ and $\tomega=-\omega$. The two presentations are related under the map of parameters
	\begin{equation}
	\begin{alignedat}{3}
		\omega&=-\tomega&\quad &\leftrightarrow&\quad &\omega^\text{\cite{Hristov:2022lcw}}\,,\\
		D_a&\equiv r_a+\fft{\ri\lambda_a}{2\pi\ri\nu_R}&\quad &\leftrightarrow&\quad &\Delta_i^\text{\cite{Hristov:2022lcw}}\,, \\
		n_a&\equiv r_a-\mn_a&\quad &\leftrightarrow&\quad &\mn_i^\text{\cite{Hristov:2022lcw}}\,,
	\end{alignedat}
	\end{equation}
	under which the holomorphic factor matches immediately, while matching the anti-holomorphic factors requires a suitable branch choice. Concretely, the anti-holomorphic factor in the second line of~\eqref{ABJM:TTI:fact} maps onto the $\text{Bi}$ term of~\cite{Hristov:2022lcw} --- up to an $N$-independent multiplicative factor and exponentially suppressed corrections --- through the Airy connection formula,
	\begin{align}
		\text{Ai}\big(e^{\pm 2\pi\ri/3}z\big)=\tfrac12 e^{\pm\ri\pi/3}\big[\text{Ai}(z)\mp\ri\,\text{Bi}(z)\big]\;\xrightarrow{z\to\infty\,(\Re[z^{3/2}]>0)}\;\mp\tfrac{\ri}{2}\,e^{\pm\ri\pi/3}\,\text{Bi}(z)\,,
	\end{align}
	provided one fixes the branch of $\tmC_{(2\nu_R)}^{-1/3}$ in the connection formula as 
	\begin{align}
		\tmC^{-1/3}_{(2\nu_R)}=\bigg(\fft{2\tomega^2}{\pi^2k\prod_a(D_a+\tomega n_a)}\bigg)^{-1/3}=e^{\pm 2\pi\ri/3}\bigg(\fft{2\omega^2}{\pi^2k\prod_a(D_a-\omega n_a)}\bigg)^{-1/3}\,,
	\end{align}
	and that the dominance conditions 
	\begin{align}
		\Re[(\mC_{(2\nu_R)}^{-1/3}(N-\mB_{(2\nu_R)}))^{3/2}],~\Re[(e^{\mp 2\pi\ri/3}\tmC_{(2\nu_R)}^{-1/3}(N-\tmB_{(2\nu_R)}))^{3/2}]>0
	\end{align}
	hold. A fully general treatment of the phases in the factorization formula --- beyond the branch choice and the dominance assumption $\Re[\,\cdots\,]>0$ adopted above --- is left for future work.
	
	\item We emphasize that the factorization~\eqref{ABJM:TTI:fact} unambiguously determines the $N$-independent contributions to the ABJM TTI in terms of the $\mA$-function of the $S^3_b$ partition function --- contributions neglected by the gravitational-block form of~\cite{Hristov:2022lcw}, which fixes only $(\mB,\mC)$. Specifically, the $N$-independent part of the logarithm of the factorized TTI~\eqref{ABJM:TTI:fact} reads
	\begin{align}
		\log \mI_{{\rm TTI}\,(2\nu_R)}^{\text{ABJM}}(\beta,k,\boldlambda,\boldmn)\Big|_\text{$N$-indept}&=\mA_{(2\nu_R)}+\tmA_{(2\nu_R)}+\log\fft{\mC_{(2\nu_R)}^{-1/4}\tmC_{(2\nu_R)}^{-1/4}}{4\pi}\,.\label{ABJM:TTI:fact:N-indept}
	\end{align}
	Although neither side admits a closed-form expression for generic flavor parameters, \eqref{ABJM:TTI:fact:N-indept} furnishes an explicit relation between the $N$-independent contributions to distinct 3d $\mN=2$ supersymmetric partition functions --- data that are typically among the hardest to determine in each case. It thereby allows the knowledge available in one observable to be carried over to the others, using whatever is known in a more tractable case to fix, or at least constrain, its less accessible counterparts.
	
	In the unrefined limit, such a comparison yields a highly non-trivial consistency test. Concretely, in the large-$k$ limit, the expansion~\eqref{ABJM:mA} reduces the r.h.s. of~\eqref{ABJM:TTI:fact:N-indept} to
	\begin{align}
		\lim_{\omega\to0}\bigg[\mA_{(2\nu_R)}+\tmA_{(2\nu_R)}+\log\fft{\mC_{(2\nu_R)}^{-1/4}\tmC_{(2\nu_R)}^{-1/4}}{4\pi}\bigg]&=-\fft{\zeta(3)}{8\pi^2}k^2\sum_{a=1}^4\hat f_{0,2,a}(\boldD)\,n_a\nn\\
		&\quad+\fft16\log k+\mO(k^0)
	\end{align}
	where
	\begin{align}
		\label{f0:general:lead}
		\hat f_{0,2,1}(\boldD)&=D_1+\fft{D_1D_3}{D_{14}}+\fft{D_1D_4}{D_{13}}+\fft{D_1D_4D_{23}}{D_{14}^2}+\fft{D_1D_3D_{24}}{D_{13}^2} \nn \\
		&\quad-\fft{2D_3D_4}{D_{13}D_{14}}-\fft{D_2^2(D_1-D_2)}{D_{23}D_{24}} +\fft{D_2D_3D_{14}}{D_{13}D_{23}}+\fft{D_2D_4D_{13}}{D_{14}D_{24}} \, , \nn \\[1mm]
		\hat f_{0,2,2}(\boldD)&=\hat f_{0,2,1}(\boldD)|_{D_1\leftrightarrow D_2} \, , \\
		\hat f_{0,2,3}(\boldD)&=\hat f_{0,2,1}(\boldD)|_{D_1\leftrightarrow D_3,D_2\leftrightarrow D_4} \, , \nn \\
		\hat f_{0,2,4}(\boldD)&=\hat f_{0,2,1}(\boldD)|_{D_1\leftrightarrow D_4,D_2\leftrightarrow D_3} \, . \nn
	\end{align}
	This perfectly matches a direct calculation of the l.h.s. of \eqref{ABJM:TTI:fact:N-indept} --- the unrefined TTI of \cite{Bobev:2022eus} --- under the identification $(D_a,n_a)\leftrightarrow(\Delta_a,\mn_a)^\text{\cite{Bobev:2022eus}}$ with the variables of \cite{Bobev:2022eus}. The apparent $\log k$ discrepancy stems from the $k$-fold degeneracy included in \cite{Bobev:2022eus} but absent from the factorization formula \eqref{ABJM:TTI:fact}, which isolates a single saddle-point contribution accounting for the Weyl degeneracy but not the $k$-fold one.

	\item Taking $N$ large, \eqref{ABJM:TTI:fact} produces the TTI analogue of~\eqref{ABJM:SCI:factorization:largeN},
	\begin{align}
		-\log \mI^{\rm ABJM}_{{\rm TTI}\,(2\nu_R)}&\approx\fft23\big(\mC_{(2\nu_R)}^{-\fft12}+\tmC_{(2\nu_R)}^{-\fft12}\big)N^\fft32-\big(\mC_{(2\nu_R)}^{-\fft12}\mB_{(2\nu_R)}+\tmC_{(2\nu_R)}^{-\fft12}\tmB_{(2\nu_R)}\big)N^\fft12\nn\\
		&\quad+\fft12\log N+\mO(N^0)\,.\label{ABJM:TTI:largeN}
	\end{align}
	At universal configuration with vanishing flavor parameters ($\boldlambda=\boldmn=0$), \eqref{ABJM:TTI:largeN} reduces to
	\begin{align}
		-\log \mI^{\rm ABJM}_{{\rm TTI}\,(2\nu_R)}\Big|_{\rm universal}\approx\fft{\pi\sqrt{2k}}{3}\,N^\fft32-\fft{\pi(k^2+32)}{24\sqrt{2k}}\,N^\fft12+\fft12\log N+\mO(N^0)\,,\label{ABJM:TTI:largeN:univ}
	\end{align}
	where all $\omega$-dependence cancels up to $\mO(N^0)$ order. This means that the refined universal ABJM TTI \eqref{ABJM:TTI:largeN:univ} dual to the magnetically charged stationary AdS$_4$ black hole of \cite{Hristov:2018spe} becomes identical in the first three leading terms in the large $N$ limit to the unrefined universal ABJM TTI \cite{Bobev:2021oku,Bobev:2022jte,Bobev:2022eus} that is dual to the static one \cite{Romans:1991nq,Bobev:2021oku}. Such an $\omega$-independence is, however, special to these leading orders: the rotation refinement reappears beyond the $\log N$ term. We refer the readers to \cite{Bobev:2021oku,Bobev:2022jte} for more detailed discussion on the structure of \eqref{ABJM:TTI:largeN:univ} in the context of holographic duality and associated black hole microstate counting. It would be desirable to extend such a supergravity analysis beyond the universal configuration, but the complete conformal-supergravity formalism with matter multiplets remains out of reach; see \cite{BenettiGenolini:2026qdm} for a promising bypass that reproduces the dual field-theory results of \cite{Hong:2024uns} via the equivariant-localization technique in higher-derivative supergravity.

	\item Although the ABJM TTI factorization formula \eqref{ABJM:TTI:fact} was obtained in the anti-periodic scheme with the $R$-symmetry holonomy choice $2\nu_R=\pm 1$ as in \eqref{nuR:half}, the resulting TTI can be identified with that evaluated in the conventional periodic scheme with $\nu_R=0$ after a suitable redefinition of the flavor holonomies. To be precise, the two TTIs coincide modulo a harmless overall phase factor --- which affects at most the imaginary part of the logarithm of the TTI --- provided $\lambda_a+2\nu_R\pi r_a$ in the anti-periodic scheme is mapped to $\lambda_a$ in the periodic scheme. This identification is manifest from the way the $R$-symmetry holonomy enters the localization formula \eqref{TTI:loc}. 
\end{itemize}

%%%%%
\subsubsection{ADHM theory}\label{sec:ex:TTI:ADHM}
%%%%%
Lastly, we consider the application of the factorization formula \eqref{TTI:main} to the ADHM theory briefly reviewed in Subsection \ref{sec:ex:ADHM}. The TTI matrix model \eqref{TTI:loc} for the ADHM theory reads 
\begin{align}
	&\mI_{{\rm TTI}\,(2\nu_R)}^{\text{ADHM}}(\beta,N_f,\boldlambda,\boldmn)\nn\\
	&=\fft{1}{N!}\sum_{\boldmm\in\mathbb{Z}^N}\int\fft{d\boldh}{(2\pi)^N}\prod_{i=1}^N e^{\ri \mn_Th_i+\ri\mm_i\lambda_T}\times\prod_{i\neq j}(e^{2\pi\ri\nu_R}q)^{-\fft{|\mm_{ij}|}{2}}\big(1-e^{\ri h_{ij}}q^{|\mm_{ij}|}\big)\label{ADHM:TTI}\\
	&\quad\times\prod_{I=1}^3\prod_{i,j=1}^N\fft{\big(e^{\ri h_{ij}+\ri\lambda_I}(e^{2\pi\ri\nu_R})^{r_I-1}\big)^{\fft{\mm_{ij}+\mn_I+(r_I-1)\mn_R}{2}}}{(e^{\ri h_{ij}+\ri\lambda_I}(e^{2\pi\ri\nu_R})^{r_I}q^{1-(\mm_{ij}+\mn_I+(r_I-1)\mn_R)};q^2)_{\mm_{ij}+\mn_I+(r_I-1)\mn_R}}\nn\\
	&\quad\times\prod_{i=1}^N\bigg[\fft{\big(e^{\ri h_i+\ri\lambda_f}(e^{2\pi\ri\nu_R})^{r_f-1}\big)^{\fft{\mm_i+\mn_f+(r_f-1)\mn_R}{2}}}{(e^{\ri h_i+\ri\lambda_f}(e^{2\pi\ri\nu_R})^{r_f}q^{1-(\mm_i+\mn_f+(r_f-1)\mn_R)};q^2)_{\mm_i+\mn_f+(r_f-1)\mn_R}}\nn\\
	&\kern4em\times\fft{\big(e^{-\ri h_i+\ri\lambda_{\tf}}(e^{2\pi\ri\nu_R})^{r_{\tf}-1}\big)^{\fft{-\mm_i+\mn_{\tf}+(r_{\tf}-1)\mn_R}{2}}}{(e^{-\ri h_i+\ri\lambda_{\tf}}(e^{2\pi\ri\nu_R})^{r_{\tf}}q^{1-(-\mm_i+\mn_{\tf}+(r_{\tf}-1)\mn_R)};q^2)_{-\mm_i+\mn_{\tf}+(r_{\tf}-1)\mn_R}}\bigg]^{N_f}\,. \nn
\end{align}
The $S^3_b$ partition function for the ADHM theory also allows for a matrix model expression \eqref{ADHM:S3b}, which in the large $N$ limit is captured by the Airy function as \eqref{ADHM:S3b:Airy}. Assuming that the Airy formula \eqref{ADHM:S3b:Airy} yields a dominant contribution to the $S^3_b$ partition function in the large $N$ limit as in the ABJM case and substituting it into the factorization formula \eqref{TTI:main}, we obtain
\begin{align}
	\mI_{{\rm TTI}\,(2\nu_R)}^{\text{ADHM}}(\beta,N_f,\boldlambda,\boldmn) &\approx \mC_{(2\nu_R)}^{-1/3}e^{\mA_{(2\nu_R)}}\text{Ai}\Big[\mC_{(2\nu_R)}^{-1/3}(N-\mB_{(2\nu_R)})\Big] \nn \\
	&\quad\times\tmC_{(2\nu_R)}^{-1/3}e^{\tmA_{(2\nu_R)}}\text{Ai}\Big[\tmC_{(2\nu_R)}^{-1/3}(N-\tmB_{(2\nu_R)})\Big] \Big(1+\mO(e^{-\#\sqrt{N}})\Big)\,. \label{ADHM:TTI:fact}
\end{align}
This is the factorized form of the ADHM TTI. The Airy parameters $(\mA_{(2\nu_R)},\mB_{(2\nu_R)},\mC_{(2\nu_R)})$ and $(\tmA_{(2\nu_R)},\tmB_{(2\nu_R)},\tmC_{(2\nu_R)})$ are obtained from \eqref{ADHM:Airy:coeffi} by replacing $(\omega,\boldvarphi)\to(\omega,\boldvarphi^{(2\nu_R)})$ and $(\omega,\boldvarphi)\to(\tomega,\boldtvarphi^{(2\nu_R)})$ respectively, where the latter are related to the TTI parameters via \eqref{complexification}, \eqref{TTI:b}, and  \eqref{ADHM:omegavarphi}. These consecutive maps of parameters can be written compactly as ($Q\equiv b+b^{-1}$, $\tQ\equiv\tb+\tb^{-1}$):
\begin{equation}
	\begin{alignedat}{3}
		\omega&=b^2&&&&=-\fft{\beta}{2\pi\ri\nu_R}\,,\\
		\tomega&=\tb^2&&&&=\fft{\beta}{2\pi\ri\nu_R}\,,\\
		\varphi_a^{(2\nu_R)}&=(1+\omega)\mbDelta_a^{(2\nu_R)}\,,&\quad \Delta_X^{(2\nu_R)}&=r_X+\fft{2\hv_X}{\ri Q}&&=r_X-\fft{\ri\lambda_X+\beta\mn_X}{\beta-2\pi\ri\nu_R}\,,\\
		\tvarphi_a^{(2\nu_R)}&=(1+\tomega)\bmbDelta_a^{(2\nu_R)}\,,&\quad \bDelta_X^{(2\nu_R)}&=r_X+\fft{2\htv_X}{\ri\tQ}&&=r_X-\fft{-\ri\lambda_X+\beta\mn_X}{\beta+2\pi\ri\nu_R}\,,
	\end{alignedat}\label{ADHM:TTI:varphi}
\end{equation}
where $\mbDelta_a^{(2\nu_R)}$'s and $\Delta_X^{(2\nu_R)}$'s (the overlined parameters as well) are related to each other through the same linear combination presented in \eqref{ADHM:mbDelta}.

\medskip

The comments on the ABJM TTI factorization formula~\eqref{ABJM:TTI:fact} apply equally to its ADHM analogue~\eqref{ADHM:TTI:fact}. First, \eqref{ADHM:TTI:fact} reproduces and improves earlier analyses --- restricted to the first two leading orders in the Cardy-like expansion~\cite{Choi:2019dfu}, the strict large-$N$ limit~\cite{Hosseini:2022vho}, and the unrefined limit to all orders in $1/N$~\cite{Bobev:2023lkx}. It is also consistent with the factorized form predicted from the gravity side via equivariant topological strings \cite{Cassia:2025aus,Cassia:2025jkr}, improving it further by determining how the $N$-independent multiplicative factors governed by the $\mA$-function enter the factorization. Lastly, both the holographic comparison at the universal configuration and the map of parameters connecting~\eqref{ADHM:TTI:fact} in the anti-periodic scheme to the result of~\cite{Bobev:2023lkx} in the periodic scheme carry over directly from the ABJM discussion, and we do not repeat them here.

%%%%%
\section{AdS/CFT and black holes}
\label{sec:holo}
%%%%%

We now proceed with a discussion of the applications of our results to holography and black hole physics. The three Euclidean SCFT partition functions we discussed above have a dual holographic description in the large $N$ limit as Euclidean asymptotically AdS$_4$ supersymmetric solutions of 10d or 11d supergravity. Typically, these solutions are constructed in 4d $\mathcal{N}=2$ gauged supergravity theories coupled to matter fields that arise as consistent truncations of string or M-theory. Such explicit consistent truncations are rare and in general it is not known how to associate a 4d $\mathcal{N}=2$ gauged supergravity theory with finitely many matter multiplets to a given 3d $\mathcal{N}=2$ holographic SCFT even in cases where the 10d/11d description of the conformal vacuum of the SCFT is known in terms of an explicit AdS$_4$ flux background. To this end our discussion below will focus on the ``universal truncation'' to the minimal 4d $\mathcal{N}=2$ gauged supergravity that arises for any AdS$_4$ flux vacuum of string/M-theory, see \cite{Gauntlett:2007ma}, the STU model of 4d $\mathcal{N}=2$ supergravity applicable to the holographic description of the ABJM theory \cite{Cvetic:1999xp,Azizi:2016noi}, or general results based on 4d $\mathcal{N}=2$ gauged supergravity coupled to matter that do not necessarily apply to any top-down AdS/CFT construction.

%%%%%%%%%%
\subsection{Minimal 4d $\mathcal{N}=2$ supergravity}
%%%%%%%%%%

We start our discussion with the 2-derivative minimal 4d $\mathcal{N}=2$ gauged supergravity. The bosonic action of the theory is simply that of the Einstein-Maxwell theory with a negative cosmological constant.\footnote{We mostly follow the supergravity conventions and notation used in~\cite{Bobev:2021oku}.} The three supersymmetric Euclidean backgrounds relevant for our discussion are the supersymmetric AdS-Taub-NUT, Kerr-Newman (KN) and Reissner-Nordstr\"om (RN) solutions of the theory. As is standard in AdS/CFT the logarithm of the Euclidean partition function of the CFT is evaluated in the large $N$ limit by computing the regularized on-shell action of the dual asymptotically locally AdS$_4$ supergravity solution. For any solution of the equations of motion of the 2-derivative 4d $\mathcal{N}=2$ gauged supergravity this on-shell action takes the simple form
\begin{equation}\label{eq:2derOSA}
	I_{2\partial} = \frac{\pi L^2}{2G_{N}} \mathcal{F}\,,
\end{equation}
where $L$ is the length scale of AdS$_4$ which is related to the cosmological constant of the theory and $G_{N}$ is the 4d Newton constant. The quantity $\mathcal{F}$ depends on the particular solution and for the three supersymmetric solutions of interest here is given by
\begin{equation}
	\mathcal{F}_{\rm TN} = \frac{1}{4}(b+b^{-1})^2\,, \qquad \mathcal{F}_{\rm KN} = \frac{(\omega+1)^2}{2\omega}\,, \qquad \mathcal{F}_{\rm RN} = (1-\mathfrak{g})\,. \qquad
\end{equation}
Here the continuous parameter $b$ is related to the NUT charge and determines the squashing deformation of the boundary $S^3$. It is directly related to the $b$ used in the QFT discussion of the squashed $S^3$ partition function. The parameter $\omega$ determines the angular velocity of the KN black hole solution and is the same $\omega$ used in the SCI discussion above. Finally, $\mathfrak{g}$ is the genus of the horizon of the RN black hole solution.

The higher-derivative corrections to 4d $\mathcal{N}=2$ gauged supergravity are in general not systematically studied. For the minimal theory however, it was shown in~\cite{Bobev:2020egg,Bobev:2021oku} that supersymmetry allows for two independent four-derivative terms in the supergravity Lagrangian and to each of them we can associate a dimensionless coupling constant $c_{1,2}$. It was also shown in~\cite{Bobev:2020egg,Bobev:2021oku} that one can evaluate the regularized 4-derivative on-shell action of the theory which takes the following simple form
\begin{equation}\label{eq:I4der}
I_{4\partial} = \left[1+ \frac{64\pi G_N}{L^2}(c_2-c_1)\right]\frac{\pi L^2}{2G_{N}} \mathcal{F} + 32 \pi^2 c_1 \chi\,.
\end{equation}
Here $\chi$ is the (appropriately regularized) Euler number of the 4d Euclidean background which for the solutions of interest is given by
\begin{equation}
\chi_{\rm TN} = 1\,, \qquad \chi_{\rm KN} = 2\,, \qquad \chi_{\rm RN} = 2(1-\mathfrak{g})\,. \qquad
\end{equation}

The logarithmic corrections to the gravitational path integral can be calculated by studying the contribution of the light, i.e. below the Planck scale, quantum fields in the gravitational background of interest. For the asymptotically AdS$_4$ backgrounds of interest here it was argued in~\cite{Hristov:2021zai,Bobev:2023dwx} that the logarithmic corrections to the gravitational free energy take the simple universal form
\begin{equation}\label{eq:Ilog}
	\delta I_{\rm log} = \frac{\chi}{6} \log \frac{L^2}{G_N}\,.
\end{equation}

One can now apply the conjugation rules we derived in the QFT analysis in Section~\ref{sec:fact} and Section~\ref{sec:TTI} to the supergravity parameters $\mathcal{F}$ and $\chi$ to find the relations
\begin{equation}\label{eq:FFbartilde}
\begin{split}
	\overline{\mathcal{F}}_{\rm TN} &= \mathcal{F}_{\rm TN}\,, \qquad \widetilde{\mathcal{F}}_{\rm TN} = \frac{1}{4}(\ri b-\ri b^{-1})^2\,, \\
	\overline{\chi}_{\rm TN} &= \chi_{\rm TN}\,, \qquad \widetilde{\chi}_{\rm TN} = \chi_{\rm TN}\,.
\end{split}
\end{equation}
With this at hand one can use \eqref{eq:I4der} and \eqref{eq:Ilog} to show that the relations we derived on the QFT side are also obeyed in supergravity up to and including the order $\log \frac{L^2}{G_N}$ in the semiclassical expansion of the gravitational path integral. In other words we find\footnote{For the RN solution we focus on $\mathfrak{g}=0$. We comment on the solutions with $\mathfrak{g}\geq 1$ further below.}
\begin{equation}\label{eq:IKNRNTN}
	I_{\rm KN} = I_{\rm TN} + \overline{I}_{\rm TN}\,, \qquad I_{\rm RN} = I_{\rm TN} + \widetilde{I}_{\rm TN}\,.
\end{equation}

This analysis establishes the validity of the relations $\mI_{\rm SCI} \sim Z_{S^3_b} \overline{Z}_{S^3_b}$ and $\mI_{\rm TTI} \sim Z_{S^3_b} \widetilde{Z}_{S^3_b}$ to order $\log \frac{L^2}{G_N}$ in the semiclassical expansion for any 3d $\mathcal{N}=2$ holographic SCFT captured by the universal 4d $\mathcal{N}=2$ minimal gauged supergravity truncation. Importantly, in the supergravity analysis we do not need to take the Cardy-like limit to establish the relations between the SCI, TTI and $S^3_b$ partition function, i.e. they are valid at finite values of the parameters $\{b,\omega,\beta\}$. It will be most interesting to understand the reason behind this extended regime of validity of the relation we derived by QFT methods assuming the Cardy-like limit.

It is worth stressing that the 4d minimal $\mathcal{N}=2$ gauged supergravity results above have appeared in various guises in the literature. At the two-derivative level they were discussed in~\cite{Azzurli:2017kxo,Bobev:2017uzs,Bobev:2019zmz} based on the universal 4d minimal $\mathcal{N}=2$ gauged supergravity truncation of 10d and 11d supergravity of~\cite{Gauntlett:2007ma}. At the four-derivative level the relations were discussed in~\cite{Bobev:2020egg,Bobev:2020zov} by using the direct evaluation of the on-shell action presented above, as well as from the point of view of localization of the on-shell action in the supergravity theory~\cite{BenettiGenolini:2019jdz,Genolini:2021urf}.

To illustrate these supergravity results in a bit more detail we can use the analysis of~\cite{Bobev:2020egg,Bobev:2021oku} and~\cite{Hristov:2021zai,Bobev:2023dwx} for the ABJM and ADHM models. For these two SCFTs arising from M2-branes one finds
\begin{equation}
	\frac{L^2}{2G_N} = AN^{\frac{3}{2}}+aN^{\frac{1}{2}}\,, \qquad c_{1,2} = \frac{v_{1,2}}{32\pi} N^{\frac{1}{2}}\,,
\end{equation}
which then leads to 
\begin{equation}
	I_{4\partial} + \delta I_{\rm log} = \pi A\, \mathcal{F} N^{\frac{3}{2}} + \pi \left[(a+v_2) \mathcal{F}  -  v_1( \mathcal{F}-\chi)\right] N^{\frac{1}{2}} + \frac{\chi}{4} \log N + \ldots
\end{equation}
The parameters $\{A,a,v_1,v_2\}$ can be determined by using supersymmetric localization as was shown in~\cite{Bobev:2020egg,Bobev:2021oku} and one finds
\begin{equation}
\begin{split}
	{\rm ABJM:} &\quad A = \frac{\sqrt{2k}}{3}\,, \qquad a+v_2 = - \frac{k^2+8}{24\sqrt{2k}}\,, \qquad v_1 = - \frac{1}{\sqrt{2k}}\,, \\
	{\rm ADHM:} &\quad A = \frac{\sqrt{2N_f}}{3}\,, \qquad a+v_2 =  \frac{N_f^2-4}{8\sqrt{2N_f}}\,, \qquad v_1 = - \frac{N_f^2+5}{6\sqrt{2N_f}}\,.
\end{split}
\end{equation}

We can perform additional holographic checks of our QFT analysis by considering the exponentially suppressed terms in the large $N$ expansion of the ABJM partition function which, as explained in \cite{Gautason:2023igo,Gautason:2025per}, are captured in the bulk by quantizing M2-branes on the AdS$_4 \times S^7/\mathbb{Z}_k$ background. The leading exponentially suppressed term on an asymptotically AdS$_4 \times S^7/\mathbb{Z}_k$ background reads~\cite{Gautason:2025per} 
\begin{equation}\label{eq:Iexp}
	\delta I_{\rm exp}= -2 \sum_{\rm f.p.} \frac{s(2/k)^{2k} s(x_{+})^{-k}s(x_{-})^{-k}}{t(x_{+})t(x_{-})}e^{-2\pi\sqrt{2N/k}}\,, \qquad x_{\pm} = \frac{2}{k}(1 \pm f)\,,
\end{equation}
where the overall minus sign reflects our convention $I=I_{4\partial}+\delta I=-\mZ_{\rm M2}$ with the M2 brane partition function $\mZ_{\rm M2}$ as in~\cite{Gautason:2023igo,Gautason:2025per,vanMuiden:2026nsp}; here, $\delta I$ incorporates higher-derivative and loop corrections as well as the $\delta I_{\rm log}$ and $\delta I_{\rm exp}$ terms presented above. This formula applies to any supersymmetric solution of the 4d $\mathcal{N}=2$ minimal gauged supergravity which can be uplifted to 11d and is asymptotic to AdS$_4 \times S^7/\mathbb{Z}_k$. The sum in~\eqref{eq:Iexp} is over all fixed points of the Killing vector constructed as a bilinear out of the Killing spinor of the 4d background. The functions in~\eqref{eq:Iexp} read 
\begin{equation}
	s(z) = {\rm exp} \left( \frac{\ri }{2\pi} {\rm Li}_{2}(e^{2\pi \ri z}) - \frac{\ri \pi}{12}+\frac{\ri \pi z^2}{2} - z \log(1-e^{2\pi \ri z})\right)\,, \quad t(z) = \frac{4}{k} \frac{\sin \pi z}{z}\,,
\end{equation}
and $f$ in~\eqref{eq:Iexp} depends on the given 4d supergravity background. For the supersymmetric solutions of interest here it is given by
\begin{equation}
	f_{\rm TN} = \frac{b^2-1}{b^2+1}\,, \qquad f_{\rm KN} = \frac{\omega-1}{\omega+1}\,, \qquad f_{\rm RN} = 1\,. \qquad
\end{equation}
For the TN solution we have only one fixed point at the center of the Euclidean AdS$_4$, while for the KN and $\mathfrak{g}=0$ RN solutions we have two fixed points at the north and south pole of the $S^2$ horizon. Using these facts, together with the map $\omega \to b^2$ for the SCI and $f_{\rm TN} \to 1$ in the large $b$ limit, it is simple to see that the familiar identities in~\eqref{eq:IKNRNTN} are obeyed by the exponentially suppressed contribution in~\eqref{eq:Iexp}. A closer inspection, however, reveals that, unlike its SCI counterpart, the TTI factorization in \eqref{eq:IKNRNTN} agrees with \eqref{eq:Iexp} through the first few orders but breaks down from $\mathcal{O}(b^4)$ onward in the small-$b$ expansion. This may signal that \eqref{eq:Iexp} itself requires improvement: for instance, the function $s(z)$ diverges at nonzero integer $z$, rendering it ill-defined for various configurations of $(k,f)$, and a suitable improvement of $\delta I_{\rm exp}$ might then restore the identity~\eqref{eq:IKNRNTN} to all orders in $b$. Alternatively, the Cardy-like
limit~\eqref{Cardy} used in deriving the factorization structure may not commute with the large-$N$ limit underlying~\eqref{eq:Iexp} in the exponentially suppressed sector, which would explain the disagreement. We leave a definitive resolution of this small puzzle for future work.

In the discussion above we focused on the TTI for $\mathfrak{g}=0$. As emphasized in~\cite{Bobev:2022eus} one can show using the BAE formulation of the TTI that $I_{S^1\times \Sigma_{\mathfrak{g}}}(\Delta,\mathfrak{n}) = (\mathfrak{g}-1)I_{S^1\times S^2}(\Delta,\mathfrak{n}/(1-\mathfrak{g}))$. This implies that we can use the results above to find the general TTI on $S^1\times \Sigma_{\mathfrak{g}}$ for $\mathfrak{g}\neq 1$.

So far we have presented results for the Euclidean semi-classical gravitational path integral. The KN solution and the RN solution for $\mathfrak{g}>1$ also admit a Lorentzian continuation to smooth supersymmetric black hole backgrounds with finite real entropy. In these cases one can employ the quantum statistical relation between the on-shell action $I$, the mass, $M$, charge, $Q$, and spin, $J$, of a given solution
\begin{equation}
	I = - S +\beta (M-\Phi Q -\omega J)\,,
\end{equation}
to find the black hole entropy $S$. The electric chemical potential $\Phi$ and the angular fugacity $\omega$ have to be treated carefully in the Lorentzian supersymmetric limit since $\beta \to \infty$, see~\cite{Cassani:2019mms,Bobev:2019zmz,Bobev:2020pjk} for further details. Moreover, the gravitational on-shell action is naturally derived in the ensemble of fixed chemical potentials and thus to find the entropy as a function of the electric charge one needs to perform an appropriate Laplace transformation.

%%%%%%%%%%
\subsection{4d $\mathcal{N}=2$ supergravity with matter}
%%%%%%%%%%

To apply our relations between the SCI, TTI and $Z_{S^3_b}$ in a holographic context for general values of the various continuous and discrete parameters, collectively denoted by $\boldm$ and $\boldxi$ in \eqref{eq:IZZbintro}-\eqref{eq:ITTIZZbintro}, one needs to find general supersymmetric solutions of 4d $\mathcal{N}=2$ gauged supergravity coupled to matter and ideally study them beyond the leading two-derivative supergravity approximation. This is a complicated technical problem and there is no general systematic procedure for constructing such solutions. To this end one either resorts to specific examples, like solutions of the 4d $\mathcal{N}=2$ STU model, or to general results often based on methods related to equivariant localization of the supergravity on-shell action.     

The 4d $\mathcal{N}=2$ STU model in gauged supergravity is obtained by adding three Abelian vector multiplets to the minimal supergravity theory with a two-derivative prepotential of the form $\mathfrak{F} \sim \sqrt{X^0 X^1 X^2 X^3}$ where $X^I$ are the complex scalars that obey a constraint and encode the three physical complex scalars $z_{1,2,3}$ that parametrize the $[{\rm SU}(1,1)/{\rm U}(1)]^3$ special K\"ahler manifold. This model can be obtained as a consistent truncation of the maximal 4d $\mathcal{N}=8$ ${\rm SO}(8)$ gauged supergravity~\cite{Cvetic:1999xp}, which in turn arises from 11d supergravity on $S^7/\mathbb{Z}_k$~\cite{Nicolai:2011cy,Varela:2015ywx}.\footnote{For $k >2$ supersymmetry is broken to $\mathcal{N}=6$ and the STU model should be viewed as a consistent truncation of the 4d $\mathcal{N}=6$ gauged supergravity.} This chain of supergravity consistent truncations therefore ensures that the solutions of the STU model admit a holographic interpretation in the ABJM theory.

As discussed in Section~\ref{sec:ex:ABJM} the Airy form of the squashed sphere partition function for the ABJM theory reads
\begin{equation}
	Z_{S^3_b}(N,b,\xi) = \mathcal{C}^{-1/3} e^{\mathcal{A}} {\rm Ai}[\mathcal{C}^{-1/3}(N-\mathcal{B})](1+\mathcal{O}(e^{-\#\sqrt{N}}))\,,
\end{equation}
where $\{\mathcal{C},\mathcal{B},\mathcal{A}\}$ are defined in~\eqref{ABJM:Airy:coeffi}-\eqref{ABJM:mA}. Expanding this result at large $N$ one finds the following QFT prediction for the bulk gravitational free energy
\begin{equation}\label{eq:IS3bABJM}
	-\log Z_{S^3_b} = \frac{2}{3\sqrt{\mathcal{C}}} N^{\frac{3}{2}} -\frac{\mathcal{B}}{\sqrt{\mathcal{C}}}N^{\frac{1}{2}} + \frac{1}{4}\log N  + \ldots \,.
\end{equation}
Combining this with the QFT relations derived in this work implies that the SCI has the following large $N$ expansion
\begin{equation}\label{eq:ISCIABJM}
	-\log \mI_{\rm SCI} = \frac{2}{3}\frac{\sqrt{\mathcal{C}}+\sqrt{\overline{\mathcal{C}}}}{\sqrt{\mathcal{C} \overline{\mathcal{C}}}} N^{\frac{3}{2}} -\frac{\mathcal{B}\sqrt{\overline{\mathcal{C}}}+\overline{\mathcal{B}}\sqrt{\mathcal{C}}}{\sqrt{\mathcal{C} \overline{\mathcal{C}}}}N^{\frac{1}{2}} + \frac{1}{2}\log N  + \ldots \,,
\end{equation}
while for the TTI we find
\begin{equation}\label{eq:ITTIABJM}
	-\log \mI_{\rm TTI} = \frac{2}{3}\frac{\sqrt{\mathcal{C}}+\sqrt{\widetilde{\mathcal{C}}}}{\sqrt{\mathcal{C} \widetilde{\mathcal{C}}}} N^{\frac{3}{2}} -\frac{\mathcal{B}\sqrt{\widetilde{\mathcal{C}}}+\widetilde{\mathcal{B}}\sqrt{\mathcal{C}}}{\sqrt{\mathcal{C} \widetilde{\mathcal{C}}}}N^{\frac{1}{2}} + \frac{1}{2}\log N  + \ldots \,.
\end{equation}
The rigorous supergravity checks of these results are restricted to the leading $N^{\frac{3}{2}}$ and the $\log N$ terms in the large $N$ expansion. For $b=1$ the leading term in~\eqref{eq:IS3bABJM} was derived in~\cite{Freedman:2013oja} by constructing explicit Euclidean supergravity solutions of the Euclidean STU model and evaluating their regularized on-shell action. Similarly, Euclidean STU model solutions with an $S^1\times \Sigma_{\mathfrak{g}}$ boundary that reproduce the leading term in~\eqref{eq:ITTIABJM} were found in~\cite{Bobev:2020pjk}. The Euclidean solutions are intimately related to the supersymmetric Lorentzian dyonic black holes found in~\cite{Cacciatori:2009iz} (with rotating extensions in~\cite{Hristov:2018spe}). The leading $N^{\frac{3}{2}}$ contribution to the entropy of these black holes was reproduced by the TTI in~\cite{Benini:2015eyy,Benini:2016rke}.\footnote{These leading $N^{\frac{3}{2}}$ results for the $S^3$ free energy and the TTI can be extended to the mABJM 3d $\mathcal{N}=2$ SCFT and its dual description in terms of the 4d $\mathcal{N}=2$ STU model coupled to a hypermultiplet, see~\cite{Bobev:2018uxk,Bobev:2018wbt}.} The general Euclidean Kerr-Newman solution of the STU model which reproduces the leading $N^{\frac{3}{2}}$ term in~\eqref{eq:ISCIABJM} has not been constructed. As discussed in~\cite{Cassani:2019mms} certain limits of the solution are known and their regularized on-shell action agrees with~\eqref{eq:ISCIABJM}. The same applies to the entropy of the Lorentzian supersymmetric black holes of the STU model constructed in~\cite{Hristov:2019mqp}, see also~\cite{Choi:2019zpz,Choi:2019dfu}. The $\log N$ term in~\eqref{eq:IS3bABJM} was derived in supergravity in~\cite{Bhattacharyya:2012ye}, see also~\cite{Hristov:2021zai,Bobev:2023dwx}. Similarly, the logarithmic terms in~\eqref{eq:ISCIABJM} and~\eqref{eq:ITTIABJM} can be obtained by supergravity calculations, see~\cite{Liu:2017vbl,Hristov:2021zai,Bobev:2023dwx}.

Further tests of the QFT predictions above are less rigorous and rely on various conjectures and educated guesses. The starting point is the observation that the STU model two-derivative prepotential $\mathfrak{F} \sim \sqrt{X^0 X^1 X^2 X^3}$ takes the same form as the leading $N^{\frac{3}{2}}$ term in~\eqref{eq:IS3bABJM} upon an appropriate identification of the real mass parameters $\Delta_a$ with the supergravity sections $X^{I}$. This then naturally leads to the conjecture that the four- and higher-derivative prepotential is fully determined by the $S^3$ ABJM partition function where the $1/N$ expansion of the free energy is identified with the derivative expansion of $\mathfrak{F}$, see~\cite{Bobev:2021oku,Bobev:2022eus,Hristov:2021qsw,Hristov:2022lcw,Hristov:2022plc}. Combining this conjecture for the prepotential with the supergravity ``gluing rules'' discussed in~\cite{Hosseini:2019iad,Hristov:2021qsw} one finds non-trivial agreement with the field theory predictions in~\eqref{eq:ISCIABJM} and~\eqref{eq:ITTIABJM} including the full series of $1/N$ corrections. Importantly, these conjectural gravitational results do not include the $N^0$ terms in the large $N$ expansion for which there are currently no supergravity calculations that agree with the QFT results. Recently these results were put on a more solid footing by applying techniques from equivariant localization to the calculation of the on-shell action of supersymmetric solutions of the 4d $\mathcal{N}=2$ STU model at the two-derivative~\cite{BenettiGenolini:2024xeo,BenettiGenolini:2024kyy,BenettiGenolini:2024lbj} and higher-derivative level~\cite{BenettiGenolini:2026qdm}.\footnote{See~\cite{Martelli:2023oqk} for a different application of equivariant localization in similar holographic contexts.} These equivariant localization calculations confirm the supergravity results for the subleading corrections to the SCI and TTI in ~\cite{Bobev:2021oku,Bobev:2022eus,Hristov:2021qsw,Hristov:2022lcw,Hristov:2022plc} and are compatible with the $Z_{\rm BH} \sim |Z_{S^3}|^2$ type relations we derived in this work.

Encouraged by these successful holographic checks of our QFT relations in the context of the ABJM theory and the 4d STU model we proceed with a more general discussion.

The QFT relation~\eqref{eq:IZZbintro} leads us to conjecture that for general AdS$_4$ supersymmetric Kerr-Newman solutions in 4d $\mathcal{N}=2$ gauged supergravity the gravitational path integral can be expressed as  
\begin{equation}\label{eq:ZZZbKN}
Z_{\rm KN} \sim Z(N,b,\xi) \overline{Z}(N,b,\xi)\,.
\end{equation}
The right hand side is determined by the partition function of the holographically dual 3d $\mathcal{N}=2$ SCFT placed on the squashed $S^3$ in the presence of general real mass deformation for the flavor symmetry multiplet. This relation should be understood as valid in the context of the Euclidean gravitational path integral for the finite $\beta$ supersymmetric supergravity background. Based on the evidence available for the ABJM theory~\eqref{eq:ZZZbKN} should be valid to all orders in the $1/N$ expansion and perhaps even non-perturbatively. The conjugation rules that define $\overline{Z}(N,b,\xi)$ are determined by the QFT results in Section~\ref{sec:fact:SCI-S3b}. The right hand side of~\eqref{eq:ZZZbKN} can also be thought of as arising from contributions to the gravitational path integral localized to the north and the south pole of the finite size $S^2$ at the ``tip of the cigar'' where the $S^1$ $\beta$-circle shrinks to zero size. In situations where the Euclidean KN solutions admit a $\beta \to \infty$ limit and Lorentzian interpretation as supersymmetric black holes, the relation \eqref{eq:ZZZbKN} determines the black hole partition function in the ensemble of fixed electric chemical potentials and angular velocity. In principle, this result  can then be Laplace transformed to obtain the black hole entropy as a function of the electric charges and angular momentum of the black hole. 

Similarly, the QFT relation~\eqref{eq:ITTIZZbintro} leads to a conjecture for the path integral of the supersymmetric $\mathfrak{g}=0$ AdS$_4$ Reissner-Nordstr\"om solutions in 4d $\mathcal{N}=2$ gauged supergravity\footnote{For the $\mathfrak{g} > 1$ Euclidean RN solutions one can use the QFT relation $\log Z_{\mathfrak{g}} = (\mathfrak{g}-1) \log Z_{\mathfrak{g}=0}$ to find the relevant partition function. The case $\mathfrak{g}=1$ appears degenerate from this point of view and it will be most interesting to understand the corresponding gravitational solutions with $T^2$ horizons from the perspective of the relations discussed in this work.}
\begin{equation}\label{eq:ZZZbRN}
	Z_{\rm RN} \sim Z(N,b,\xi) \widetilde{Z}(N,b,\xi)\,.
\end{equation}
Again, this relation should be understood in the context of the Euclidean gravitational path integral and the right hand side is defined by the SCFT squashed sphere partition function together with the conjugation rules spelled out in Section~\ref{sec:TTI}. If the Euclidean RN solutions allow for a smooth $\beta \to \infty$ limit and a Lorentzian interpretation as supersymmetric black holes, then~\eqref{eq:ZZZbRN} determines the black hole partition function in the ensemble of fixed electric chemical potentials and magnetic charges. This can then be Laplace transformed and subjected to $\mathcal{I}$-extremization, see~\cite{Benini:2015eyy,Benini:2016rke}, to obtain the dyonic black hole entropy as a function of the quantized electric and magnetic charges. 

The relations~\eqref{eq:ZZZbKN} and~\eqref{eq:ZZZbRN} clearly resemble the OSV conjecture for asymptotically flat supersymmetric black holes arising from string theory on compact CY 3-folds~\cite{Ooguri:2004zv}. In the OSV setup the right-hand side of the relation is determined by the topological string partition function which by the results of~\cite{Bershadsky:1993cx,Antoniadis:1993ze} determines the higher-derivative 4d $\mathcal{N}=2$ ungauged supergravity prepotential. This bears a structural similarity to~\eqref{eq:ZZZbKN} and~\eqref{eq:ZZZbRN} since the right hand side is determined by the $S^3$ partition function, which as shown in~\cite{Zan:2021ftf} at the two-derivative level and extended in some cases to higher-derivatives~\cite{Bobev:2021oku,Bobev:2022eus,Hristov:2021qsw,Hristov:2022lcw,Hristov:2022plc,BenettiGenolini:2026qdm}, is intimately related to the gauged supergravity prepotential. Yet another similarity with OSV is related to the results of~\cite{Beasley:2006us} where it is shown that the right-hand side of the OSV relation can be interpreted as arising from a localized string theory contribution at the north and south pole of the $S^2$ black hole horizon. This clearly resonates with the discussion in~\cite{Hristov:2021qsw,BenettiGenolini:2026qdm} for the higher-derivative on-shell action in 4d $\mathcal{N}=2$ gauged supergravity. A major difference with OSV is that there is no systematic construction of 4d $\mathcal{N}=2$ gauged supergravity theories coupled to matter fields for general AdS$_4$ vacua of gauged supergravity and moreover the relation to topological strings is more tenuous.\footnote{The $S^3$ partition function of the ABJM theory in the absence of real masses and for the special values of the squashing parameter $b^2=1$ and $b^2=3$ is related to topological string theory on non-compact CY manifolds but it is not known whether this can be generalized to other theories or more general values of the parameters, see \cite{Fuji:2011km,Marino:2011eh,Hatsuda:2016uqa}.}

Clearly it will be most interesting to subject the relations in~\eqref{eq:ZZZbKN} and~\eqref{eq:ZZZbRN} to more tests and apply them to other holographic 3d $\mathcal{N}=2$ SCFTs arising from string and M-theory. Holographic SCFTs similar to ABJM that arise from M2-branes should provide a direct generalization of our discussion. In many of these theories there are, conjectural or rigorously derived, Airy formulas for the $S^3$ partition function with squashing and mass deformations, see \cite{Fuji:2011km,Marino:2011eh,Mezei:2013gqa,Nosaka:2015iiw,Hatsuda:2016uqa,Nosaka:2024gle,Kubo:2024qhq,Geukens:2024zmt,Kubo:2025dot,He:2025zxk} and~\cite{Bobev:2025ltz} for a summary and additional evidence for these Airy formulae. We can use these results to find the 4d $\mathcal{N}=2$ supergravity prepotential and then test and use the relation derived in this work in conjunction with the supergravity equivariant localization results in~\cite{BenettiGenolini:2026qdm}. Extension to other 3d $\mathcal{N}=2$ holographic SCFTs described by AdS$_4$ flux vacua in (massive) IIA or IIB supergravity, or the more exotic class $\mathcal{R}$ SCFTs arising from M5-branes wrapped on hyperbolic 3-manifolds, will also be interesting to pursue.

%%%%%
\section{Discussion}
\label{sec:discussion}
%%%%%

In this work we showed that the SCI and TTI of 3d $\mathcal{N}=2$ SCFTs in the Cardy-like limit factorize into a product of two squashed sphere partition functions in the limit of large squashing. In the context of large $N$ holographic SCFTs these relations provide a new perspective on the path integral of supersymmetric AdS$_4$ black holes and bear similarities to the OSV conjecture.

Our results lead to many open questions and suggest several avenues for generalization, some of which we discuss below: 

\begin{itemize}

\item Our QFT results are derived for SCFTs admitting a UV Lagrangian formulation by direct manipulations of the corresponding supersymmetric localization matrix models. It will be most interesting to derive these relations with different methods, potentially applicable to non-Lagrangian theories. For instance, one can imagine a derivation that employs a ``supersymmetric thermal EFT'' approach similar to the one used in~\cite{Cassani:2021fyv} for the analysis of the SCI of 4d $\mathcal{N}=1$ SCFTs (see also~\cite{ArabiArdehali:2019orz,ArabiArdehali:2021nsx,Ardehali:2021irq}), or perhaps a derivation based on the ``holomorphic blocks'' of~\cite{Pasquetti:2011fj,Beem:2012mb,Hwang:2012jh}. In addition to offering new insights into the structure of these relations, such a derivation will also have interesting implications for the 3d $\mathcal{N}=2$ theories of class $\mathcal{R}$ and their dual description in terms of $\mathbf{SL}(N)$ Chern-Simons theory on hyperbolic 3-manifolds.

\item We derived our results using the Cardy-like limit of small $\beta$. The comparison with bulk holographic calculations suggests that the relations between the TTI, SCI, and the 3-sphere partition function may also be true in the large $N$ limit without taking the Cardy-like limit. It will be very interesting to investigate this further and to understand whether there are any interesting phase transitions at large $N$ as one varies the continuous parameters, like fugacities and real masses, on which the SCI, TTI, and  $Z_{S^3}$ depend. 

\item 3d $\mathcal{N}=2$ SCFTs can be placed on more general compact Euclidean manifolds while preserving supersymmetry, see~\cite{Closset:2019hyt} for a review and further references. It is prudent to systematically investigate relations of the type we studied here in these more general setups and to explore their implications for AdS/CFT. 

\item Another interesting question for further study is to explore the generalization of our results to other dimensions. In Appendix~\ref{app:5d} we briefly discuss the generalization to 5d SCFTs where a relation of the form $Z_{S^1\times S^4} \sim |Z_{S^5_{b_1,b_2,b_3}}|^2$ between the 5d SCI and the squashed $S^5$ partition function appears to hold. It is important to study this in more detail and uncover its implications for holography and black holes. More generally, there are universal relations between various partition functions of holographic SCFTs that can be found in the leading two-derivative supergravity, see~\cite{Bobev:2017uzs} for a summary and further references. It is desirable to understand which of these relations, if any, can be promoted to fully-fledged QFT identities.

\item The original OSV conjecture relates the asymptotically flat supersymmetric black hole partition function to topological strings. In contrast, our results show a relation between the path integral of AdS black holes and the $S^3$ partition function of the dual SCFT. Nevertheless, there are still intriguing connections to topological strings. For some holographic SCFTs on $S^3$ at special values of the squashing parameter the squashed $S^3$ partition function is equivalent to that of topological strings on a non-compact CY manifold, see~\cite{Marino:2016new} for a review. Moreover, it was recently suggested that the squashed $S^3$ partition function of holographic SCFTs can be related to the equivariant generalization of certain topological string path integrals~\cite{Cassia:2025aus,Cassia:2025jkr,Hristov:2026zjh}, see also~\cite{Bobev:2025ltz}. Clearly, it is important to understand this connection between AdS$_4$ black holes, holography, and topological strings better and to put it on a firmer footing.

\item A useful technique for the explicit calculation of large $N$ $S^3$ partition functions is to introduce a chemical potential $\mu$ for the rank of the gauge group $N$ and compute the corresponding partition function in the fixed $\mu$, i.e. grand canonical, ensemble, see~\cite{Marino:2016new}. Recently, it was emphasized in~\cite{Gautason:2025plx,vanMuiden:2026nsp} that this $\mu$-ensemble has a natural interpretation for AdS$_4$ backgrounds in M-theory. It will be very interesting to investigate the TTI and SCI partition functions in the $\mu$-ensemble especially in the context of the relations to the $S^3$ partition function and holography that we studied in this work. A particularly interesting problem is to understand whether the simple cubic polynomial in the large $\mu$ expansion of the $S^3$ free energy of holographic SCFTs has a direct analog for the TTI and SCI in the $\mu$-ensemble.

\item From the vantage point of supergravity our work raises several open questions. Clearly, it is important to understand better the role of  ``gravitational blocks''~\cite{Hosseini:2019iad} and ``gluing rules'' that have appeared in studies of precision holography at the two- and higher-derivative level~\cite{Hristov:2022lcw}.  Additionally, it is important to systematically understand the relation between the $S^3$ partition function of 3d $\mathcal{N}=2$ holographic SCFTs in the planar limit and the prepotential of a dual 4d $\mathcal{N}=2$ gauged supergravity theory. This has been established in~\cite{Zan:2021ftf} at the two-derivative level and there are a number of impressive results that suggest it may be true also in the presence of higher-derivative corrections, see~\cite{Bobev:2021oku,Bobev:2022eus,Hristov:2021qsw,Hristov:2022lcw,Hristov:2022plc,BenettiGenolini:2026qdm}. Given the paucity of direct string theory and supergravity methods to derive lower-dimensional gauged supergravity effective actions, it is important to make progress on this question.

\item Finally, it is clearly important to devise methods to test or derive the relation $Z_{\rm BH} \sim |Z_{S^3}|^2$ for AdS$_4$ supersymmetric black holes that we discussed in this work. This is a hard technical problem, but perhaps one can employ supersymmetric localization techniques in supergravity, along the lines of~\cite{Dabholkar:2010uh,Hristov:2018lod,Hristov:2019xku}, or maybe even directly in string/M-theory to make progress in this direction.

\end{itemize}

%%%%%%%%%%%%%%%%%%%%%%
\section*{Acknowledgments}
%%%%%%%%%%%%%%%%%%%%%%
We are grateful to Davide Cassani, Pieter-Jan De Smet, Fri\dh rik Freyr Gautason, Seok Kim, Zohar Komargodski, Siyul Lee, Jesse van Muiden, and Xuao Zhang for valuable discussions. NB is supported in part by FWO projects G003523N, G094523N, and G0E2723N, as well as by the KU Leuven C1 project C16/25/01. SC is supported in part by WPI Initiative, MEXT, Japan at Kavli IPMU, the University of Tokyo as well as by Basic Science Research Program through the National Research Foundation of Korea (NRF) funded by the Ministry of Education (RS-2025-02663044). JH is supported by the National Research Foundation of Korea (NRF) grant funded by the Korean government (MSIT), Grant No. RS-2024-00449284; by the Sogang University Research Grant No. 202410008.01; and by the Basic Science Research Program of the NRF funded by the Ministry of Education through the Center for Quantum Spacetime (CQUeST), Grant No.
RS-2020-NR049598. VR is partly supported by a Visibilit\'e Scientifique Junior Fellowship from LabEx LMH. NB, SC, and JH are grateful to the Galileo Galilei Institute, Florence, for warm hospitality during the final stages of the completion of this work.

%JH is grateful to KIAS and Seoul National University for the warm hospitality during parts of this project. SC is grateful to KU Leuven for kind hospitality during part of this project.

\appendix

%%%%%%%%%
\section{Special functions}
\label{app:special}
%%%%%%%%%
The $\infty$-Pochhammer symbol is defined within the unit disk as
\begin{equation}
	(a;x)=\prod_{n=0}^\infty(1-ax^n)\,, \qquad |x|<1\,.\label{def:poch}
\end{equation}
It can be defined outside the unit disk following the prescription of \cite{Narukawa:2003ltf} as
\begin{align}
	(a;x)\equiv\begin{cases}
		(a;x) & |x|<1 \\
		(ax^{-1};x^{-1})^{-1} & |x|>1
	\end{cases}=(ax^{-1};x^{-1})^{-1}\,.\label{poch:continuation}
\end{align}
The $\infty$-Pochhammer symbol satisfies the following useful identity: 
\begin{equation}
	(-x)^{\fft{\mm}{2}}\fft{(xq^{1+\mm};q^2)}{(x^{-1}q^{1+\mm};q^2)}=(-x)^{-\fft{\mm}{2}}\fft{(xq^{1-\mm};q^2)}{(x^{-1}q^{1-\mm};q^2)}\,,\qquad \mm\in\mathbb{Z}\,.\label{poch:identity}
\end{equation}

The $q$-Pochhammer symbol is defined as
\begin{align}
	(a;x)_n=\begin{cases}
		\prod_{i=0}^{n-1}(1-ax^i) & (n\geq1) \\
		\prod_{i=n}^{-1}(1-ax^i)^{-1} & (n\leq -1) \\
		1 & (n=0)
	\end{cases}\,.
\end{align}
It satisfies the identity
\begin{align}
	(a;x)_n=\fft{1}{(ax^n;x)_{-n}}\qquad\&\qquad (a;x)_n=\fft{(a;x)}{(ax^n;x)}\,.\label{q-poch:identity}
\end{align}
%

%\medskip

The double sine function is defined by \cite{Narukawa:2003ltf,Bobev:2025ltz}
\begin{equation}
	s_b(z)=\fft{\Gamma_2(\fft{Q}{2}+\ri z;b,b^{-1})}{\Gamma_2(\fft{Q}{2}-\ri z;b,b^{-1})}=\prod_{m,n=0}^\infty\fft{mb+nb^{-1}+\fft{Q}{2}-\ri z}{mb+nb^{-1}+\fft{Q}{2}+\ri z}\,,\label{bsine}
\end{equation}
where $Q=b+b^{-1}$ and $\Re[b^{\pm1}]>0$. The multiple Hurwitz zeta and Gamma functions are defined in their domains of convergence as 
\begin{equation}
	\begin{split}
		\zeta_r(s,z;\vec{\omega})&\equiv\sum_{n_1,\cdots,n_r=0}^\infty\fft{1}{(\vec{n}\cdot\vec{\omega}+z)^s}\,,\\
		\Gamma_r(z;\vec{\omega})&\equiv\exp[\fft{\partial}{\partial s}\zeta_r(s,z;\vec{\omega})\bigg|_{s=0}]=\prod_{n_1,\cdots,n_r=0}^\infty\fft{1}{(\vec{n}\cdot\vec{\omega}+z)}
	\end{split}\label{HurwitzZetaGamma}
\end{equation}
for a given vector $\vec{\omega}=(\omega_1,\cdots,\omega_r)$, and can be extended outside the domain of convergence by analytic continuation. The double sine function satisfies the identities \cite{Hatsuda:2016uqa,Bobev:2025ltz}
\begin{subequations}
\begin{align}
	s_b(z)s_b(-z)&=1\,,\\
	\overline{s_b(z)}&=s_b(-\bar{z})\,,\\
	s_b\left(\fft{\ri}{2}b^{\pm1}+z\right)s_b\left(\fft{\ri}{2}b^{\pm1}-z\right)&=\fft{1}{2\cosh(\pi b^{\pm1}z)}\,,
\end{align}\label{bsine:property}
\end{subequations}
from which one also finds
\begin{equation}
	s_b\bigg(\fft{\ri Q}{2}+z\bigg)s_b\bigg(\fft{\ri Q}{2}-z\bigg)=\fft{1}{2\sinh(\pi bz)}\fft{1}{2\sinh(\pi b^{-1}z)}\,.\label{bsine:property:2}
\end{equation}
The last identity was used to rewrite the vector multiplet contribution to the $S^3_b$ partition function in (\ref{S3b}).

%\medskip

The double sine function can be expressed in terms of the $\infty$-Pochhammer symbols, following Proposition 5 of \cite{Narukawa:2003ltf}, as
\begin{align}
	s_b(\fft{\ri Q}{2}-\hu)&=e^{-\mfs\fft{\pi\ri}{2}((\fft{\ri Q}{2}-\hu)^2+\fft{b^2+b^{-2}}{12})}\fft{(e^{-2\mfs\pi b^{-1}\hu+2\mfs\pi\ri b^{-2}};e^{2\mfs\pi\ri b^{-2}})}{(e^{-2\mfs\pi b\hu};e^{-2\mfs\pi\ri b^2})}\,,\label{sb:poch}
\end{align}
for either sign choice $\mfs\in\{\pm1\}$, provided that $\Im[b^2]\neq0$. To derive the expressions in (\ref{Z:bZ:squash}), it is convenient to choose the sign $\mfs$ to be the same as that of $\beta/(\pi \ri b^2)$ for the holomorphic part, and opposite to it for the anti-holomorphic part.

%%%%%%%%%
\section{Comments on the 5d SCI and $Z_{S^5}$}
\label{app:5d}
%%%%%%%%%

In this appendix, we briefly illustrate a conjectural relation between two supersymmetric partition functions of 5d $\mathcal{N}=1$ superconformal field theories: the superconformal index (SCI) $\mathcal{I}_\textrm{SCI}^{\textrm{5d}}$ and the squashed five-sphere partition function $Z_{S^5_{b_1,b_2,b_3}}$.

We consider 5d $\mathcal{N}=1$ supersymmetric gauge theories whose UV fixed points are interacting SCFTs \cite{Seiberg:1996bd,Intriligator:1997pq}. The SCI on $S^1 \times S^4$ is defined as \cite{Bhattacharya:2008zy,Kim:2012gu}
\begin{equation}\label{5d-index-tr}
    \mathcal{I}_\textrm{SCI}^{\textrm{5d}} (\omega_1,\omega_2, \boldsymbol{\lambda},q) = \textrm{Tr} \left[ (-1)^F e^{2\pi \ri \, \omega_1 (J_1+R)} e^{2\pi \ri\, \omega_2(J_2+R)}  \prod_{x=1}^{r_F}e^{\ri\lambda_x F_x} \, q^k\right]\ ,
\end{equation}
where $F$ is the fermion number operator, $J_1$ and $J_2$ are the Cartan charges of the ${\rm SO}(5)$ rotation symmetry, $R$ is the Cartan charge of the SU(2)$_R$ R-symmetry, and $F_x$ are the flavor charges of the theory. In addition, there is a U(1)$_I$ topological symmetry associated with the current $j_\mu \sim \star_5 \textrm{tr} (f \wedge f)_\mu$, where $f$ is the gauge field strength. The corresponding conserved charge is the instanton number $k$ \cite{Nekrasov:2002qd}. The trace is taken over the Hilbert space of the radially quantized SCFT on $\mathbb{R} \times S^4$, and the SCI receives contributions from $\frac{1}{8}$-BPS states annihilated by both the Poincar\'e supercharge $Q$ and its conjugate conformal supercharge $S=Q^\dagger$. These states satisfy
\begin{equation}
    \{Q,S\} = \Delta  - J_1 - J_2 - 3R = 0\ ,
\end{equation}
where $\Delta$ is the energy. The SCI can also be viewed as a Euclidean path integral on $S^1 \times S^4$, depending on various continuous parameters corresponding to the chemical potentials in \eqref{5d-index-tr} \cite{Kim:2012gu}. Importantly, it is a nontrivial function of $\omega_1$ and $\omega_2$, which are the equivariant parameters for the ${\rm SO}(5)$ rotations on $S^4$. We are interested in the Cardy-like limit in which $\omega_{1,2} \to \ri \, 0^+$, corresponding to the regime where the radius of $S^1$ becomes much smaller than that of $S^4$.

On the squashed five-sphere, the supersymmetric partition function can be computed via supersymmetric localization \cite{Kallen:2012cs,Hosomichi:2012ek,Kallen:2012va,Kim:2012ava,Imamura:2012xg,Imamura:2012efi,Lockhart:2012vp,Kim:2012qf,Pasquetti:2016dyl}. The squashed $S^5$ partition function $Z$ depends on real mass parameters, as well as on the three squashing parameters $b_1, b_2, b_3$ of $S^5$.\footnote{One combination of these parameters can be viewed as an overall change of the radius of $S^5$, to which the supersymmetric partition function is insensitive.} Here, the Cardy-like limit of the SCI is mapped to a large-squashing limit of $S^5$, after an appropriate analytic continuation of the squashing parameters.

In the Cardy-like limit $\omega_1, \omega_2 \to \ri \, 0^+$, we expect the following relation to hold:
\begin{equation}\label{5d-rel}
    \mathcal{I}_\textrm{SCI}^{\textrm{5d}} (\omega_1,\omega_2;\boldsymbol{\xi}) \approx Z(b_1,b_2,b_3;\boldsymbol{m}) \overline{Z}(b_1,b_2,b_3;\boldsymbol{m}) \ ,
\end{equation}
where $\boldsymbol{\xi}$ and $\boldsymbol{m}$ collectively denote, respectively, the chemical potentials and real mass parameters for the flavor symmetry, including the topological symmetry associated with the instanton number. Here, $\approx$ means equality to all orders in the perturbative expansion at small $\omega_1,\omega_2$ with fixed $\boldsymbol{\xi}$, up to nonperturbative corrections of order $\mathcal{O}\left(e^{-\#/|\omega_i|}\right)$, with $\#$ a positive constant. The parameters on the two sides are mapped as
\begin{equation}\label{5d-map}
    \omega_1 = \frac{b_1}{b_3}\ , \quad 
    \omega_2 = \frac{b_2}{b_3}\ , \quad 
    \boldsymbol{\xi} = 2\pi \, \frac{\boldsymbol{m}}{b_3}\ ,
\end{equation}
and the conjugate factor is defined by
\begin{equation}
    \overline{Z}(b_1,b_2,b_3;\boldsymbol{m})
    \equiv
    Z(b_1,b_2,-b_3;\boldsymbol{m})\ .
\end{equation}

A useful way to understand \eqref{5d-rel} is through equivariant localization. Both the SCI on $S^1\times S^4$ \cite{Kim:2012gu} and the supersymmetric partition function on squashed $S^5$ \cite{Kallen:2012cs,Hosomichi:2012ek,Kallen:2012va,Kim:2012ava,Imamura:2012xg,Imamura:2012efi,Lockhart:2012vp,Kim:2012qf} can be written, at least for theories admitting a Lagrangian description, in terms of local building blocks associated with the fixed points of the relevant equivariant action. These building blocks are closely related to the 5d Nekrasov instanton partition function on the $\Omega$-deformed background $S^1\times\mathbb{C}^2_{\epsilon_1,\epsilon_2}$ \cite{Nekrasov:2002qd}. Thus, \eqref{5d-rel} can be viewed as an asymptotic statement about two different gluings of essentially the same local data.

For concreteness, we test \eqref{5d-rel} in two examples. Before doing so, we introduce two useful special functions, the $\infty$-double-Pochhammer symbol and the elliptic gamma function:
\begin{equation}
    \begin{aligned}
        &(z;q_1,q_2) \equiv \left\{ \begin{array}{ll}
             \prod_{k_1,k_2\geq 0} (1-zq_1^{k_1}q_2^{k_2})&  |q_1|<1,|q_2|<1 \\
             \prod_{k_1,k_2\geq 0} (1-zq_1^{-k_1-1}q_2^{-k_2-1})&  |q_1|>1,|q_2|>1 \\
             \prod_{k_1,k_2\geq 0} (1-zq_1^{k_1}q_2^{-k_2-1})^{-1}&  |q_1|<1,|q_2|>1 \\
             \prod_{k_1,k_2\geq 0} (1-zq_1^{-k_1-1}q_2^{k_2})^{-1}& |q_1|>1,|q_2|<1 
        \end{array} \right.\ ,\\
        &\Gamma(z;q_1,q_2) \equiv \frac{(z^{-1}q_1q_2;q_1,q_2)}{(z;q_1,q_2)}\ .
    \end{aligned}
\end{equation}

Our first example is the free theory of a single hypermultiplet. The squashed $S^5$ partition function of the hypermultiplet is given by
\begin{equation}
\begin{aligned}
        Z = \prod_{\rho}& \frac{(-y^{-\rho}q_1^{-1/2}q_2^{1/2};q_1^{-1},q_2)_{(1)}}{(-y^{-\rho}q_1^{-1/2}q_2^{-1/2};q_1^{-1},q_2^{-1})_{(2)}(-y^{-\rho}q_1^{1/2}q_2^{1/2};q_1,q_2)_{(3)}} \\
        &\times \left[\frac{\Gamma(-y^{\rho}q_1^{-1/2}q_2^{-1/2};q_1^{-1},q_2^{-1})_{(2)}\Gamma(-y^{\rho}q_1^{1/2}q_2^{1/2};q_1,q_2)_{(3)}}{\Gamma(-y^{\rho}q_1^{-1/2}q_2^{1/2};q_1^{-1},q_2)_{(1)}}\right]^{1/2}\ ,
\end{aligned}
\end{equation}
where the subscript $(l)$, with $l=1,2,3$, indicates that the parameters inside the corresponding parentheses are evaluated as
\begin{equation}
    \begin{aligned}
        {y^{(l)}}^\rho= e^{2\pi \ri \rho(m) / b_l}\ , \quad q_1^{(l)} = e^{2\pi \ri \, b_{l-1}/b_l}\ , \quad q_2^{(l)} = e^{2\pi \ri \, b_{l+1}/b_l}\ ,
    \end{aligned}
\end{equation}
with $b_0 \equiv b_3$ and $b_4 \equiv b_1$. Here, $\rho$ runs over the weights of the flavor representation. The SCI of the hypermultiplet is given by
\begin{equation}
    \mathcal{I}_\textrm{SCI}^{\textrm{5d}} = \prod_{\rho} \frac{\Gamma (-y^\rho q_1^{1/2}q_2^{1/2} ;q_1,q_2)}{(-y^{-\rho} q_1^{1/2}q_2^{1/2} ;q_1,q_2)^2}\ ,
\end{equation}
where $q_{1,2} = e^{2\pi \ri \, \omega_{1,2}}$ and $y^\rho = e^{\ri \rho(\xi)}$. Then, in the Cardy-like limit $\omega_1, \omega_2 \to \ri \, 0^+$, it is straightforward to show that
\begin{equation}
    \begin{aligned}
        \mathcal{I}_\textrm{SCI}^{\textrm{5d}}(\omega_1,\omega_2;\xi) \approx Z(b_1,b_2,b_3;m)\overline{Z}(b_1,b_2,b_3;m)\ ,
    \end{aligned}
\end{equation}
which is consistent with \eqref{5d-rel}. Here we have assumed a condition analogous to \eqref{chiral-con}.

Another example is the large $N$ limit of a broad class of holographic SCFTs arising from the strong coupling limit of 5d gauge theories arising from D4-D8-O8 intersections in massive type IIA string theory \cite{Seiberg:1996bd,Intriligator:1997pq}, as well as those engineered by $(p,q)$ five-brane webs in type IIB string theory \cite{Aharony:1997ju,Kol:1997fv,Aharony:1997bh}. In this case, the SCI receives contributions from a large $N$ saddle point \cite{Choi:2019miv,Crichigno:2020ouj} corresponding to supersymmetric Kerr-Newman black holes in the AdS$_6$ dual \cite{Chow:2008ip}. In the universal case, where all flavor chemical potentials are turned off, the universal spinning black holes contribute to the SCI through their Euclidean on-shell action \cite{Choi:2018fdc,Cassani:2019mms}
\begin{equation}
    I =  \frac{2\pi^2L^4}{81G_N} \frac{(\omega_1+\omega_2+1)^3}{\omega_1\omega_2}\ ,
\end{equation}
where $L$ is the radius of AdS$_6$ and $G_N$ is the six-dimensional Newton constant. These parameters can be related to $N$ and other gauge theory data by embedding the six-dimensional solution into massive type IIA \cite{Cvetic:1999un} or type IIB string theory \cite{Jeong:2013jfc,Hong:2018amk,Malek:2018zcz,Malek:2019ucd}. The dependence on the angular momentum chemical potentials has precisely the homogeneous form expected from the squashed sphere side. Indeed, using \eqref{5d-map}, one obtains
\begin{equation}
    \frac{(\omega_1+\omega_2+1)^3}{\omega_1\omega_2}
    =
    \frac{(b_1+b_2+b_3)^3}{b_1b_2b_3}\ .
\end{equation}
This is the same universal squashing dependence that appears in the large $N$ squashed $S^5$ free energy. For studies of the (squashed) $S^5$ partition function in the large $N$ limit, see \cite{Jafferis:2012iv,DHoker:2016ysh,Uhlemann:2019ypp,Chang:2017mxc,Chang:2017cdx,Gutperle:2018axv,Hosseini:2019and,Crichigno:2018adf}. More generally, one can show that \eqref{5d-rel} continues to hold in this holographic large $N$ setting, including the dependence on flavor chemical potentials and real mass parameters \cite{Crichigno:2020ouj}.

We close this appendix with a few comments.
\begin{itemize}
    \item It would be desirable to give a rigorous derivation of \eqref{5d-rel} for general 5d $\mathcal{N}=1$ superconformal gauge theories. Since the relation has been checked explicitly for the hypermultiplet, extending the argument to general perturbative contributions should be straightforward. The main subtlety lies in nonperturbative instanton contributions. In the large $N$ gauge theories considered above, such instanton contributions are suppressed \cite{Jafferis:2012iv,Choi:2019miv}. However, there are many theories in which instanton contributions are not suppressed, such as pure $\mathcal{N}=1$ ${\rm SU}(N)$ SYM theory, understood as a relevant deformation of a UV SCFT.

    Since the proposed relation originates from equivariant localization on both $S^5$ and $S^1\times S^4$, we expect \eqref{5d-rel} to continue to hold even after including instanton contributions. A key step would be to understand the behavior of the $k$-instanton contribution to the instanton partition function on the $\Omega$-deformed background $S^1\times\mathbb{C}^2_{\epsilon_1,\epsilon_2}$, especially in the limit $|e^{2\pi \ri \epsilon_2}| \to 0^+$. Showing that such instanton contributions are either suppressed or canceled within a conjugate pair of blocks in the squashed $S^5$ partition function would be a crucial step toward a full nonperturbative proof of the conjecture.

    \item In this appendix, we considered only the squashed $S^5$ partition function and the superconformal index. However, one can also consider topologically twisted indices on $S_b^3\times S^2$ or $S^1\times S^2\times S^2$ \cite{Crichigno:2018adf,Hosseini:2018uzp,Hosseini:2021mnn}. These TTIs also admit an equivariant localization description and can be obtained by appropriately gluing copies of the instanton partition function on $S^1\times\mathbb{C}^2_{\epsilon_1,\epsilon_2}$. It is therefore natural to expect relations analogous to \eqref{5d-rel} for these observables as well. In the large $N$ limit, the TTI on $S^1 \times S^2 \times S^2$ receives a contribution from a saddle point corresponding to the supersymmetric static magnetically charged black holes/strings in the AdS$_6$ dual \cite{Hosseini:2018usu,Suh:2018tul,Suh:2018szn}.

    \item There also exist 5d SCFTs that do not admit conventional gauge theory descriptions; see e.g. \cite{Bao:2013pwa,Kim:2023qwh}. It would be interesting to understand whether \eqref{5d-rel} can be established for such non-Lagrangian SCFTs, for instance using an effective field theory description on $S^4$. A successful derivation in such examples would indicate that \eqref{5d-rel} is a universal property of 5d SCFTs, rather than a special consequence of supersymmetric localization for theories admitting Lagrangian gauge theory descriptions.
\end{itemize}

%%%%%%%%%%%%%%%%%%%%%%%%%%%%
\bibliography{OSVAdS}

@article{Kim:2023qwh,
    author = "Kim, Hee-Cheol and Kim, Minsung and Kim, Sung-Soo and Zafrir, Gabi",
    title = "{Superconformal indices for non-Lagrangian theories in five dimensions}",
    eprint = "2307.03231",
    archivePrefix = "arXiv",
    primaryClass = "hep-th",
    doi = "10.1007/JHEP03(2024)164",
    journal = "JHEP",
    volume = "03",
    pages = "164",
    year = "2024"
}

@article{Bao:2013pwa,
    author = "Bao, Ling and Mitev, Vladimir and Pomoni, Elli and Taki, Masato and Yagi, Futoshi",
    title = "{Non-Lagrangian Theories from Brane Junctions}",
    eprint = "1310.3841",
    archivePrefix = "arXiv",
    primaryClass = "hep-th",
    reportNumber = "DESY-13-176, HU-Mathematik-17-13, HU-EP-13-50, KIAS-P-13058, RIKEN-MP-78",
    doi = "10.1007/JHEP01(2014)175",
    journal = "JHEP",
    volume = "01",
    pages = "175",
    year = "2014"
}

@article{Suh:2018szn,
    author = "Suh, Minwoo",
    title = "{Supersymmetric $AdS_6$ black holes from matter coupled $F(4)$ gauged supergravity}",
    eprint = "1810.00675",
    archivePrefix = "arXiv",
    primaryClass = "hep-th",
    doi = "10.1007/JHEP02(2019)108",
    journal = "JHEP",
    volume = "02",
    pages = "108",
    year = "2019"
}

@article{Suh:2018tul,
    author = "Suh, Minwoo",
    title = "{Supersymmetric AdS$_{6}$ black holes from F(4) gauged supergravity}",
    eprint = "1809.03517",
    archivePrefix = "arXiv",
    primaryClass = "hep-th",
    doi = "10.1007/JHEP01(2019)035",
    journal = "JHEP",
    volume = "01",
    pages = "035",
    year = "2019"
}

@article{Hosseini:2018usu,
    author = "Hosseini, Seyed Morteza and Hristov, Kiril and Passias, Achilleas and Zaffaroni, Alberto",
    title = "{6D attractors and black hole microstates}",
    eprint = "1809.10685",
    archivePrefix = "arXiv",
    primaryClass = "hep-th",
    reportNumber = "IPMU18-0155, UUITP-41/18",
    doi = "10.1007/JHEP12(2018)001",
    journal = "JHEP",
    volume = "12",
    pages = "001",
    year = "2018"
}

@article{Hosseini:2018uzp,
    author = "Hosseini, Seyed Morteza and Yaakov, Itamar and Zaffaroni, Alberto",
    title = "{Topologically twisted indices in five dimensions and holography}",
    eprint = "1808.06626",
    archivePrefix = "arXiv",
    primaryClass = "hep-th",
    reportNumber = "IPMU18-0133",
    doi = "10.1007/JHEP11(2018)119",
    journal = "JHEP",
    volume = "11",
    pages = "119",
    year = "2018"
}

@article{Crichigno:2018adf,
    author = "Crichigno, P. Marcos and Jain, Dharmesh and Willett, Brian",
    title = "{5d Partition Functions with A Twist}",
    eprint = "1808.06744",
    archivePrefix = "arXiv",
    primaryClass = "hep-th",
    doi = "10.1007/JHEP11(2018)058",
    journal = "JHEP",
    volume = "11",
    pages = "058",
    year = "2018"
}

@article{Gutperle:2018axv,
    author = "Gutperle, Michael and Kaidi, Justin and Raj, Himanshu",
    title = "{Mass deformations of 5d SCFTs via holography}",
    eprint = "1801.00730",
    archivePrefix = "arXiv",
    primaryClass = "hep-th",
    doi = "10.1007/JHEP02(2018)165",
    journal = "JHEP",
    volume = "02",
    pages = "165",
    year = "2018"
}

@article{Chang:2017cdx,
    author = "Chang, Chi-Ming and Fluder, Martin and Lin, Ying-Hsuan and Wang, Yifan",
    title = "{Spheres, Charges, Instantons, and Bootstrap: A Five-Dimensional Odyssey}",
    eprint = "1710.08418",
    archivePrefix = "arXiv",
    primaryClass = "hep-th",
    reportNumber = "CALT-TH-2017-030, PUPT-2539",
    doi = "10.1007/JHEP03(2018)123",
    journal = "JHEP",
    volume = "03",
    pages = "123",
    year = "2018"
}

@article{Chang:2017mxc,
    author = "Chang, Chi-Ming and Fluder, Martin and Lin, Ying-Hsuan and Wang, Yifan",
    title = "{Romans Supergravity from Five-Dimensional Holograms}",
    eprint = "1712.10313",
    archivePrefix = "arXiv",
    primaryClass = "hep-th",
    reportNumber = "CALT-TH-2017-058, PUPT-2545",
    doi = "10.1007/JHEP05(2018)039",
    journal = "JHEP",
    volume = "05",
    pages = "039",
    year = "2018"
}

@article{Uhlemann:2019ypp,
    author = "Uhlemann, Christoph F.",
    title = "{Exact results for 5d SCFTs of long quiver type}",
    eprint = "1909.01369",
    archivePrefix = "arXiv",
    primaryClass = "hep-th",
    doi = "10.1007/JHEP11(2019)072",
    journal = "JHEP",
    volume = "11",
    pages = "072",
    year = "2019"
}

@article{DHoker:2016ysh,
    author = "D'Hoker, Eric and Gutperle, Michael and Uhlemann, Christoph F.",
    title = "{Holographic duals for five-dimensional superconformal quantum field theories}",
    eprint = "1611.09411",
    archivePrefix = "arXiv",
    primaryClass = "hep-th",
    doi = "10.1103/PhysRevLett.118.101601",
    journal = "Phys. Rev. Lett.",
    volume = "118",
    number = "10",
    pages = "101601",
    year = "2017"
}

@article{Jafferis:2012iv,
    author = "Jafferis, Daniel L. and Pufu, Silviu S.",
    title = "{Exact results for five-dimensional superconformal field theories with gravity duals}",
    eprint = "1207.4359",
    archivePrefix = "arXiv",
    primaryClass = "hep-th",
    reportNumber = "MIT-CTP-4387",
    doi = "10.1007/JHEP05(2014)032",
    journal = "JHEP",
    volume = "05",
    pages = "032",
    year = "2014"
}

@article{Malek:2019ucd,
    author = "Malek, Emanuel and Samtleben, Henning and Vall Camell, Valent{\'\i}",
    title = "{Supersymmetric AdS$_7$ and AdS$_6$ vacua and their consistent truncations with vector multiplets}",
    eprint = "1901.11039",
    archivePrefix = "arXiv",
    primaryClass = "hep-th",
    doi = "10.1007/JHEP04(2019)088",
    journal = "JHEP",
    volume = "04",
    pages = "088",
    year = "2019"
}

@article{Malek:2018zcz,
    author = "Malek, Emanuel and Samtleben, Henning and Vall Camell, Valent{\'\i}",
    title = "{Supersymmetric AdS$_{7}$ and AdS$_6$ vacua and their minimal consistent truncations from exceptional field theory}",
    eprint = "1808.05597",
    archivePrefix = "arXiv",
    primaryClass = "hep-th",
    reportNumber = "LMU-ASC 54/18, MPP-2018-212",
    doi = "10.1016/j.physletb.2018.09.037",
    journal = "Phys. Lett. B",
    volume = "786",
    pages = "171--179",
    year = "2018"
}

@article{Hong:2018amk,
    author = "Hong, Junho and Liu, James T. and Mayerson, Daniel R.",
    title = "{Gauged Six-Dimensional Supergravity from Warped IIB Reductions}",
    eprint = "1808.04301",
    archivePrefix = "arXiv",
    primaryClass = "hep-th",
    doi = "10.1007/JHEP09(2018)140",
    journal = "JHEP",
    volume = "09",
    pages = "140",
    year = "2018"
}

@article{Jeong:2013jfc,
    author = "Jeong, Jaehoon and Kelekci, Ozgur and O Colgain, Eoin",
    title = "{An alternative IIB embedding of F(4) gauged supergravity}",
    eprint = "1302.2105",
    archivePrefix = "arXiv",
    primaryClass = "hep-th",
    reportNumber = "CQUEST--2013-0583, FPAUO-13-02",
    doi = "10.1007/JHEP05(2013)079",
    journal = "JHEP",
    volume = "05",
    pages = "079",
    year = "2013"
}

@article{Cvetic:1999un,
    author = "Cvetic, Mirjam and Lu, Hong and Pope, C. N.",
    title = "{Gauged six-dimensional supergravity from massive type IIA}",
    eprint = "hep-th/9906221",
    archivePrefix = "arXiv",
    reportNumber = "CTP-TAMU-27-99, UPR-851-T",
    doi = "10.1103/PhysRevLett.83.5226",
    journal = "Phys. Rev. Lett.",
    volume = "83",
    pages = "5226--5229",
    year = "1999"
}

@article{Chow:2008ip,
    author = "Chow, David D. K.",
    title = "{Charged rotating black holes in six-dimensional gauged supergravity}",
    eprint = "0808.2728",
    archivePrefix = "arXiv",
    primaryClass = "hep-th",
    reportNumber = "DAMTP-2008-73",
    doi = "10.1088/0264-9381/27/6/065004",
    journal = "Class. Quant. Grav.",
    volume = "27",
    pages = "065004",
    year = "2010"
}

@article{Aharony:1997ju,
    author = "Aharony, Ofer and Hanany, Amihay",
    title = "{Branes, superpotentials and superconformal fixed points}",
    eprint = "hep-th/9704170",
    archivePrefix = "arXiv",
    reportNumber = "RU-97-25, IASSNS-HEP-97-38",
    doi = "10.1016/S0550-3213(97)00472-0",
    journal = "Nucl. Phys. B",
    volume = "504",
    pages = "239--271",
    year = "1997"
}

@article{Kol:1997fv,
    author = "Kol, Barak",
    title = "{5-D field theories and M theory}",
    eprint = "hep-th/9705031",
    archivePrefix = "arXiv",
    reportNumber = "SU-ITP-97-22",
    doi = "10.1088/1126-6708/1999/11/026",
    journal = "JHEP",
    volume = "11",
    pages = "026",
    year = "1999"
}

@article{Aharony:1997bh,
    author = "Aharony, Ofer and Hanany, Amihay and Kol, Barak",
    title = "{Webs of (p,q) five-branes, five-dimensional field theories and grid diagrams}",
    eprint = "hep-th/9710116",
    archivePrefix = "arXiv",
    reportNumber = "IASSNS-HEP-97-113, RU-97-81, SU-ITP-97-40",
    doi = "10.1088/1126-6708/1998/01/002",
    journal = "JHEP",
    volume = "01",
    pages = "002",
    year = "1998"
}

@article{Kallen:2012cs,
    author = {K{\"a}ll{\'e}n, Johan and Zabzine, Maxim},
    title = "{Twisted supersymmetric 5D Yang-Mills theory and contact geometry}",
    eprint = "1202.1956",
    archivePrefix = "arXiv",
    primaryClass = "hep-th",
    reportNumber = "UUITP-04-12",
    doi = "10.1007/JHEP05(2012)125",
    journal = "JHEP",
    volume = "05",
    pages = "125",
    year = "2012"
}

@article{Hosomichi:2012ek,
    author = "Hosomichi, Kazuo and Seong, Rak-Kyeong and Terashima, Seiji",
    title = "{Supersymmetric Gauge Theories on the Five-Sphere}",
    eprint = "1203.0371",
    archivePrefix = "arXiv",
    primaryClass = "hep-th",
    reportNumber = "YITP-12-10, IMPERIAL-TP-12-RS-02",
    doi = "10.1016/j.nuclphysb.2012.08.007",
    journal = "Nucl. Phys. B",
    volume = "865",
    pages = "376--396",
    year = "2012"
}

@article{Kallen:2012va,
    author = {K{\"a}ll{\'e}n, Johan and Qiu, Jian and Zabzine, Maxim},
    title = "{The perturbative partition function of supersymmetric 5D Yang-Mills theory with matter on the five-sphere}",
    eprint = "1206.6008",
    archivePrefix = "arXiv",
    primaryClass = "hep-th",
    reportNumber = "UUITP-17-12",
    doi = "10.1007/JHEP08(2012)157",
    journal = "JHEP",
    volume = "08",
    pages = "157",
    year = "2012"
}

@article{Kim:2012ava,
    author = "Kim, Hee-Cheol and Kim, Seok",
    title = "{M5-branes from gauge theories on the 5-sphere}",
    eprint = "1206.6339",
    archivePrefix = "arXiv",
    primaryClass = "hep-th",
    reportNumber = "SNUTP12-002, KIAS-P12038",
    doi = "10.1007/JHEP05(2013)144",
    journal = "JHEP",
    volume = "05",
    pages = "144",
    year = "2013"
}

@article{Imamura:2012xg,
    author = "Imamura, Yosuke",
    title = "{Supersymmetric theories on squashed five-sphere}",
    eprint = "1209.0561",
    archivePrefix = "arXiv",
    primaryClass = "hep-th",
    reportNumber = "TIT-HEP-620",
    doi = "10.1093/ptep/pts052",
    journal = "PTEP",
    volume = "2013",
    pages = "013B04",
    year = "2013"
}

@article{Imamura:2012efi,
    author = "Imamura, Yosuke",
    title = "{Perturbative partition function for squashed $S^5$}",
    eprint = "1210.6308",
    archivePrefix = "arXiv",
    primaryClass = "hep-th",
    reportNumber = "TIT-HEP-622",
    doi = "10.1093/ptep/ptt044",
    journal = "PTEP",
    volume = "2013",
    number = "7",
    pages = "073B01",
    year = "2013"
}

@article{Lockhart:2012vp,
    author = "Lockhart, Guglielmo and Vafa, Cumrun",
    title = "{Superconformal Partition Functions and Non-perturbative Topological Strings}",
    eprint = "1210.5909",
    archivePrefix = "arXiv",
    primaryClass = "hep-th",
    doi = "10.1007/JHEP10(2018)051",
    journal = "JHEP",
    volume = "10",
    pages = "051",
    year = "2018"
}

@article{Kim:2012qf,
    author = "Kim, Hee-Cheol and Kim, Joonho and Kim, Seok",
    title = "{Instantons on the 5-sphere and M5-branes}",
    eprint = "1211.0144",
    archivePrefix = "arXiv",
    primaryClass = "hep-th",
    reportNumber = "KIAS-P12070, SNUTP12-004",
    month = "11",
    year = "2012"
}

@article{Nekrasov:2002qd,
    author = "Nekrasov, Nikita A.",
    title = "{Seiberg-Witten prepotential from instanton counting}",
    eprint = "hep-th/0206161",
    archivePrefix = "arXiv",
    reportNumber = "ITEP-TH-22-02, IHES-P-04-22",
    doi = "10.4310/ATMP.2003.v7.n5.a4",
    journal = "Adv. Theor. Math. Phys.",
    volume = "7",
    number = "5",
    pages = "831--864",
    year = "2003"
}

@article{Intriligator:1997pq,
    author = "Intriligator, Kenneth A. and Morrison, David R. and Seiberg, Nathan",
    title = "{Five-dimensional supersymmetric gauge theories and degenerations of Calabi-Yau spaces}",
    eprint = "hep-th/9702198",
    archivePrefix = "arXiv",
    reportNumber = "RU-96-99, IASSNS-HEP-96-112",
    doi = "10.1016/S0550-3213(97)00279-4",
    journal = "Nucl. Phys. B",
    volume = "497",
    pages = "56--100",
    year = "1997"
}

@article{Kim:2012gu,
    author = "Kim, Hee-Cheol and Kim, Sung-Soo and Lee, Kimyeong",
    title = "{5-dim Superconformal Index with Enhanced En Global Symmetry}",
    eprint = "1206.6781",
    archivePrefix = "arXiv",
    primaryClass = "hep-th",
    reportNumber = "KIAS-P12033",
    doi = "10.1007/JHEP10(2012)142",
    journal = "JHEP",
    volume = "10",
    pages = "142",
    year = "2012"
}

@article{Nicolai:2011cy,
	author = "Nicolai, Hermann and Pilch, Krzysztof",
	title = "{Consistent Truncation of d = 11 Supergravity on AdS$_4 \times S^7$}",
	eprint = "1112.6131",
	archivePrefix = "arXiv",
	primaryClass = "hep-th",
	reportNumber = "AEI-2011-093",
	doi = "10.1007/JHEP03(2012)099",
	journal = "JHEP",
	volume = "03",
	pages = "099",
	year = "2012"
}

@article{Pasquetti:2011fj,
	author = "Pasquetti, Sara",
	title = "{Factorisation of N = 2 Theories on the Squashed 3-Sphere}",
	eprint = "1111.6905",
	archivePrefix = "arXiv",
	primaryClass = "hep-th",
	doi = "10.1007/JHEP04(2012)120",
	journal = "JHEP",
	volume = "04",
	pages = "120",
	year = "2012"
}

@article{Varela:2015ywx,
	author = "Varela, Oscar",
	title = "{Complete $D=11$ embedding of SO(8) supergravity}",
	eprint = "1512.04943",
	archivePrefix = "arXiv",
	primaryClass = "hep-th",
	doi = "10.1103/PhysRevD.97.045010",
	journal = "Phys. Rev. D",
	volume = "97",
	number = "4",
	pages = "045010",
	year = "2018"
}

@article{Azizi:2016noi,
	author = "Azizi, Arash and Godazgar, Hadi and Godazgar, Mahdi and Pope, C. N.",
	title = "{Embedding of gauged STU supergravity in eleven dimensions}",
	eprint = "1606.06954",
	archivePrefix = "arXiv",
	primaryClass = "hep-th",
	reportNumber = "MI-TH-1613, DAMTP-2016-28",
	doi = "10.1103/PhysRevD.94.066003",
	journal = "Phys. Rev. D",
	volume = "94",
	number = "6",
	pages = "066003",
	year = "2016"
}

@article{Cvetic:1999xp,
	author = "Cvetic, Mirjam and Duff, M. J. and Hoxha, P. and Liu, James T. and Lu, Hong and Lu, J. X. and Martinez-Acosta, R. and Pope, C. N. and Sati, H. and Tran, Tuan A.",
	title = "{Embedding AdS black holes in ten-dimensions and eleven-dimensions}",
	eprint = "hep-th/9903214",
	archivePrefix = "arXiv",
	reportNumber = "UPR-0840-T, CTP-TAMU-11-99, RU-99-4-B",
	doi = "10.1016/S0550-3213(99)00419-8",
	journal = "Nucl. Phys. B",
	volume = "558",
	pages = "96--126",
	year = "1999"
}

@article{BenettiGenolini:2024kyy,
    author = {Benetti Genolini, Pietro and Gauntlett, Jerome P. and Jiao, Yusheng and L{\"u}scher, Alice and Sparks, James},
    title = "{Localization and attraction}",
    eprint = "2401.10977",
    archivePrefix = "arXiv",
    primaryClass = "hep-th",
    doi = "10.1007/JHEP05(2024)152",
    journal = "JHEP",
    volume = "05",
    pages = "152",
    year = "2024"
}

@article{Bobev:2018uxk,
    author = "Bobev, Nikolay and Min, Vincent S. and Pilch, Krzysztof",
    title = "{Mass-deformed ABJM and black holes in AdS$_{4}$}",
    eprint = "1801.03135",
    archivePrefix = "arXiv",
    primaryClass = "hep-th",
    doi = "10.1007/JHEP03(2018)050",
    journal = "JHEP",
    volume = "03",
    pages = "050",
    year = "2018"
}

@article{Beasley:2006us,
    author = "Beasley, Chris and Gaiotto, Davide and Guica, Monica and Huang, Lisa and Strominger, Andrew and Yin, Xi",
    title = "{Why Z(BH) = |Z(top)|**2}",
    eprint = "hep-th/0608021",
    archivePrefix = "arXiv",
    month = "8",
    year = "2006"
}

@article{Hristov:2019xku,
	author = "Hristov, Kiril and Lodato, Ivano and Reys, Valentin",
	title = "{One-loop determinants for black holes in 4d gauged supergravity}",
	eprint = "1908.05696",
	archivePrefix = "arXiv",
	primaryClass = "hep-th",
	doi = "10.1007/JHEP11(2019)105",
	journal = "JHEP",
	volume = "11",
	pages = "105",
	year = "2019"
}

@article{Hristov:2018lod,
	author = "Hristov, Kiril and Lodato, Ivano and Reys, Valentin",
	title = "{On the quantum entropy function in 4d gauged supergravity}",
	eprint = "1803.05920",
	archivePrefix = "arXiv",
	primaryClass = "hep-th",
	doi = "10.1007/JHEP07(2018)072",
	journal = "JHEP",
	volume = "07",
	pages = "072",
	year = "2018"
}

@article{Romans:1991nq,
	author = "Romans, L. J.",
	title = "{Supersymmetric, cold and lukewarm black holes in cosmological Einstein-Maxwell theory}",
	eprint = "hep-th/9203018",
	archivePrefix = "arXiv",
	reportNumber = "PRINT-92-0114 (JPL,CAL-TECH)",
	doi = "10.1016/0550-3213(92)90684-4",
	journal = "Nucl. Phys. B",
	volume = "383",
	pages = "395--415",
	year = "1992"
}

@article{Hristov:2018spe,
	author = "Hristov, Kiril and Katmadas, Stefanos and Toldo, Chiara",
	title = "{Rotating attractors and BPS black holes in $AdS_4$}",
	eprint = "1811.00292",
	archivePrefix = "arXiv",
	primaryClass = "hep-th",
	doi = "10.1007/JHEP01(2019)199",
	journal = "JHEP",
	volume = "01",
	pages = "199",
	year = "2019"
}

@article{Hosseini:2022vho,
	author = "Hosseini, Seyed Morteza and Zaffaroni, Alberto",
	title = "{The large N limit of topologically twisted indices: a direct approach}",
	eprint = "2209.09274",
	archivePrefix = "arXiv",
	primaryClass = "hep-th",
	doi = "10.1007/JHEP12(2022)025",
	journal = "JHEP",
	volume = "12",
	pages = "025",
	year = "2022"
}

@article{Martelli:2023oqk,
    author = "Martelli, Dario and Zaffaroni, Alberto",
    title = "{Equivariant localization and holography}",
    eprint = "2306.03891",
    archivePrefix = "arXiv",
    primaryClass = "hep-th",
    doi = "10.1007/s11005-023-01752-1",
    journal = "Lett. Math. Phys.",
    volume = "114",
    number = "1",
    pages = "15",
    year = "2024"
}

@article{BenettiGenolini:2024xeo,
    author = {Benetti Genolini, Pietro and Gauntlett, Jerome P. and Jiao, Yusheng and L{\"u}scher, Alice and Sparks, James},
    title = "{Localization of the Free Energy in Supergravity}",
    eprint = "2407.02554",
    archivePrefix = "arXiv",
    primaryClass = "hep-th",
    doi = "10.1103/PhysRevLett.133.141601",
    journal = "Phys. Rev. Lett.",
    volume = "133",
    number = "14",
    pages = "141601",
    year = "2024"
}

@article{BenettiGenolini:2024lbj,
    author = {Benetti Genolini, Pietro and Gauntlett, Jerome P. and Jiao, Yusheng and L{\"u}scher, Alice and Sparks, James},
    title = "{Equivariant localization for D = 4 gauged supergravity}",
    eprint = "2412.07828",
    archivePrefix = "arXiv",
    primaryClass = "hep-th",
    doi = "10.1007/JHEP08(2025)211",
    journal = "JHEP",
    volume = "08",
    pages = "211",
    year = "2025"
}

@article{BenettiGenolini:2019jdz,
    author = "Benetti Genolini, Pietro and Perez Ipi{\~n}a, Juan Manuel and Sparks, James",
    title = "{Localization of the action in AdS/CFT}",
    eprint = "1906.11249",
    archivePrefix = "arXiv",
    primaryClass = "hep-th",
    doi = "10.1007/JHEP10(2019)252",
    journal = "JHEP",
    volume = "10",
    pages = "252",
    year = "2019"
}

@article{Genolini:2021urf,
    author = "Genolini, Pietro Benetti and Richmond, Paul",
    title = "{Supersymmetry of higher-derivative supergravity in AdS4 holography}",
    eprint = "2107.04590",
    archivePrefix = "arXiv",
    primaryClass = "hep-th",
    doi = "10.1103/PhysRevD.104.L061902",
    journal = "Phys. Rev. D",
    volume = "104",
    number = "6",
    pages = "L061902",
    year = "2021"
}

@article{Bobev:2018wbt,
    author = "Bobev, Nikolay and Min, Vincent S. and Pilch, Krzysztof and Rosso, Felipe",
    title = "{Mass Deformations of the ABJM Theory: The Holographic Free Energy}",
    eprint = "1812.01026",
    archivePrefix = "arXiv",
    primaryClass = "hep-th",
    doi = "10.1007/JHEP03(2019)130",
    journal = "JHEP",
    volume = "03",
    pages = "130",
    year = "2019"
}

@article{Benini:2016rke,
    author = "Benini, Francesco and Hristov, Kiril and Zaffaroni, Alberto",
    title = "{Exact microstate counting for dyonic black holes in AdS4}",
    eprint = "1608.07294",
    archivePrefix = "arXiv",
    primaryClass = "hep-th",
    reportNumber = "SISSA-41-2016-FISI, SISSA 41/2016/FISI",
    doi = "10.1016/j.physletb.2017.05.076",
    journal = "Phys. Lett. B",
    volume = "771",
    pages = "462--466",
    year = "2017"
}

@article{Cacciatori:2009iz,
    author = "Cacciatori, Sergio L. and Klemm, Dietmar",
    title = "{Supersymmetric AdS(4) black holes and attractors}",
    eprint = "0911.4926",
    archivePrefix = "arXiv",
    primaryClass = "hep-th",
    reportNumber = "IFUM-947-FT",
    doi = "10.1007/JHEP01(2010)085",
    journal = "JHEP",
    volume = "01",
    pages = "085",
    year = "2010"
}

@article{Bobev:2020pjk,
    author = "Bobev, Nikolay and Charles, Anthony M. and Min, Vincent S.",
    title = "{Euclidean black saddles and AdS$_{4}$ black holes}",
    eprint = "2006.01148",
    archivePrefix = "arXiv",
    primaryClass = "hep-th",
    doi = "10.1007/JHEP10(2020)073",
    journal = "JHEP",
    volume = "10",
    pages = "073",
    year = "2020"
}

@article{Fuji:2011km,
    author = "Fuji, Hiroyuki and Hirano, Shinji and Moriyama, Sanefumi",
    title = "{Summing Up All Genus Free Energy of ABJM Matrix Model}",
    eprint = "1106.4631",
    archivePrefix = "arXiv",
    primaryClass = "hep-th",
    doi = "10.1007/JHEP08(2011)001",
    journal = "JHEP",
    volume = "08",
    pages = "001",
    year = "2011"
}

@article{Marino:2011eh,
    author = "Marino, Marcos and Putrov, Pavel",
    title = "{ABJM theory as a Fermi gas}",
    eprint = "1110.4066",
    archivePrefix = "arXiv",
    primaryClass = "hep-th",
    doi = "10.1088/1742-5468/2012/03/P03001",
    journal = "J. Stat. Mech.",
    volume = "1203",
    pages = "P03001",
    year = "2012"
}

@article{Dabholkar:2010uh,
    author = "Dabholkar, Atish and Gomes, Joao and Murthy, Sameer",
    title = "{Quantum black holes, localization and the topological string}",
    eprint = "1012.0265",
    archivePrefix = "arXiv",
    primaryClass = "hep-th",
    doi = "10.1007/JHEP06(2011)019",
    journal = "JHEP",
    volume = "06",
    pages = "019",
    year = "2011"
}

@article{Marino:2016new,
    author = "Marino, Marcos",
    title = "{Localization at large N in Chern{\textendash}Simons-matter theories}",
    eprint = "1608.02959",
    archivePrefix = "arXiv",
    primaryClass = "hep-th",
    doi = "10.1088/1751-8121/aa5f69",
    journal = "J. Phys. A",
    volume = "50",
    number = "44",
    pages = "443007",
    year = "2017"
}

@article{Zan:2021ftf,
    author = "Zan, Bernardo and Freedman, Daniel Z. and Pufu, Silviu S.",
    title = "{The $ \mathcal{N} $ = 2 prepotential and the sphere free energy}",
    eprint = "2112.06931",
    archivePrefix = "arXiv",
    primaryClass = "hep-th",
    reportNumber = "PUPT-2628, MIT-CTP/5379",
    doi = "10.1007/JHEP06(2022)045",
    journal = "JHEP",
    volume = "06",
    pages = "045",
    year = "2022"
}

@article{Bershadsky:1993cx,
    author = "Bershadsky, M. and Cecotti, S. and Ooguri, H. and Vafa, C.",
    title = "{Kodaira-Spencer theory of gravity and exact results for quantum string amplitudes}",
    eprint = "hep-th/9309140",
    archivePrefix = "arXiv",
    reportNumber = "HUTP-93-A025, RIMS-946, SISSA-142-93-EP",
    doi = "10.1007/BF02099774",
    journal = "Commun. Math. Phys.",
    volume = "165",
    pages = "311--428",
    year = "1994"
}

@article{Antoniadis:1993ze,
    author = "Antoniadis, Ignatios and Gava, E. and Narain, K. S. and Taylor, T. R.",
    title = "{Topological amplitudes in string theory}",
    eprint = "hep-th/9307158",
    archivePrefix = "arXiv",
    reportNumber = "NUB-3071, IC-93-202, CPTH-A258-0793",
    doi = "10.1016/0550-3213(94)90617-3",
    journal = "Nucl. Phys. B",
    volume = "413",
    pages = "162--184",
    year = "1994"
}

@article{Gautason:2025per,
    author = "Gautason, Fridrik Freyr and van Muiden, Jesse",
    title = "{Localization of the M2-Brane}",
    eprint = "2503.16597",
    archivePrefix = "arXiv",
    primaryClass = "hep-th",
    doi = "10.1103/67bh-xd42",
    journal = "Phys. Rev. Lett.",
    volume = "135",
    number = "10",
    pages = "101601",
    year = "2025"
}

@article{BenettiGenolini:2026qdm,
    author = "Benetti Genolini, Pietro and Gaar, Florian and Gauntlett, Jerome P. and Sparks, James",
    title = "{Equivariant localization for higher derivative supergravity}",
    eprint = "2604.08656",
    archivePrefix = "arXiv",
    primaryClass = "hep-th",
    month = "4",
    year = "2026"
}

@article{Cassia:2025jkr,
    author = "Cassia, Luca and Hristov, Kiril",
    title = "{M2-brane partition functions and HD supergravity from equivariant volumes}",
    eprint = "2508.21619",
    archivePrefix = "arXiv",
    primaryClass = "hep-th",
    doi = "10.1007/JHEP03(2026)100",
    journal = "JHEP",
    volume = "03",
    pages = "100",
    year = "2026"
}

@article{Gautason:2025plx,
    author = "Gautason, Fridrik Freyr and van Muiden, Jesse",
    title = "{Ensembles in M-theory and holography}",
    eprint = "2505.21633",
    archivePrefix = "arXiv",
    primaryClass = "hep-th",
    doi = "10.1007/JHEP11(2025)078",
    journal = "JHEP",
    volume = "11",
    pages = "078",
    year = "2025"
}

@inproceedings{vanMuiden:2026nsp,
    author = "van Muiden, Jesse",
    title = "{Quantum M2-branes and Holography}",
    eprint = "2603.14544",
    archivePrefix = "arXiv",
    primaryClass = "hep-th",
    month = "3",
    year = "2026"
}

@article{He:2025zxk,
    author = "He, Bin and Nosaka, Tomoki",
    title = "{New recursion relations for M2-brane matrix models}",
    eprint = "2509.15801",
    archivePrefix = "arXiv",
    primaryClass = "hep-th",
    doi = "10.1007/JHEP02(2026)243",
    journal = "JHEP",
    volume = "02",
    pages = "243",
    year = "2026"
}

@article{Hristov:2026zjh,
    author = "Hristov, Kiril and Kubo, Naotaka and Pang, Yi",
    title = "{$S^3$ partition functions and Equivariant CY$_4 $/ CY$_3$ correspondence from Quantum curves}",
    eprint = "2603.19159",
    archivePrefix = "arXiv",
    primaryClass = "hep-th",
    month = "3",
    year = "2026"
}

@article{Choi:2019miv,
    author = "Choi, Sunjin and Kim, Seok",
    title = "{Large AdS$_{6}$ black holes from CFT$_{5}$}",
    eprint = "1904.01164",
    archivePrefix = "arXiv",
    primaryClass = "hep-th",
    reportNumber = "SNUTP19-001",
    doi = "10.1007/JHEP08(2024)228",
    journal = "JHEP",
    volume = "08",
    pages = "228",
    year = "2024"
}

@article{Choi:2026,
	author = "Choi, Sunjin and Hong, Junho and Lee, Jehyun and Lee, Suhyun",
	title = "{To appear}",
	eprint = "26xx.xxxxx",
	archivePrefix = "arXiv",
	primaryClass = "hep-th",
	month = "x",
	year = "2026"
}

@article{Atiyah:1978ri,
	author = "Atiyah, M. F. and Hitchin, Nigel J. and Drinfeld, V. G. and Manin, Yu. I.",
	editor = "Shifman, Mikhail A.",
	title = "{Construction of Instantons}",
	doi = "10.1016/0375-9601(78)90141-X",
	journal = "Phys. Lett. A",
	volume = "65",
	pages = "185--187",
	year = "1978"
}

@article{Geukens:2024zmt,
	author = "Geukens, Seppe and Hong, Junho",
	title = "{Subleading analysis for S$^{3}$ partition functions of $ \mathcal{N} $ = 2 holographic SCFTs}",
	eprint = "2405.00845",
	archivePrefix = "arXiv",
	primaryClass = "hep-th",
	doi = "10.1007/JHEP06(2024)190",
	journal = "JHEP",
	volume = "06",
	pages = "190",
	year = "2024"
}

@article{ArabiArdehali:2025bub,
	author = "Arabi Ardehali, Arash and Boisvert, Mathieu and Fadda, Shehab Hossam",
	title = "{Cardy limit of the 3d superconformal index}",
	eprint = "2509.18285",
	archivePrefix = "arXiv",
	primaryClass = "hep-th",
	month = "9",
	year = "2025"
}

@article{Freedman:2013oja,
	author = "Freedman, Daniel Z. and Pufu, Silviu S.",
	title = "{The holography of $F$-maximization}",
	eprint = "1302.7310",
	archivePrefix = "arXiv",
	primaryClass = "hep-th",
	reportNumber = "SU-ITP-13-01, MIT-CTP-4443",
	doi = "10.1007/JHEP03(2014)135",
	journal = "JHEP",
	volume = "03",
	pages = "135",
	year = "2014"
}

@article{Crichigno:2020ouj,
    author = "Crichigno, P. Marcos and Jain, Dharmesh",
    title = "{The 5d Superconformal Index at Large $N$ and Black Holes}",
    eprint = "2005.00550",
    archivePrefix = "arXiv",
    primaryClass = "hep-th",
    doi = "10.1007/JHEP09(2020)124",
    journal = "JHEP",
    volume = "09",
    pages = "124",
    year = "2020"
}

@article{Kubo:2024qhq,
    author = "Kubo, Naotaka and Nosaka, Tomoki and Pang, Yi",
    title = "{Exact large N expansion of mass deformed ABJM theory on squashed sphere}",
    eprint = "2411.07334",
    archivePrefix = "arXiv",
    primaryClass = "hep-th",
    doi = "10.1007/JHEP02(2025)106",
    journal = "JHEP",
    volume = "02",
    pages = "106",
    year = "2025"
}

@article{Kubo:2025dot,
    author = "Kubo, Naotaka and Nosaka, Tomoki and Pang, Yi",
    title = "{Exact large N expansion of N=4 circular quiver Chern-Simons theories}",
    eprint = "2504.04402",
    archivePrefix = "arXiv",
    primaryClass = "hep-th",
    reportNumber = "USTC-ICTS/PCFT-25-15",
    doi = "10.1103/x5h6-r3dd",
    journal = "Phys. Rev. D",
    volume = "112",
    number = "4",
    pages = "046023",
    year = "2025"
}

@article{Nosaka:2024gle,
    author = "Nosaka, Tomoki",
    title = "{Large N expansion of mass deformed ABJM matrix model: M2-instanton condensation and beyond}",
    eprint = "2401.11484",
    archivePrefix = "arXiv",
    primaryClass = "hep-th",
    doi = "10.1007/JHEP03(2024)087",
    journal = "JHEP",
    volume = "03",
    pages = "087",
    year = "2024"
}

@article{Kapustin:2010xq,
    author = "Kapustin, Anton and Willett, Brian and Yaakov, Itamar",
    title = "{Nonperturbative Tests of Three-Dimensional Dualities}",
    eprint = "1003.5694",
    archivePrefix = "arXiv",
    primaryClass = "hep-th",
    doi = "10.1007/JHEP10(2010)013",
    journal = "JHEP",
    volume = "10",
    pages = "013",
    year = "2010"
}

@article{Jafferis:2010un,
    author = "Jafferis, Daniel L.",
    title = "{The Exact Superconformal R-Symmetry Extremizes Z}",
    eprint = "1012.3210",
    archivePrefix = "arXiv",
    primaryClass = "hep-th",
    doi = "10.1007/JHEP05(2012)159",
    journal = "JHEP",
    volume = "05",
    pages = "159",
    year = "2012"
}

@article{Imamura:2011wg,
    author = "Imamura, Yosuke and Yokoyama, Daisuke",
    title = "{N=2 supersymmetric theories on squashed three-sphere}",
    eprint = "1109.4734",
    archivePrefix = "arXiv",
    primaryClass = "hep-th",
    reportNumber = "TIT-HEP-613",
    doi = "10.1103/PhysRevD.85.025015",
    journal = "Phys. Rev. D",
    volume = "85",
    pages = "025015",
    year = "2012"
}

@article{Willett:2016adv,
    author = "Willett, Brian",
    title = "{Localization on three-dimensional manifolds}",
    eprint = "1608.02958",
    archivePrefix = "arXiv",
    primaryClass = "hep-th",
    doi = "10.1088/1751-8121/aa612f",
    journal = "J. Phys. A",
    volume = "50",
    number = "44",
    pages = "443006",
    year = "2017"
}

@article{Chester:2021gdw,
    author = "Chester, Shai M. and Kalloor, Rohit R. and Sharon, Adar",
    title = "{Squashing, Mass, and Holography for 3d Sphere Free Energy}",
    eprint = "2102.05643",
    archivePrefix = "arXiv",
    primaryClass = "hep-th",
    doi = "10.1007/JHEP04(2021)244",
    journal = "JHEP",
    volume = "04",
    pages = "244",
    year = "2021"
}

@article{Hatsuda:2016uqa,
    author = "Hatsuda, Yasuyuki",
    title = "{ABJM on ellipsoid and topological strings}",
    eprint = "1601.02728",
    archivePrefix = "arXiv",
    primaryClass = "hep-th",
    doi = "10.1007/JHEP07(2016)026",
    journal = "JHEP",
    volume = "07",
    pages = "026",
    year = "2016"
}

@article{Hosseini:2019and,
    author = "Hosseini, Seyed Morteza and Toldo, Chiara and Yaakov, Itamar",
    title = "{Supersymmetric R{\'e}nyi entropy and charged hyperbolic black holes}",
    eprint = "1912.04868",
    archivePrefix = "arXiv",
    primaryClass = "hep-th",
    reportNumber = "IPMU19-0180, CPHT-045/11 2019",
    doi = "10.1007/JHEP07(2020)131",
    journal = "JHEP",
    volume = "07",
    pages = "131",
    year = "2020"
}

@article{Martelli:2011qj,
    author = "Martelli, Dario and Sparks, James",
    title = "{The large N limit of quiver matrix models and Sasaki-Einstein manifolds}",
    eprint = "1102.5289",
    archivePrefix = "arXiv",
    primaryClass = "hep-th",
    doi = "10.1103/PhysRevD.84.046008",
    journal = "Phys. Rev. D",
    volume = "84",
    pages = "046008",
    year = "2011"
}

@article{Nosaka:2015iiw,
    author = "Nosaka, Tomoki",
    title = "{Instanton effects in ABJM theory with general R-charge assignments}",
    eprint = "1512.02862",
    archivePrefix = "arXiv",
    primaryClass = "hep-th",
    reportNumber = "YITP-15-105",
    doi = "10.1007/JHEP03(2016)059",
    journal = "JHEP",
    volume = "03",
    pages = "059",
    year = "2016"
}

@article{Narukawa:2003ltf,
    author = "Narukawa, Atsushi",
    title = "{The modular properties and the integral representations of the multiple elliptic gamma functions}",
    eprint = "math/0306164",
    archivePrefix = "arXiv",
    month = "6",
    year = "2003"
}

@article{Hristov:2022plc,
    author = "Hristov, Kiril",
    title = "{Maximally symmetric nuts in 4d $\mathcal{N}=2$ higher derivative supergravity}",
    eprint = "2212.10590",
    archivePrefix = "arXiv",
    primaryClass = "hep-th",
    doi = "10.1007/JHEP02(2023)110",
    journal = "JHEP",
    volume = "02",
    pages = "110",
    year = "2023"
}

@article{Guica:2007wd,
    author = "Guica, Monica and Strominger, Andrew",
    editor = "Baulieu, Laurent and de Boer, Jan and Douglas, Michael R. and Rabinovici, Eliezer and Vanhove, Pierre and Windey, Paul",
    title = "{Cargese lectures on string theory with eight supercharges}",
    eprint = "0704.3295",
    archivePrefix = "arXiv",
    primaryClass = "hep-th",
    doi = "10.1016/j.nuclphysbps.2007.06.007",
    journal = "Nucl. Phys. B Proc. Suppl.",
    volume = "171",
    pages = "39--68",
    year = "2007"
}

@article{Pioline:2006ni,
    author = "Pioline, Boris",
    title = "{Lectures on black holes, topological strings and quantum attractors}",
    eprint = "hep-th/0607227",
    archivePrefix = "arXiv",
    reportNumber = "LPTENS-06-27",
    doi = "10.1088/0264-9381/23/21/S05",
    journal = "Class. Quant. Grav.",
    volume = "23",
    pages = "S981",
    year = "2006"
}

@article{Ooguri:2004zv,
    author = "Ooguri, Hirosi and Strominger, Andrew and Vafa, Cumrun",
    title = "{Black hole attractors and the topological string}",
    eprint = "hep-th/0405146",
    archivePrefix = "arXiv",
    reportNumber = "HUTP-04-A020, CALT-68-2501",
    doi = "10.1103/PhysRevD.70.106007",
    journal = "Phys. Rev. D",
    volume = "70",
    pages = "106007",
    year = "2004"
}

@article{Hama:2011ea,
    author = "Hama, Naofumi and Hosomichi, Kazuo and Lee, Sungjay",
    title = "{SUSY Gauge Theories on Squashed Three-Spheres}",
    eprint = "1102.4716",
    archivePrefix = "arXiv",
    primaryClass = "hep-th",
    reportNumber = "DAMTP-2011-6, YITP-11-1",
    doi = "10.1007/JHEP05(2011)014",
    journal = "JHEP",
    volume = "05",
    pages = "014",
    year = "2011"
}

@article{Hama:2010av,
    author = "Hama, Naofumi and Hosomichi, Kazuo and Lee, Sungjay",
    title = "{Notes on SUSY Gauge Theories on Three-Sphere}",
    eprint = "1012.3512",
    archivePrefix = "arXiv",
    primaryClass = "hep-th",
    reportNumber = "DAMTP-2010-129, YITP-10-100",
    doi = "10.1007/JHEP03(2011)127",
    journal = "JHEP",
    volume = "03",
    pages = "127",
    year = "2011"
}

@article{Kapustin:2009kz,
    author = "Kapustin, Anton and Willett, Brian and Yaakov, Itamar",
    title = "{Exact Results for Wilson Loops in Superconformal Chern-Simons Theories with Matter}",
    eprint = "0909.4559",
    archivePrefix = "arXiv",
    primaryClass = "hep-th",
    reportNumber = "CALT-68-2750",
    doi = "10.1007/JHEP03(2010)089",
    journal = "JHEP",
    volume = "03",
    pages = "089",
    year = "2010"
}

@article{ArabiArdehali:2019orz,
    author = "Arabi Ardehali, Arash and Hong, Junho and Liu, James T.",
    title = "{Asymptotic growth of the 4d $ \mathcal{N} $ = 4 index and partially deconfined phases}",
    eprint = "1912.04169",
    archivePrefix = "arXiv",
    primaryClass = "hep-th",
    reportNumber = "LCTP 19-32",
    doi = "10.1007/JHEP07(2020)073",
    journal = "JHEP",
    volume = "07",
    pages = "073",
    year = "2020"
}

@article{Bobev:2020zov,
    author = "Bobev, Nikolay and Charles, Anthony M. and Gang, Dongmin and Hristov, Kiril and Reys, Valentin",
    title = "{Higher-derivative supergravity, wrapped M5-branes, and theories of class $ \mathrm{\mathcal{R}} $}",
    eprint = "2011.05971",
    archivePrefix = "arXiv",
    primaryClass = "hep-th",
    doi = "10.1007/JHEP04(2021)058",
    journal = "JHEP",
    volume = "04",
    pages = "058",
    year = "2021"
}

@article{Bobev:2024mqw,
    author = "Bobev, Nikolay and Choi, Sunjin and Hong, Junho and Reys, Valentin",
    title = "{Superconformal indices of 3d $ \mathcal{N} $ = 2 SCFTs and holography}",
    eprint = "2407.13177",
    archivePrefix = "arXiv",
    primaryClass = "hep-th",
    doi = "10.1007/JHEP10(2024)121",
    journal = "JHEP",
    volume = "10",
    pages = "121",
    year = "2024"
}

@article{Bobev:2025ltz,
    author = "Bobev, Nikolay and De Smet, Pieter-Jan and Hong, Junho and Reys, Valentin and Zhang, Xuao",
    title = "{An Airy tale at large N}",
    eprint = "2502.04606",
    archivePrefix = "arXiv",
    primaryClass = "hep-th",
    doi = "10.1007/JHEP07(2025)123",
    journal = "JHEP",
    volume = "07",
    pages = "123",
    year = "2025"
}

@article{Cassia:2025aus,
    author = "Cassia, Luca and Hristov, Kiril",
    title = "{Constant maps in equivariant topological strings and geometric modeling of fluxes}",
    eprint = "2502.20444",
    archivePrefix = "arXiv",
    primaryClass = "hep-th",
    doi = "10.1088/1751-8121/ae22ac",
    journal = "J. Phys. A",
    volume = "58",
    number = "49",
    pages = "495201",
    year = "2025"
}

@article{Ardehali:2021irq,
    author = "Ardehali, Arash Arabi and Hong, Junho",
    title = "{Decomposition of BPS moduli spaces and asymptotics of supersymmetric partition functions}",
    eprint = "2110.01538",
    archivePrefix = "arXiv",
    primaryClass = "hep-th",
    doi = "10.1007/JHEP01(2022)062",
    journal = "JHEP",
    volume = "01",
    pages = "062",
    year = "2022"
}

@article{Hristov:2021zai,
    author = "Hristov, Kiril and Reys, Valentin",
    title = "{Factorization of log-corrections in AdS$_{4}$/CFT$_{3}$ from supergravity localization}",
    eprint = "2107.12398",
    archivePrefix = "arXiv",
    primaryClass = "hep-th",
    doi = "10.1007/JHEP12(2021)031",
    journal = "JHEP",
    volume = "12",
    pages = "031",
    year = "2021"
}

@article{Bobev:2023dwx,
    author = "Bobev, Nikolay and David, Marina and Hong, Junho and Reys, Valentin and Zhang, Xuao",
    title = "{A compendium of logarithmic corrections in AdS/CFT}",
    eprint = "2312.08909",
    archivePrefix = "arXiv",
    primaryClass = "hep-th",
    doi = "10.1007/JHEP04(2024)020",
    journal = "JHEP",
    volume = "04",
    pages = "020",
    year = "2024"
}

@article{Cassani:2021fyv,
    author = "Cassani, Davide and Komargodski, Zohar",
    title = "{EFT and the SUSY Index on the 2nd Sheet}",
    eprint = "2104.01464",
    archivePrefix = "arXiv",
    primaryClass = "hep-th",
    doi = "10.21468/SciPostPhys.11.1.004",
    journal = "SciPost Phys.",
    volume = "11",
    pages = "004",
    year = "2021"
}

@article{ArabiArdehali:2021nsx,
    author = "Arabi Ardehali, Arash and Murthy, Sameer",
    title = "{The 4d superconformal index near roots of unity and 3d Chern-Simons theory}",
    eprint = "2104.02051",
    archivePrefix = "arXiv",
    primaryClass = "hep-th",
    doi = "10.1007/JHEP10(2021)207",
    journal = "JHEP",
    volume = "10",
    pages = "207",
    year = "2021"
}

@article{Hristov:2022lcw,
    author = "Hristov, Kiril",
    title = "{ABJM at finite N via 4d supergravity}",
    eprint = "2204.02992",
    archivePrefix = "arXiv",
    primaryClass = "hep-th",
    doi = "10.1007/JHEP10(2022)190",
    journal = "JHEP",
    volume = "10",
    pages = "190",
    year = "2022"
}

@inproceedings{Hristov:2024cgj,
    author = "Hristov, Kiril",
    title = "{Equivariant localization and gluing rules in 4d $\mathcal{N}=2$ higher derivative supergravity}",
    eprint = "2406.18648",
    archivePrefix = "arXiv",
    primaryClass = "hep-th",
    month = "6",
    year = "2024"
}

@article{Hong:2024uns,
	author = "Hong, Junho",
	title = "{Perturbatively exact supersymmetric partition functions of ABJM theory on Seifert manifolds and holography}",
	eprint = "2411.09006",
	archivePrefix = "arXiv",
	primaryClass = "hep-th",
	doi = "10.1007/JHEP01(2025)194",
	journal = "JHEP",
	volume = "01",
	pages = "194",
	year = "2025"
}

@article{Nian:2019pxj,
	author = "Nian, Jun and Pando Zayas, Leopoldo A.",
	title = "{Microscopic entropy of rotating electrically charged AdS$_{4}$ black holes from field theory localization}",
	eprint = "1909.07943",
	archivePrefix = "arXiv",
	primaryClass = "hep-th",
	reportNumber = "LCTP-19-22",
	doi = "10.1007/JHEP03(2020)081",
	journal = "JHEP",
	volume = "03",
	pages = "081",
	year = "2020"
}

@article{Grassi:2014vwa,
	author = "Grassi, Alba and Marino, Marcos",
	title = "{M-theoretic matrix models}",
	eprint = "1403.4276",
	archivePrefix = "arXiv",
	primaryClass = "hep-th",
	doi = "10.1007/JHEP02(2015)115",
	journal = "JHEP",
	volume = "02",
	pages = "115",
	year = "2015"
}

@article{Mezei:2013gqa,
	author = "Mezei, M\'ark and Pufu, Silviu S.",
	title = "{Three-sphere free energy for classical gauge groups}",
	eprint = "1312.0920",
	archivePrefix = "arXiv",
	primaryClass = "hep-th",
	reportNumber = "MIT-CTP-4520",
	doi = "10.1007/JHEP02(2014)037",
	journal = "JHEP",
	volume = "02",
	pages = "037",
	year = "2014"
}

@article{Aharony:2008ug,
	author = "Aharony, Ofer and Bergman, Oren and Jafferis, Daniel Louis and Maldacena, Juan",
	title = "{N=6 superconformal Chern-Simons-matter theories, M2-branes and their gravity duals}",
	eprint = "0806.1218",
	archivePrefix = "arXiv",
	primaryClass = "hep-th",
	reportNumber = "WIS-12-08-JUN-DPP",
	doi = "10.1088/1126-6708/2008/10/091",
	journal = "JHEP",
	volume = "10",
	pages = "091",
	year = "2008"
}

@article{Choi:2018hmj,
	author = "Choi, Sunjin and Kim, Joonho and Kim, Seok and Nahmgoong, June",
	title = "{Large AdS black holes from QFT}",
	eprint = "1810.12067",
	archivePrefix = "arXiv",
	primaryClass = "hep-th",
	reportNumber = "SNUTP18-005, KIAS-P18097",
	month = "10",
	year = "2018"
}

@article{Hosseini:2021mnn,
	author = "Hosseini, Seyed Morteza and Yaakov, Itamar and Zaffaroni, Alberto",
	title = "{The joy of factorization at large N: five-dimensional indices and AdS black holes}",
	eprint = "2111.03069",
	archivePrefix = "arXiv",
	primaryClass = "hep-th",
	doi = "10.1007/JHEP02(2022)097",
	journal = "JHEP",
	volume = "02",
	pages = "097",
	year = "2022"
}

@article{Hosseini:2019iad,
	author = "Hosseini, Seyed Morteza and Hristov, Kiril and Zaffaroni, Alberto",
	title = "{Gluing gravitational blocks for AdS black holes}",
	eprint = "1909.10550",
	archivePrefix = "arXiv",
	primaryClass = "hep-th",
	reportNumber = "IPMU19-0132",
	doi = "10.1007/JHEP12(2019)168",
	journal = "JHEP",
	volume = "12",
	pages = "168",
	year = "2019"
}

@article{Pasquetti:2016dyl,
	author = "Pasquetti, Sara",
	title = "{Holomorphic blocks and the 5d AGT correspondence}",
	eprint = "1608.02968",
	archivePrefix = "arXiv",
	primaryClass = "hep-th",
	doi = "10.1088/1751-8121/aa60fe",
	journal = "J. Phys. A",
	volume = "50",
	number = "44",
	pages = "443016",
	year = "2017"
}

@article{Gautason:2023igo,
	author = "Gautason, Fridrik Freyr and Puletti, Valentina Giangreco M. and van Muiden, Jesse",
	title = "{Quantized strings and instantons in holography}",
	eprint = "2304.12340",
	archivePrefix = "arXiv",
	primaryClass = "hep-th",
	doi = "10.1007/JHEP08(2023)218",
	journal = "JHEP",
	volume = "08",
	pages = "218",
	year = "2023"
}

@article{Drukker:2010nc,
	author = "Drukker, Nadav and Marino, Marcos and Putrov, Pavel",
	title = "{From weak to strong coupling in ABJM theory}",
	eprint = "1007.3837",
	archivePrefix = "arXiv",
	primaryClass = "hep-th",
	reportNumber = "HU-EP-10-39",
	doi = "10.1007/s00220-011-1253-6",
	journal = "Commun. Math. Phys.",
	volume = "306",
	pages = "511--563",
	year = "2011"
}

@article{Liu:2017vbl,
	author = "Liu, James T. and Pando Zayas, Leopoldo A. and Rathee, Vimal and Zhao, Wenli",
	title = "{One-Loop Test of Quantum Black Holes in anti\textendash{}de Sitter Space}",
	eprint = "1711.01076",
	archivePrefix = "arXiv",
	primaryClass = "hep-th",
	reportNumber = "LCTP-17-02",
	doi = "10.1103/PhysRevLett.120.221602",
	journal = "Phys. Rev. Lett.",
	volume = "120",
	number = "22",
	pages = "221602",
	year = "2018"
}

@article{Bhattacharyya:2012ye,
	author = "Bhattacharyya, Sayantani and Grassi, Alba and Marino, Marcos and Sen, Ashoke",
	title = "{A One-Loop Test of Quantum Supergravity}",
	eprint = "1210.6057",
	archivePrefix = "arXiv",
	primaryClass = "hep-th",
	doi = "10.1088/0264-9381/31/1/015012",
	journal = "Class. Quant. Grav.",
	volume = "31",
	pages = "015012",
	year = "2014"
}

@article{Seiberg:1996bd,
	author = "Seiberg, Nathan",
	title = "{Five-dimensional SUSY field theories, nontrivial fixed points and string dynamics}",
	eprint = "hep-th/9608111",
	archivePrefix = "arXiv",
	reportNumber = "RU-96-69",
	doi = "10.1016/S0370-2693(96)01215-4",
	journal = "Phys. Lett. B",
	volume = "388",
	pages = "753--760",
	year = "1996"
}

@article{Closset:2019hyt,
	author = "Closset, Cyril and Kim, Heeyeon",
	title = "{Three-dimensional $\mathcal{N}=2$ supersymmetric gauge theories and partition functions on Seifert manifolds: A review}",
	eprint = "1908.08875",
	archivePrefix = "arXiv",
	primaryClass = "hep-th",
	doi = "10.1142/S0217751X19300114",
	journal = "Int. J. Mod. Phys. A",
	volume = "34",
	number = "23",
	pages = "1930011",
	year = "2019"
}

@article{Closset:2018ghr,
	author = "Closset, Cyril and Kim, Heeyeon and Willett, Brian",
	title = "{Seifert fibering operators in 3d $\mathcal{N}=2$ theories}",
	eprint = "1807.02328",
	archivePrefix = "arXiv",
	primaryClass = "hep-th",
	reportNumber = "CERN-TH-2018-156",
	doi = "10.1007/JHEP11(2018)004",
	journal = "JHEP",
	volume = "11",
	pages = "004",
	year = "2018"
}

@article{Benini:2016hjo,
	author = "Benini, Francesco and Zaffaroni, Alberto",
	editor = "Li, Si and Lian, Bong H. and Song, Wei and Yau, Shing-Tung",
	title = "{Supersymmetric partition functions on Riemann surfaces}",
	eprint = "1605.06120",
	archivePrefix = "arXiv",
	primaryClass = "hep-th",
	reportNumber = "SISSA-28-2016-FISI",
	journal = "Proc. Symp. Pure Math.",
	volume = "96",
	pages = "13--46",
	year = "2017"
}

@article{Hristov:2019mqp,
	author = "Hristov, Kiril and Katmadas, Stefanos and Toldo, Chiara",
	title = "{Matter-coupled supersymmetric Kerr-Newman-AdS$_4$ black holes}",
	eprint = "1907.05192",
	archivePrefix = "arXiv",
	primaryClass = "hep-th",
	doi = "10.1103/PhysRevD.100.066016",
	journal = "Phys. Rev. D",
	volume = "100",
	number = "6",
	pages = "066016",
	year = "2019"
}

@article{Choi:2018fdc,
	author = "Choi, Sunjin and Hwang, Chiung and Kim, Seok and Nahmgoong, June",
	title = "{Entropy Functions of BPS Black Holes in AdS$_{4}$ and AdS$_{6}$}",
	eprint = "1811.02158",
	archivePrefix = "arXiv",
	primaryClass = "hep-th",
	reportNumber = "SNUTP18-006, KIAS-P18076",
	doi = "10.3938/jkps.76.101",
	journal = "J. Korean Phys. Soc.",
	volume = "76",
	number = "2",
	pages = "101--108",
	year = "2020"
}

@article{BenettiGenolini:2023rkq,
	author = "Benetti Genolini, Pietro and Cabo-Bizet, Alejandro and Murthy, Sameer",
	title = "{Supersymmetric phases of AdS$_{4}$/CFT$_{3}$}",
	eprint = "2301.00763",
	archivePrefix = "arXiv",
	primaryClass = "hep-th",
	doi = "10.1007/JHEP06(2023)125",
	journal = "JHEP",
	volume = "06",
	pages = "125",
	year = "2023"
}

@article{GonzalezLezcano:2022hcf,
	author = "Gonz\'alez Lezcano, Alfredo and Jerdee, Maximilian and Pando Zayas, Leopoldo A.",
	title = "{Cardy expansion of 3d superconformal indices and corrections to the dual black hole entropy}",
	eprint = "2210.12065",
	archivePrefix = "arXiv",
	primaryClass = "hep-th",
	reportNumber = "LCTP-22-11",
	doi = "10.1007/JHEP01(2023)044",
	journal = "JHEP",
	volume = "01",
	pages = "044",
	year = "2023"
}

@article{Bobev:2019zmz,
	author = "Bobev, Nikolay and Crichigno, P. Marcos",
	title = "{Universal spinning black holes and theories of class $ \mathcal{R} $}",
	eprint = "1909.05873",
	archivePrefix = "arXiv",
	primaryClass = "hep-th",
	doi = "10.1007/JHEP12(2019)054",
	journal = "JHEP",
	volume = "12",
	pages = "054",
	year = "2019"
}

@article{Cassani:2019mms,
	author = "Cassani, Davide and Papini, Lorenzo",
	title = "{The BPS limit of rotating AdS black hole thermodynamics}",
	eprint = "1906.10148",
	archivePrefix = "arXiv",
	primaryClass = "hep-th",
	doi = "10.1007/JHEP09(2019)079",
	journal = "JHEP",
	volume = "09",
	pages = "079",
	year = "2019"
}

@article{Gauntlett:2007ma,
	author = "Gauntlett, Jerome P. and Varela, Oscar",
	title = "{Consistent Kaluza-Klein reductions for general supersymmetric AdS solutions}",
	eprint = "0707.2315",
	archivePrefix = "arXiv",
	primaryClass = "hep-th",
	doi = "10.1103/PhysRevD.76.126007",
	journal = "Phys. Rev. D",
	volume = "76",
	pages = "126007",
	year = "2007"
}

@article{Bobev:2017uzs,
	author = "Bobev, Nikolay and Crichigno, P. Marcos",
	title = "{Universal RG Flows Across Dimensions and Holography}",
	eprint = "1708.05052",
	archivePrefix = "arXiv",
	primaryClass = "hep-th",
	doi = "10.1007/JHEP12(2017)065",
	journal = "JHEP",
	volume = "12",
	pages = "065",
	year = "2017"
}

@article{Azzurli:2017kxo,
	author = "Azzurli, Francesco and Bobev, Nikolay and Crichigno, P. Marcos and Min, Vincent S. and Zaffaroni, Alberto",
	title = "{A universal counting of black hole microstates in AdS$_{4}$}",
	eprint = "1707.04257",
	archivePrefix = "arXiv",
	primaryClass = "hep-th",
	doi = "10.1007/JHEP02(2018)054",
	journal = "JHEP",
	volume = "02",
	pages = "054",
	year = "2018"
}

@article{Bobev:2020egg,
	author = "Bobev, Nikolay and Charles, Anthony M. and Hristov, Kiril and Reys, Valentin",
	title = "{The Unreasonable Effectiveness of Higher-Derivative Supergravity in AdS$_4$ Holography}",
	eprint = "2006.09390",
	archivePrefix = "arXiv",
	primaryClass = "hep-th",
	doi = "10.1103/PhysRevLett.125.131601",
	journal = "Phys. Rev. Lett.",
	volume = "125",
	number = "13",
	pages = "131601",
	year = "2020"
}

@article{Beem:2012mb,
	author = "Beem, Christopher and Dimofte, Tudor and Pasquetti, Sara",
	title = "{Holomorphic Blocks in Three Dimensions}",
	eprint = "1211.1986",
	archivePrefix = "arXiv",
	primaryClass = "hep-th",
	reportNumber = "DMUS-MP-12-08, DMUS-MP-12/08",
	doi = "10.1007/JHEP12(2014)177",
	journal = "JHEP",
	volume = "12",
	pages = "177",
	year = "2014"
}

@article{Bobev:2022jte,
	author = "Bobev, Nikolay and Hong, Junho and Reys, Valentin",
	title = "{Large N Partition Functions, Holography, and Black Holes}",
	eprint = "2203.14981",
	archivePrefix = "arXiv",
	primaryClass = "hep-th",
	doi = "10.1103/PhysRevLett.129.041602",
	journal = "Phys. Rev. Lett.",
	volume = "129",
	number = "4",
	pages = "041602",
	year = "2022"
}

@article{Closset:2016arn,
	author = "Closset, Cyril and Kim, Heeyeon",
	title = "{Comments on twisted indices in 3d supersymmetric gauge theories}",
	eprint = "1605.06531",
	archivePrefix = "arXiv",
	primaryClass = "hep-th",
	doi = "10.1007/JHEP08(2016)059",
	journal = "JHEP",
	volume = "08",
	pages = "059",
	year = "2016"
}

@article{Kim:2009wb,
	author = "Kim, Seok",
	title = "{The Complete superconformal index for N=6 Chern-Simons theory}",
	eprint = "0903.4172",
	archivePrefix = "arXiv",
	primaryClass = "hep-th",
	reportNumber = "IMPERIAL-TP-09-SK-01",
	doi = "10.1016/j.nuclphysb.2009.06.025",
	journal = "Nucl. Phys. B",
	volume = "821",
	pages = "241--284",
	year = "2009",
	note = "[Erratum: Nucl.Phys.B 864, 884 (2012)]"
}

@article{Bhattacharya:2008bja,
	author = "Bhattacharya, Jyotirmoy and Minwalla, Shiraz",
	title = "{Superconformal Indices for N = 6 Chern Simons Theories}",
	eprint = "0806.3251",
	archivePrefix = "arXiv",
	primaryClass = "hep-th",
	doi = "10.1088/1126-6708/2009/01/014",
	journal = "JHEP",
	volume = "01",
	pages = "014",
	year = "2009"
}

@article{Bhattacharya:2008zy,
	author = "Bhattacharya, Jyotirmoy and Bhattacharyya, Sayantani and Minwalla, Shiraz and Raju, Suvrat",
	title = "{Indices for Superconformal Field Theories in 3,5 and 6 Dimensions}",
	eprint = "0801.1435",
	archivePrefix = "arXiv",
	primaryClass = "hep-th",
	reportNumber = "TIFR-TH-08-01, HUTP-08-A0001",
	doi = "10.1088/1126-6708/2008/02/064",
	journal = "JHEP",
	volume = "02",
	pages = "064",
	year = "2008"
}

@article{Hwang:2012jh,
	author = "Hwang, Chiung and Kim, Hee-Cheol and Park, Jaemo",
	title = "{Factorization of the 3d superconformal index}",
	eprint = "1211.6023",
	archivePrefix = "arXiv",
	primaryClass = "hep-th",
	doi = "10.1007/JHEP08(2014)018",
	journal = "JHEP",
	volume = "08",
	pages = "018",
	year = "2014"
}

@article{Closset:2017zgf,
	author = "Closset, Cyril and Kim, Heeyeon and Willett, Brian",
	title = "{Supersymmetric partition functions and the three-dimensional A-twist}",
	eprint = "1701.03171",
	archivePrefix = "arXiv",
	primaryClass = "hep-th",
	reportNumber = "CERN-TH-2017-006",
	doi = "10.1007/JHEP03(2017)074",
	journal = "JHEP",
	volume = "03",
	pages = "074",
	year = "2017"
}

@article{Benini:2015noa,
	author = "Benini, Francesco and Zaffaroni, Alberto",
	title = "{A topologically twisted index for three-dimensional supersymmetric theories}",
	eprint = "1504.03698",
	archivePrefix = "arXiv",
	primaryClass = "hep-th",
	reportNumber = "IMPERIAL-TP-2015-FB-01",
	doi = "10.1007/JHEP07(2015)127",
	journal = "JHEP",
	volume = "07",
	pages = "127",
	year = "2015"
}

@article{Aharony:2013kma,
	author = "Aharony, Ofer and Razamat, Shlomo S. and Seiberg, Nathan and Willett, Brian",
	title = "{3$d$ dualities from 4$d$ dualities for orthogonal groups}",
	eprint = "1307.0511",
	archivePrefix = "arXiv",
	primaryClass = "hep-th",
	reportNumber = "WIS-07-13-JUN-DPPA",
	doi = "10.1007/JHEP08(2013)099",
	journal = "JHEP",
	volume = "08",
	pages = "099",
	year = "2013"
}

@article{Bobev:2022eus,
	author = "Bobev, Nikolay and Hong, Junho and Reys, Valentin",
	title = "{Large N partition functions of the ABJM theory}",
	eprint = "2210.09318",
	archivePrefix = "arXiv",
	primaryClass = "hep-th",
	doi = "10.1007/JHEP02(2023)020",
	journal = "JHEP",
	volume = "02",
	pages = "020",
	year = "2023"
}

@article{Aharony:2013dha,
	author = "Aharony, Ofer and Razamat, Shlomo S. and Seiberg, Nathan and Willett, Brian",
	title = "{3d dualities from 4d dualities}",
	eprint = "1305.3924",
	archivePrefix = "arXiv",
	primaryClass = "hep-th",
	reportNumber = "WIS-04-13-APR-DPPA",
	doi = "10.1007/JHEP07(2013)149",
	journal = "JHEP",
	volume = "07",
	pages = "149",
	year = "2013"
}

@article{Kapustin:2011jm,
	author = "Kapustin, Anton and Willett, Brian",
	title = "{Generalized Superconformal Index for Three Dimensional Field Theories}",
	eprint = "1106.2484",
	archivePrefix = "arXiv",
	primaryClass = "hep-th",
	reportNumber = "68-2840",
	month = "6",
	year = "2011"
}

@article{Krattenthaler:2011da,
	author = "Krattenthaler, C. and Spiridonov, V. P. and Vartanov, G. S.",
	title = "{Superconformal indices of three-dimensional theories related by mirror symmetry}",
	eprint = "1103.4075",
	archivePrefix = "arXiv",
	primaryClass = "hep-th",
	reportNumber = "AEI-2011-015",
	doi = "10.1007/JHEP06(2011)008",
	journal = "JHEP",
	volume = "06",
	pages = "008",
	year = "2011"
}

@article{Imamura:2011uj,
	author = "Imamura, Yosuke and Yokoyama, Daisuke and Yokoyama, Shuichi",
	title = "{Superconformal index for large N quiver Chern-Simons theories}",
	eprint = "1102.0621",
	archivePrefix = "arXiv",
	primaryClass = "hep-th",
	reportNumber = "UT-11-03, TIT-HEP-608",
	doi = "10.1007/JHEP08(2011)011",
	journal = "JHEP",
	volume = "08",
	pages = "011",
	year = "2011"
}

@article{Imamura:2011su,
	author = "Imamura, Yosuke and Yokoyama, Shuichi",
	title = "{Index for three dimensional superconformal field theories with general R-charge assignments}",
	eprint = "1101.0557",
	archivePrefix = "arXiv",
	primaryClass = "hep-th",
	reportNumber = "UT-11-01, TIT-HEP-607",
	doi = "10.1007/JHEP04(2011)007",
	journal = "JHEP",
	volume = "04",
	pages = "007",
	year = "2011"
}

@article{Bobev:2023lkx,
	author = "Bobev, Nikolay and Hong, Junho and Reys, Valentin",
	title = "{Large N partition functions of 3d holographic SCFTs}",
	eprint = "2304.01734",
	archivePrefix = "arXiv",
	primaryClass = "hep-th",
	doi = "10.1007/JHEP08(2023)119",
	journal = "JHEP",
	volume = "08",
	pages = "119",
	year = "2023"
}

@article{PandoZayas:2020iqr,
	author = "Pando Zayas, Leopoldo A. and Xin, Yu",
	title = "{Universal logarithmic behavior in microstate counting and the dual one-loop entropy of $AdS_4$ black holes}",
	eprint = "2008.03239",
	archivePrefix = "arXiv",
	primaryClass = "hep-th",
	reportNumber = "LCTP-20-18",
	doi = "10.1103/PhysRevD.103.026003",
	journal = "Phys. Rev. D",
	volume = "103",
	number = "2",
	pages = "026003",
	year = "2021"
}

@article{Bobev:2022wem,
	author = "Bobev, Nikolay and Choi, Sunjin and Hong, Junho and Reys, Valentin",
	title = "{Large N superconformal indices for 3d holographic SCFTs}",
	eprint = "2210.15326",
	archivePrefix = "arXiv",
	primaryClass = "hep-th",
	reportNumber = "KIAS-P22068",
	doi = "10.1007/JHEP02(2023)027",
	journal = "JHEP",
	volume = "02",
	pages = "027",
	year = "2023"
}

@article{Pasquetti:2019uop,
	author = "Pasquetti, Sara and Sacchi, Matteo",
	title = "{From 3$d$ dualities to 2$d$ free field correlators and back}",
	eprint = "1903.10817",
	archivePrefix = "arXiv",
	primaryClass = "hep-th",
	doi = "10.1007/JHEP11(2019)081",
	journal = "JHEP",
	volume = "11",
	pages = "081",
	year = "2019"
}

@article{Bobev:2021oku,
	author = "Bobev, Nikolay and Charles, Anthony M. and Hristov, Kiril and Reys, Valentin",
	title = "{Higher-derivative supergravity, AdS$_{4}$ holography, and black holes}",
	eprint = "2106.04581",
	archivePrefix = "arXiv",
	primaryClass = "hep-th",
	doi = "10.1007/JHEP08(2021)173",
	journal = "JHEP",
	volume = "08",
	pages = "173",
	year = "2021"
}

@article{Benini:2015eyy,
	author = "Benini, Francesco and Hristov, Kiril and Zaffaroni, Alberto",
	title = "{Black hole microstates in AdS$_{4}$ from supersymmetric localization}",
	eprint = "1511.04085",
	archivePrefix = "arXiv",
	primaryClass = "hep-th",
	reportNumber = "IMPERIAL-TP-2015-FB-03",
	doi = "10.1007/JHEP05(2016)054",
	journal = "JHEP",
	volume = "05",
	pages = "054",
	year = "2016"
}

@article{Hristov:2021qsw,
	author = "Hristov, Kiril",
	title = "{4d $ \mathcal{N} $ = 2 supergravity observables from Nekrasov-like partition functions}",
	eprint = "2111.06903",
	archivePrefix = "arXiv",
	primaryClass = "hep-th",
	doi = "10.1007/JHEP02(2022)079",
	journal = "JHEP",
	volume = "02",
	pages = "079",
	year = "2022"
}

@article{Choi:2019dfu,
	author = "Choi, Sunjin and Hwang, Chiung",
	title = "{Universal 3d Cardy Block and Black Hole Entropy}",
	eprint = "1911.01448",
	archivePrefix = "arXiv",
	primaryClass = "hep-th",
	reportNumber = "SNUTP19-004",
	doi = "10.1007/JHEP03(2020)068",
	journal = "JHEP",
	volume = "03",
	pages = "068",
	year = "2020"
}

@article{Choi:2019zpz,
    author = "Choi, Sunjin and Hwang, Chiung and Kim, Seok",
    title = "{Quantum vortices, M2-branes and black holes}",
    eprint = "1908.02470",
    archivePrefix = "arXiv",
    primaryClass = "hep-th",
    reportNumber = "SNUTP18-003",
    doi = "10.1007/JHEP09(2024)096",
    journal = "JHEP",
    volume = "09",
    pages = "096",
    year = "2024"
}
\bibliographystyle{JHEP}

%%%%%%%%%%%%%%%%%%%%%%%%%%%%

\end{document}